\documentclass[twocolumn]{aastex63}

\usepackage[utf8]{inputenc}
\usepackage[T1]{fontenc}
\usepackage{ae,aecompl}

\usepackage{natbib}
\usepackage{float}
\usepackage{graphicx}
\usepackage{amssymb}
\usepackage{rotating,times,pictex,graphicx,latexsym}
\usepackage{color}
\usepackage{amsmath}
\usepackage{threeparttable}
\usepackage{lipsum}

\usepackage[all]{hypcap}
\usepackage[title,titletoc]{appendix}
\usepackage[]{hyperref}
\PassOptionsToPackage{pdfpagelabels=false}{hyperref} 
\usepackage[T1]{fontenc}

\newcommand{\footremember}

\newcommand{\zp}{$z_{\rm {P1}}$}
\newcommand{\ip}{$i_{\rm {P1}}$}

\newcommand{\msun}{\mbox{M$_{\sun}$}}
\newcommand{\lsun}{\mbox{$L_{\sun}$}}

\newcommand{\hal}{\mbox{{\rm H$\alpha$}~}}

\newcommand{\arcs}{\hbox{arcsec}}
\newcommand{\arcm}{\hbox{arcmin}}
\newcommand{\into}{\mbox{$\times$}}

\newcommand{\ra}{\mbox{$\alpha_{2000}$}}
\newcommand{\dec}{\mbox{$\delta_{2000}$}}

\newcommand{\degree}{\mbox{$^{\circ}$}}
\newcommand{\av}{\mbox{$A_{\it V}$~}}

\newcommand{\hii}{\mbox{H~{\sc ii}~}}
\newcommand{\heii}{\mbox{He~{\sc ii}~}}

\newcommand{\tco}{\hbox{$^{13}$CO}~}

\newcommand{\mum}{\hbox{$\mu$m}~}

\newcommand{\q}{\hbox{$\sim$}}

\newcommand{\kms}{\hbox{km~s$^{-1}$}}

\newcommand{\vlsr}{\hbox{V$_{\rm LSR}$}}

\graphicspath{{./}{fig/}}

\revised{X, 2021}
\accepted{Y, 2021}

\submitjournal{ApJ}

\shorttitle{Star formation activities around IRAS 05100+3723}
\shortauthors{Yadav et al.}

\begin{document}

\title{A Comprehensive Study of the Young Cluster IRAS 05100+3723: Properties, Surrounding Interstellar Matter, and Associated Star Formation}

\correspondingauthor{R. K. Yadav}
\email{ram$\_$kesh@narit.or.th}

\author[0000-0002-6740-7425]{R. K. Yadav}
\affiliation{National Astronomical Research Institute of Thailand (NARIT), Sirindhorn AstroPark, 260 Moo 4, T. Donkaew, A. Maerim, Chiangmai 50180, Thailand}

\author[0000-0002-9431-6297]{M. R. Samal}
\affiliation{Physical Research Laboratory, Navrangpura, Ahmedabad, Gujarat 380009, India}

\author[0000-0002-1912-1342]{E. Semenko}
\affiliation{National Astronomical Research Institute of Thailand (NARIT), Sirindhorn AstroPark, 260 Moo 4, T. Donkaew, A. Maerim, Chiangmai 50180, Thailand}

\author[0000-0001-9509-7316]{A. Zavagno}
\affiliation{Aix-Marseille Universite, CNRS, CNES, LAM, Marseille, France}
\affiliation{Institut Universitaire de France, Paris, France}

\author[0000-0003-3295-6595]{S. Vaddi}
\affiliation{National Centre for Radio Astrophysics (TIFR), Post Bag 3, Ganeshkhind, Pune 411007, India}

\author[0000-0002-3094-1077]{P. Prajapati}
\affiliation{Physical Research Laboratory, Navrangpura, Ahmedabad, Gujarat 380009, India}

\author[0000-0001-9312-3816]{D.K. Ojha}
\affiliation{Department  of  Astronomy  and  Astrophysics,  Tata  Institute  of  Fundamental  Research,  Homi  Bhabha  Road,  Mumbai  400005, India}

\author{A. K. Pandey}
\altaffiliation{deceased}
\affiliation{Aryabhatta Research Institute of Observational Sciences (ARIES), Nainital 263129, India}

\author{M. Ridsdill-Smith}
\affiliation{National Astronomical Research Institute of Thailand (NARIT), Sirindhorn AstroPark, 260 Moo 4, T. Donkaew, A. Maerim, Chiangmai 50180, Thailand}

\author{J. Jose}
\affiliation{Indian Institute of Science Education and Research (IISER) Tirupati, Rami Reddy Nagar, Karakambadi Road, Mangalam (P.O.), Tirupati 517 507, India}

\author{S. Patra}
\affiliation{Indian Institute of Science Education and Research (IISER) Tirupati, Rami Reddy Nagar, Karakambadi Road, Mangalam (P.O.), Tirupati 517 507, India}

\author[0000-0002-2338-4583]{S. Dutta}
\affiliation{Institute of Astronomy and Astrophysics, Academia Sinica, Taipei 10617, Taiwan}

\author{P. Irawati}
\affiliation{National Astronomical Research Institute of Thailand (NARIT), Sirindhorn AstroPark, 260 Moo 4, T. Donkaew, A. Maerim, Chiangmai 50180, Thailand}

\author[0000-0001-5731-3057]{S. Sharma}
\affiliation{Aryabhatta Research Institute of Observational Sciences (ARIES), Nainital  263129, India}

\author{D. K. Sahu}
\affiliation{Indian Institute of Astrophysics, II Block, Koramangala, Bengaluru 560 034, Karnataka, India}

\author[0000-0002-0151-2361]{N. Panwar}
\affiliation{Aryabhatta Research Institute of Observational Sciences (ARIES), Nainital 263129, India}

\begin{abstract} 

We present a  comprehensive multiwavelength investigation of a likely massive 
young cluster `IRAS 05100+3723' and its environment with the aim to understand 
its formation history and feedback effects. We find that IRAS 05100+3723 is a 
distant ($\sim$3.2 kpc), moderate mass ($\sim$500 \msun), young ($\sim$3 Myr) 
cluster with its most massive star being an O8.5V-type. From spectral modeling, 
we estimate the effective  temperature and log $g$ of the star as $\sim$33,000 K 
and $\sim$3.8, respectively. Our radio continuum observations reveal that the 
star has ionized its environment forming an \hii region of size $\sim$2.7 
pc, temperature $\sim$5,700 K, and electron density  $\sim$165 cm$^{-3}$. 
However, our large-scale dust maps reveal that it has heated the dust up to 
several parsecs ($\sim$10 pc) in the range 17$-$28 K and the morphology of warm 
dust emission resembles a bipolar \hii region. From  dust and \tco gas analyses, 
we find evidences that the formation of the \hii region has occurred at the 
very end of a long filamentary cloud around 3 Myr ago, likely due to edge collapse 
of the filament. We show that the \hii region is currently compressing a clump 
of mass $\sim$2700 \msun~ at its western outskirts, at the junction of 
the \hii region and filament. We observe several 70 \mum point sources 
of intermediate-mass and class 0 nature within the clump. We attribute 
these sources as the second generation stars of the complex.  We propose 
that the star formation in the clump is either induced or being facilitated 
by the compression of the expanding \hii region onto the inflowing filamentary material.
\end{abstract}

\keywords{Stars: formation, pre-main-sequence, protostars, clusters $-$ ISM: molecular clouds, clumps $-$ \hii regions $-$ dust, extinction}

\section{Introduction}\label{intro}

\begin{figure*}
\centering
	\includegraphics[width=\textwidth]{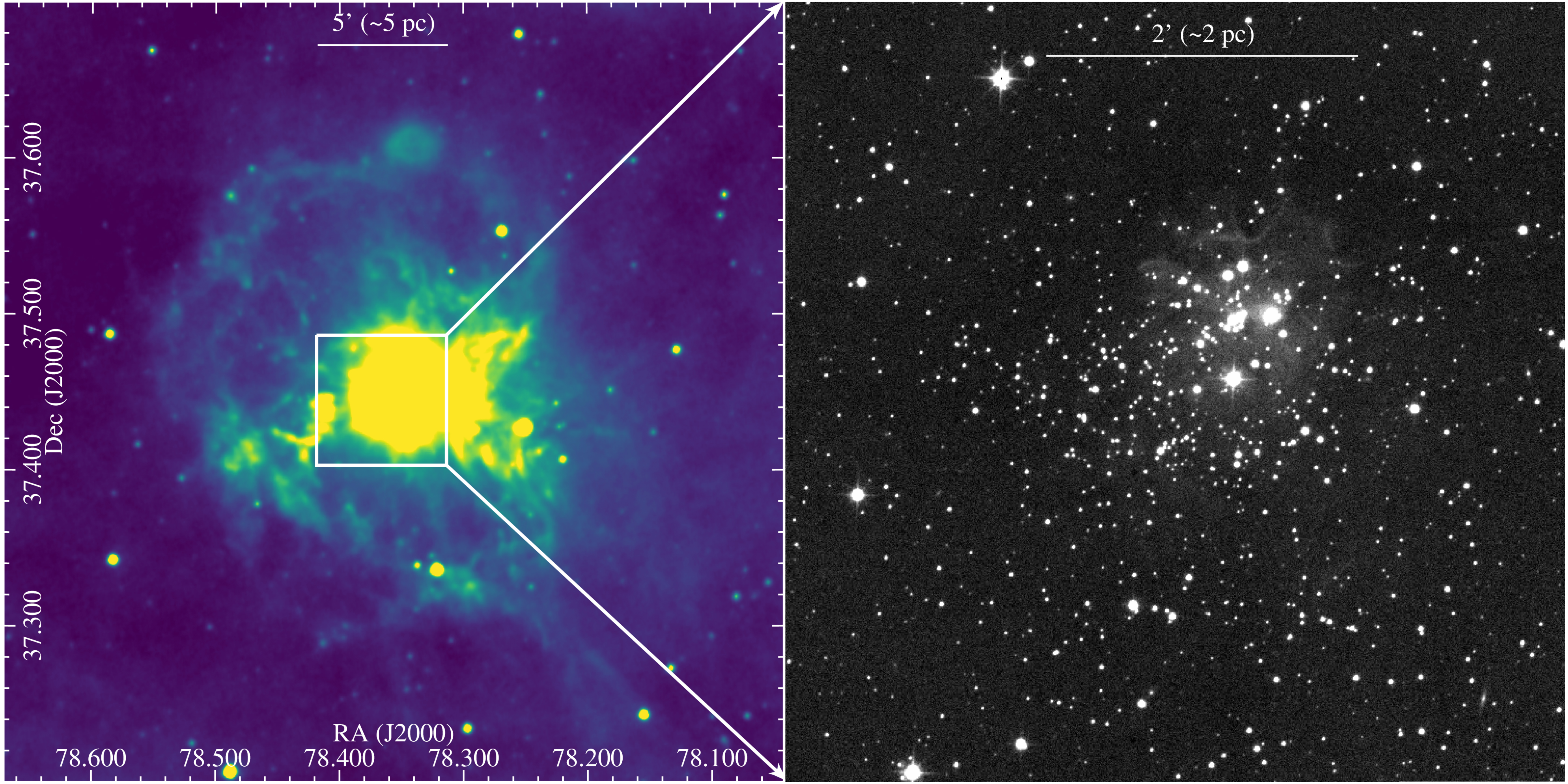}
\caption{The left panel displays the WISE 12 \mum image dust distribution 
around the cluster having an area of 30\into30\, \arcm$^2$, while the right 
panel displays the UKDISS $H$-band image of the central 5\into5\, \arcm$^2$ area.}
\label{fig_hband}
\end{figure*}

 It is believed that most, if not all, stars form  in embedded clusters  \citep{lada03}. 
 In general,  parsec-scale young clusters (age $>$ 2 Myr) have 
 smooth, centrally condensed and nearly spherical structure \citep{asc07,har08},     
 while molecular clouds, in general, have irregular and much extended (up to tens 
 of parsecs) structure \citep{and08,mol10}. In this context, a key question is 
 ``How centrally condensed clusters form from such molecular clouds?'' 
 Although, there are various proposed mechanisms of cluster formation  such as 
 the monolithic collapse of molecular clouds, hierarchical collapse and merger of 
 gas and star(s), and global non-isotropic collapse 
 \citep{long14,ban15,mot18, vaz19}, yet it is unclear that which process plays 
 a dominant role under which circumstances. Moreover, it is not known if 
 other dynamical processes such as two-body relaxation and mass-segregation 
 are also responsible to shape the structure of the newly born clusters  
 \citep{ban17,sills18}. These questions can be addressed by studying young 
 clusters while they are still associated with gas and dust. Another key question  
 related to cluster formation, in particular associated with the formation of 
 massive to intermediate-mass clusters, is whether such clusters play a 
 constructive or destructive role in the future star formation processes of  the 
 host cloud. Because massive members of such clusters, on one hand, can switch 
 off the star formation by blowing off the gas and dust from their immediate 
vicinity, while on the other hand, they can compress the cold parental gas to 
promote the star formation \citep{elm77, deh10} in the cloud. And the latter 
can prolong the star formation activity of a molecular cloud up to several 
million years \citep[e.g.][]{pre08}. In the past, studying the early evolution 
of young clusters had been  an observational challenge as they are buried 
inside molecular clouds. Recent sensitive surveys over the last decade, however, 
have provided a wealth of data in optical to mm domain 
\citep[e.g.][]{jack06, law07, minn10, mol10, pans16, gaia16}. These surveys allow 
 us to better understand the properties of the young clusters, and characterize 
 the physical conditions of the surrounding interstellar medium (ISM). 
 For example, by using parallax and proper motion (PM) measurements obtained 
 from the \textit{Gaia} \citep{gaia16, gaia18} survey data, distance to the 
 star clusters can precisely be estimated \citep[][]{2018AJ....156...58B} 
 and membership of the bright stars can be well constrained. Similarly, using 
 deep infrared data sets from surveys such as UKIDSS (UKIRT Infrared Deep Sky 
 Survey) Galactic  Plane Survey \citep[UKIDSS-GPS;][]{luc08}, {\it Spitzer} 
 Galactic Legacy Infrared Mid-Plane Survey Extraordinaire 
 \citep[GLIMPSE;][]{2009AAS...21421001W} survey and Wide-field Infrared Survey 
  Explorer \citep[WISE;][]{wis10} pre-main-sequence (PMS) members of the young 
   clusters can be identified \citep[e.g.][]{gut09,das21}. Moreover, for 
   moderately extincted nearby clusters, age determination is possible by  
   constructing color-magnitude diagrams (CMD) using data from optical surveys such 
   as the $Gaia$ \citep{gaia18},  the Panoramic Survey Telescope and Rapid 
   Response System \citep[Pan-STARRS1;][]{pans16} and Sloan Digital Sky Survey 
   \citep[SDSS;][]{sds15}. In addition to that, using wide area maps provided 
   by surveys such as {\it Herschel} infrared Galactic Plane Survey 
   \citep[Hi-GAL][]{2010A&A...518L...1P,mol10} and FCRAO  \citep[Five College 
   Radio Astronomical Observatory; ][]{jack06},  connections between 
   large-scale environment and the cluster forming clumps can be made. 

In the present work, we characterize the young cluster IRAS 05100+3723 
\citep{bica03} associated  with the \hii region Sharpless 228 
\citep[hereafter S228,][]{sha59}, also known as LBN 784, and  RAFGL 5137 
(\ra=78.$\!\!$\degree356250, \dec=+37.$\!\!$\degree450278). The \hii region is 
 believed  to be powered by a massive star ALS 19710 of spectral type between 
 O8V to B0V, whose  reddening ($E(B-V)$),  lies in the range 1.2$-$1.3 
 mag  \citep{chini84,hun90}.  We study the physical conditions of dust and gas 
 around the cluster by taking advantage of the key strengths of the 
 aforementioned multiwavelength surveys.  Figure \ref{fig_hband} shows the  
 WISE 12 $\mu$m image around the cluster over 30$\times$30 \arcm$^2$ area as well 
 as the central 5$\times$5 \arcm$^2$ area in the near-infrared (NIR) $H$-band 
 from the UKIDSS-GPS \citep[][]{luc08} wherein the cluster is clearly visible. 
 There are several shallow optical and infrared studies on the cluster suggesting 
that the cluster is located at a distance between 2.2 and 6.8 kpc and its age 
lies somewhere between 1 and 25 Myr \citep{lah85,hun90,kha13,yu18}. To date, the 
most detailed work on the cluster has been carried out by \citet{bor03} using 
NIR observations. Their results suggest that the age of the cluster is $\sim$3 
Myr and its total stellar mass is $\sim$1800~ \msun. This makes the cluster one 
of the potential  young massive clusters in our Galaxy such as Orion Nebula 
Cluster \citep[ONC, mass $\sim$2000 \msun\, and age $\sim$3 Myr, 
see][]{hill97,dar10}, albeit at a farther distance. \citet{bor03} estimated 
the cluster radius to be 1.5 \arcm\,  while \citet{yu18} reported the cluster 
radius to be 2.5 \arcm. 

Intermediate-mass to massive young clusters are rare in our Galaxy. For example, 
 \citet{lada03}, in their study of young ($<$ 3 Myr) embedded clusters within 2 
 kpc of the solar neighborhood, found that the only cluster that has stellar 
 mass $>$ 1000 \msun\, is the ONC. Thus, a cluster like IRAS 05100+3723 serves as 
 a potential candidate for understanding the formation and early evolution of 
  intermediate-mass to massive clusters. Despite the fact that IRAS 05100+3723 is 
  a potentially young massive cluster, its  detailed stellar content, physical properties, 
  and surrounding ISM are scarcely explored. In this work, we examine 
  the stellar, gaseous, and dust components of this likely massive cluster and its 
  environment, with the  aim to understand its formation history and feedback 
  effects on the star formation processes of the host cloud.

We organize this paper as follows. In Section \ref{s28_obse} we present the 
 observations, archival data sets and data reduction procedures. Section  \ref{s28_anal} 
 discusses the stellar content and properties of the cluster, and 
 the physical conditions and kinematics of the large-scale environment. Section  
 \ref{s28_disc} discusses the dynamical status of the cluster, its formation 
 history, and the effect of stellar feedback on the star formation scenario of 
 the complex, followed by a summary in Section \ref{s28_conc}. 

\section{Observations and Data Sets}\label{s28_obse}

\subsection{Optical Spectroscopic Observations}\label{ospec}
We obtained medium-resolution spectra for three bright stars in the S228 
complex using the Medium Resolution Spectrograph (MRES) mounted on the 2.4 m 
Thai National Telescope (TNT). The MRES is a  fiber-fed echelle spectrograph 
designed to work in spectral range of ~3900$-$8800\AA\, with a spectral resolution 
of R $\sim$16,000$-$19,000. This instrument is equipped with a 2048$\times$512 
pixels Andor CCD camera with a pixel size of 13.5 $\mu$m.

 The data were acquired on the night of 2019 March 18 with a lunar illumination 
 of $\sim$80\% during partially  cloudy weather conditions. The log of the 
  observations is given in Table \ref{obs_log}. In addition to the science 
  spectra, we obtained standard calibration frames such as Bias, Flat and Th-Ar 
  lamp. The data reduction was carried out using  the {\sc echelle}  standard 
  package of IRAF. The spectra were extracted using optimal extraction methods. 
  The wavelength calibration of the spectra was done using Th-Ar lamp source. 
  The wavelength calibrated, normalized spectra of the three stars are shown in Figure 
  \ref{final_spec}. We discuss the spectral classification of these stars  
  in Section \ref{s28_ion}. It should be noted that only a part of the spectra, 
  i.e. in the range 4400$-$6800 \AA,\, is being used in this work. The spectra in 
  the range 3900$-$4400 \AA\, are dominated by noise due to poor sensitivity 
in this range and spectra beyond 6800 \AA\, are dominated by multiple telluric 
lines/bands and hence not used in our analysis.

\begin{table}[]\label{obs_log}
    \centering
    \caption{Log of the MRES observations.}
    \begin{tabular}{cccccc}
        \hline
             ID &   \ra  & \dec& Date of    & V & Exposure   \\
                & ($\degree$) & ($\degree$) & observation & (mag) & time (sec)\\
        \hline
        1 & 078.356250 & +37.458222 & 2019-03-18  & 12.6  & 2,700 \\
        2 & 078.356667 & +37.458167 & 2019-03-18  & 13.2  & 3,000 \\
        3 & 078.353333 & +37.453250 & 2019-03-18  & 13.1  & 3,000 \\
    \end{tabular}
\end{table}

\subsection{Radio Continuum Observations}
Radio continuum observations of the S228 region were obtained at 610 and 1280 
MHz using the GMRT array (PI: M.R. Samal, ID: 13MRS01) with the aim to trace 
the ionized gas content of the cluster. The GMRT array consists of 30 antennae 
 arranged in an approximate Y-shaped configuration, with each antenna having a 
  diameter of 45\,m. Details about the GMRT can be found in \citet{swa91}. 

The Very Large Array (VLA) phase and flux calibrators `0555+398' and `3C48', 
respectively, were used for these observations.  We carried out the 
data reduction using the {\sc aips} software and followed the procedure described 
in \citet{mal13}. Briefly, we used various {\sc aips}  tasks for flagging the 
bad data and calibrating the data with standard phase and flux calibrators. 
 Thereafter, we run the {\sc aips} task `IMAGR' to make  maps after splitting 
 the source data from the whole observations. We applied a few iterations of 
 (phase) self-calibration  to remove ionospheric  phase distortion effects. 
 The resultant maps are discussed in Section \ref{sect_ion}.  

\subsection{Ancillary Archival Data Sets} \label{s28_anc}
For the present work, we have also used the following archival data sets covering 
 various wavelengths: 

i)  $Gaia$ Early Data Release 3 \citep[$Gaia$ EDR3][]{gaiadr3} from  the 
European Space Agency  {\it Gaia} mission \citep[][]{gaia16}. We used
the kinematic information of Gaia data to estimate the distance 
to the cluster.  The effective angular resolution of  the survey is $\sim$0.4 \arcs.

ii) The {\it Spitzer}-IRAC warm mission data at 3.6 and 4.5\,\mum  data
(PI: Barbara Whitney, Program ID: 61070)  from the $Spitzer$ Heritage 
Archive (SHA). We acquired the corrected basic calibrated data (cbcd), 
uncertainty (cbunc), and imask (bimsk) files. In order to create the final 
mosaic images with a pixel scale of 1.2 \arcs\, pixel$^{-1}$, and to obtain 
the aperture photometry of the point sources, we followed the steps mentioned 
in \citet[]{yad16} and used sources  with uncertainty $<$ 0.2 mag for our analysis.

iii) The NIR ($J, H,$ and $K$) photometric data, from the UKIDSS-GPS 
\citep[][]{luc08}, with uncertainty $<$ 0.2 mag in all three bands. The 
spatial resolution of the UKIDSS-GPS data is in the range 0.8$-$1 \arcs. 

iv) The INT/WFC Photometric \hal Survey of the Northern Galactic Plane (IPHAS) 
data release 2 \citep[][]{bar14} photometric data with uncertainty $<$ 0.2 mag 
in Sloan $r$, $i$ broad-band and \hal narrow-band filters. The spatial resolution 
of the IPHAS data is in the range 0.8$-$1 \arcs.

v) The Pan-STARRS1 \citep[hereafter; PS1][]{pans16} survey photometric data 
with uncertainty $<$ 0.2 mag in \ip\, and \zp\, bands were used to estimate the 
age of the cluster. The spatial resolution of the PS1 data is in the range 0.8$-$1 \arcs.  

vi)  The far-infrared images from the \textit{Herschel} infrared Galactic Plane 
Survey (Hi-GAL)  \citep{mol10} images at 70, 160, 250, 350, and 500 \mum were used.
The spatial resolution of the Hi-Gal observation at 70, 160, 250, 350, and 500 
\mum bands is 8.5, 13.5, 18.2, 24.9, and 36.3 \arcs, respectively.

vii) The radio continuum image at  8700 from the National Radio Astronomy 
Observatory (NRAO) data archive (Project ID: AR390). The beam size of the image 
is $\sim$10 \arcs $\times$ 8 \arcs\ while its rms noise is $\sim$0.1 $\mu$Jy beam$^{-1}$.

viii) The radio continuum image at  150 MHz from the  TIFR GMRT Sky Survey (TGSS) 
\citep{inte17}. The typical resolution of the TGSS images is 
$\sim$25 \arcs$\times$25 \arcs\, with a median noise of $\sim$3.5 mJy beam$^{-1}$.

\subsubsection{Completeness of Photometric Data}\label{phot_comp}
\begin{figure}
	\begin{center}
	\includegraphics[width=8.5cm, trim=0.2cm 0.4cm 0cm 0cm, clip=true]{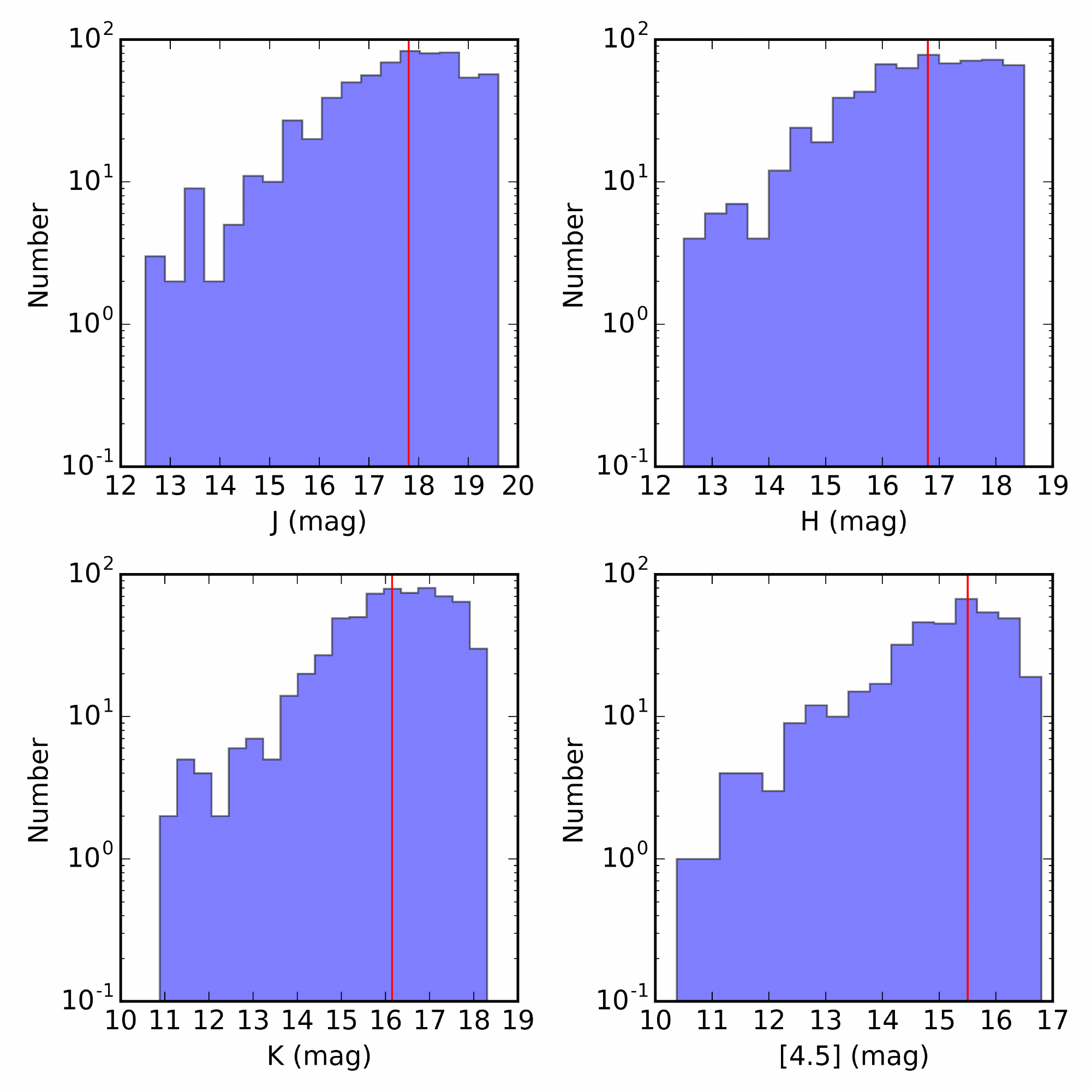}
\caption{Histograms of the sources detected in $J, H, K$, and [4.5] \mum 
photometric bands. The vertical red lines indicate the 90\% completeness limit. 
See the text for details.}
	\end{center}
\label{fig_comp}
\end{figure}

In this work, we used UKDISS, {\it Spitzer}-IRAC and IPHAS data sets to access 
the young stellar contents of the  cluster. We  obtained the completeness limits 
of these bands using the histogram turnover method. Although this method is not 
a formal tool to measure the completeness, it serves as a proxy to give the 
typical value of  completeness limit across the field \citep[e.g.][]{sam15,dam21}. 
 In this approach, the magnitude at which the histogram deviates from the linear  
 distribution is, in general, considered as 90\% complete. Figure \ref{fig_comp} 
 shows  histograms of sources detected in various bands over a radius 
 $\sim$2.5\arcmin (see Section \ref{intro}). The above approach suggest, in 
 general, our photometry is $\sim$90\% complete  down to $J$ = 17.8 mag, 
 $H$ = 16.8 mag,  $K$ = 16.2 mag and [4.5] = 15.5 mag, these are marked by 
 vertical  lines in Figure \ref{fig_comp}.  Similarly, we also estimated the 
  completeness limits of IPHAS $r$, \hal, and $i$; and PS1 \ip\, and \zp\, bands 
  as 20.3 mag, 19.8 mag, and 19.0 mag; and 20.4, and 19.7 mag, respectively. 
 
\begin{figure*}
\centering
	\includegraphics[height=9cm]{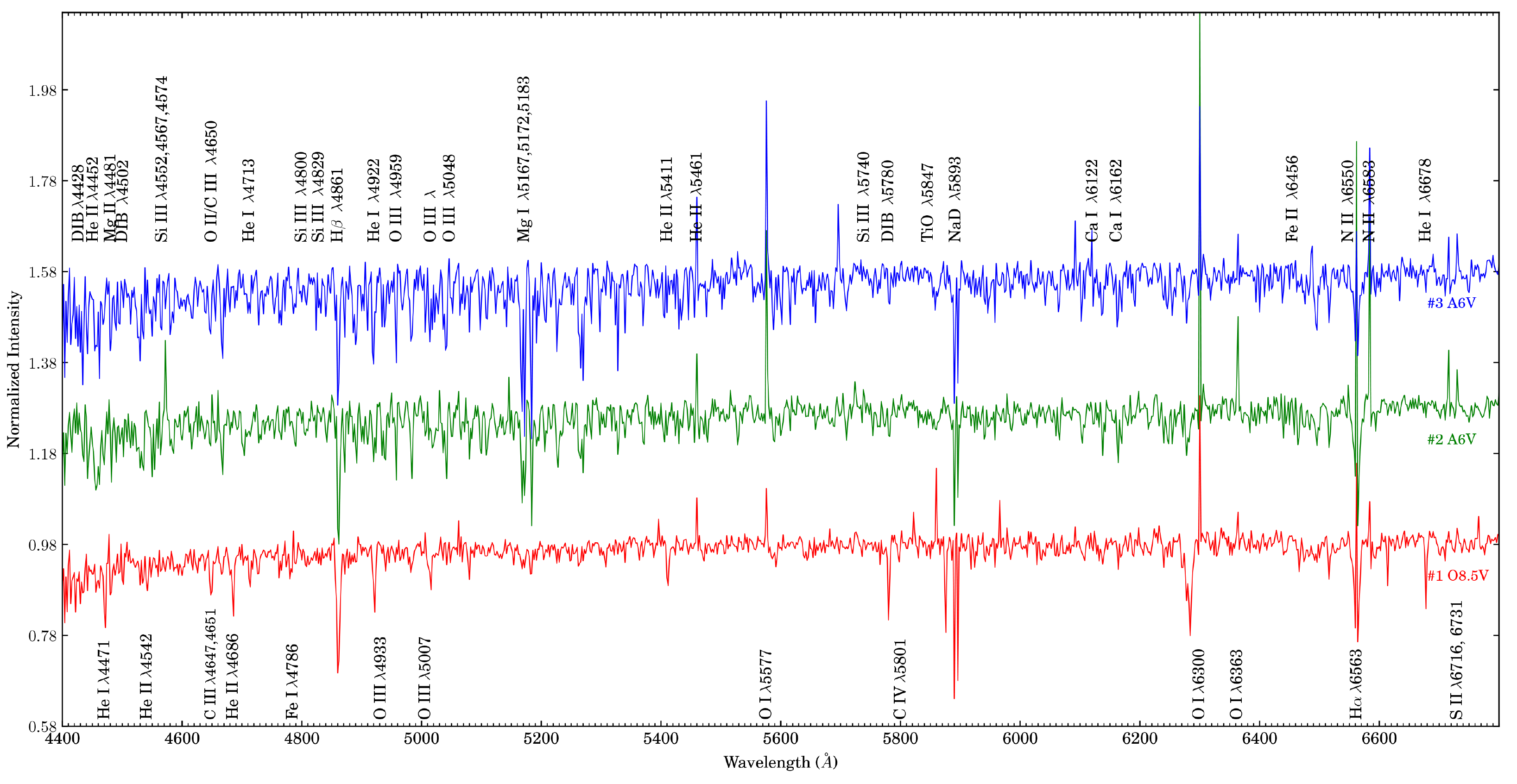}
\caption{Continuum normalized and scaled spectra of three bright stars \#1, \#2, 
and \#3 observed with MRES mounted on TNT. Important stellar lines are marked.}
\label{final_spec}
\end{figure*}

\section{Analysis and Results}\label{s28_anal}

\subsection{Stellar Content and Properties of the Cluster}
In this section, we access massive and low-mass stellar contents of the cluster 
and derive cluster properties using various photometric  catalogs.
\subsubsection{Spectral Classification and Modeling of the Bright Sources} \label{s28_ion}
\begin{figure}
	\begin{center}
\includegraphics[width=8.5cm, trim=0.2cm 1.4cm 0.2cm 1.4cm, clip=true]{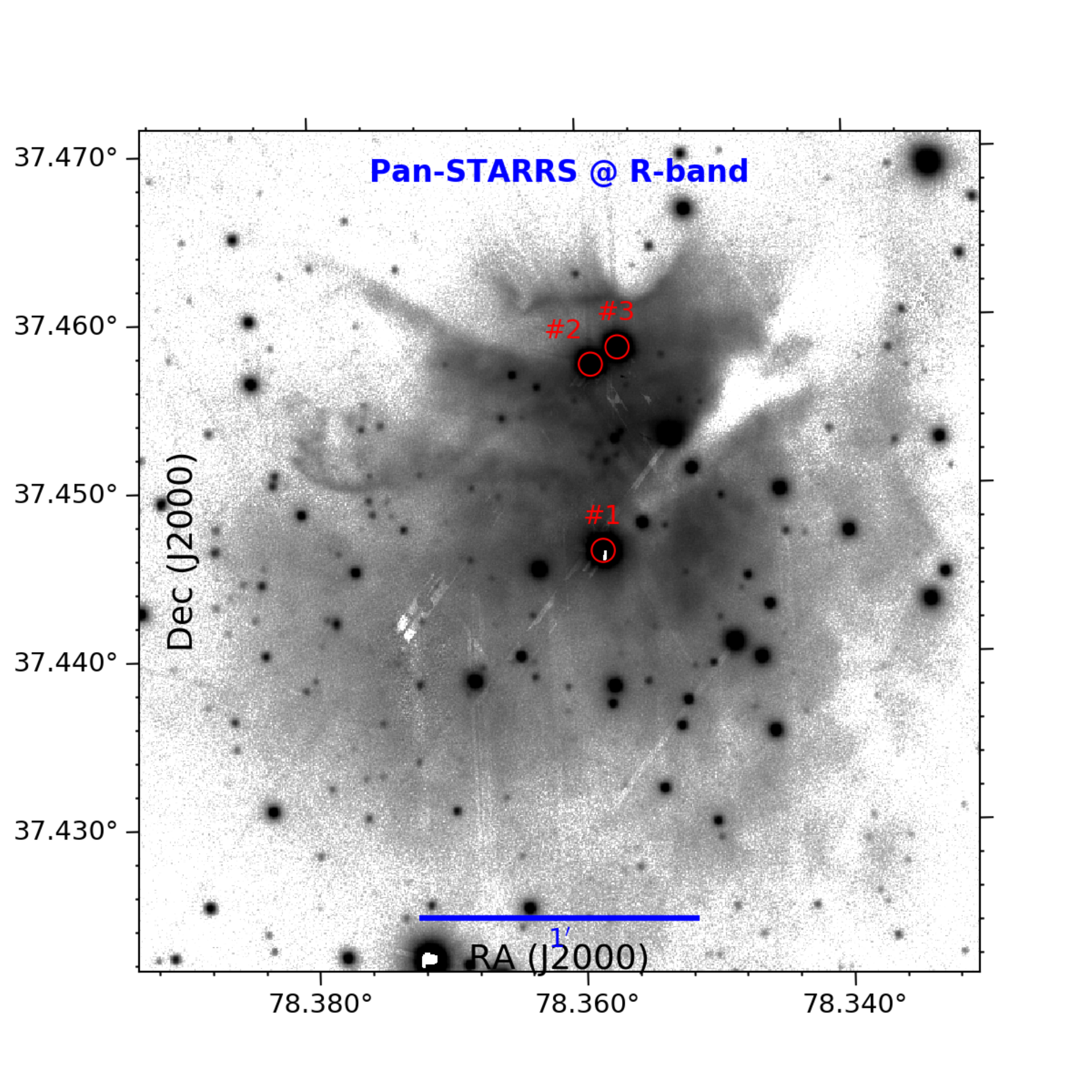}
\caption{Pan-STARRS $r$-band image of the cluster along  with the sources for which
spectroscopic observations have been conducted. The ID and the coordinates of the
sources are given in Table \ref{obs_log}.}
	\end{center}
\label{fig_obstar}
\end{figure}

 The cluster IRAS 05100+3723 is associated to an \hii region, implying that 
 its massive members must be responsible for the ionization of the region. 
To identify the  ionizing source(s) of the \hii region, we selected three bright 
sources located within the region of  strong \hal emission. Following 
 \citet{2005A&A...430..541C}, we  use the infrared 
reddening-free pseudo color index, Q = ($J - H$) $-$ 1.70$\times$($H - K$) and 
select sources with Q $<$  0.1 as probable OB stars of the region. These sources 
are marked in Figure~\ref{fig_obstar} as \#1, \#2, and \#3. However, selected 
sources may be  contaminated  by  objects  such  as  AGB  stars,  carbon stars,  
and  A-type  giants \citep{2005A&A...430..541C}. We thus performed optical 
 spectroscopic analyses of these three bright sources. 

\begin{figure}
\centering{
\includegraphics[width=8cm]{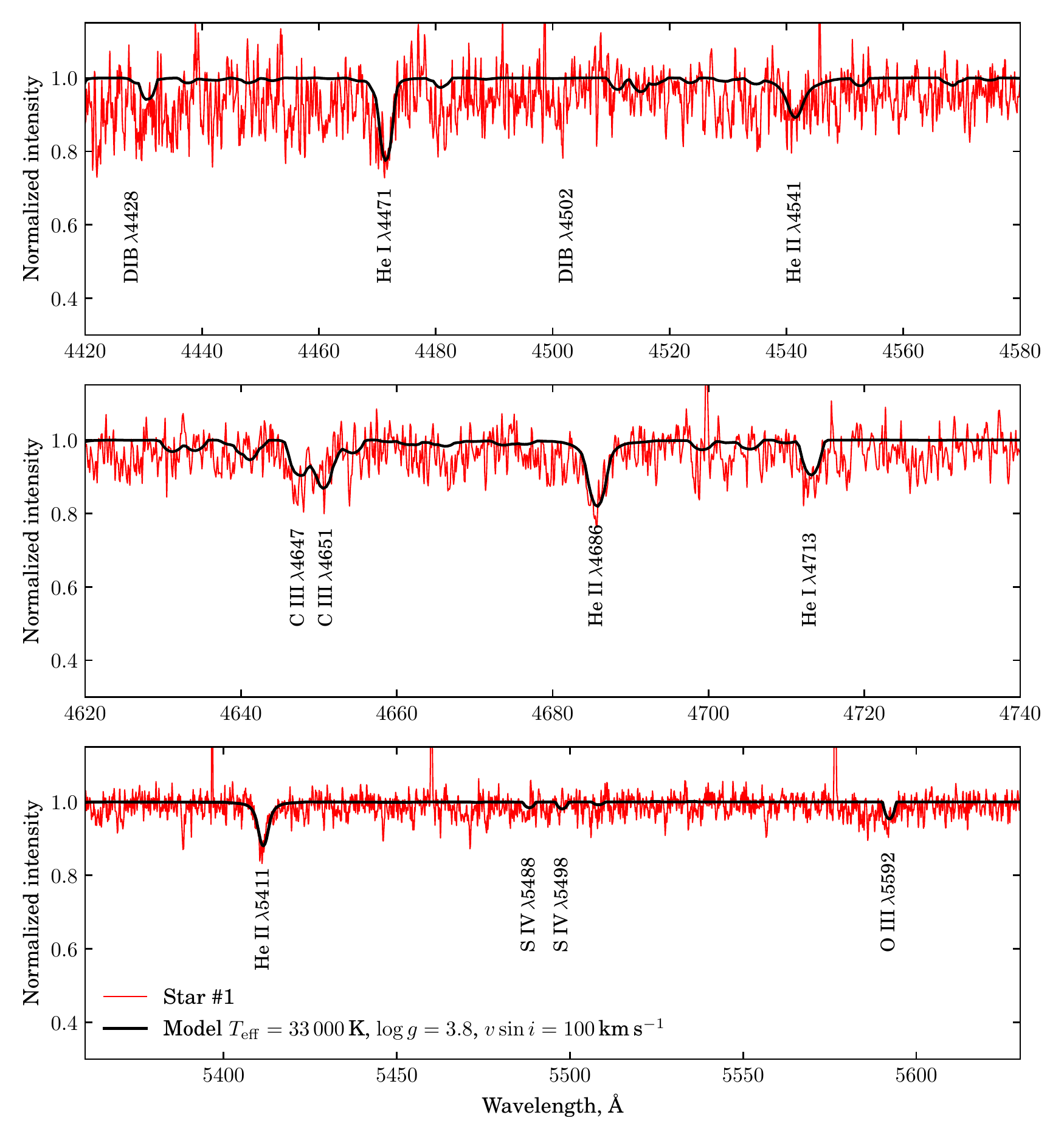}}
\caption{Part of the MRES spectrum (red) for the source \#1 fitted with a 
synthetic profile (black). The synthetic spectrum was computed using the 
parameters specified in the text.} 

\label{fig_ob_model}
\end{figure}

As mentioned in Section \ref{ospec}, observations were carried out under the 
cloudy weather conditions during the waxing Moon with ~80\% illumination. This 
caused the stellar spectra to be heavily contaminated with the solar lines. 
The contamination becomes more obvious upon close examination of the spectra 
~(Figures~\ref{final_spec}, and \ref{fig_ob_model}). The variable contamination 
 of stellar spectra restricts our capabilities to find the accurate 
 physical parameters  of the stars, but the spectra still contain enough markers 
 of the effective temperature. 

The abundance of neutral and/or ionized helium lines in absorption is 
indicative of O or early B type main sequence stars. These lines may be 
in emission in case of giants/super giants. The strength of \heii 4686 
gets weaker for late O-type stars, and this line is last seen in B0.5 stars 
\citep{1990PASP..102..379W}. We could find such lines only in the 
spectrum of the star \#1. To further constrain this, we compare our 
data with synthetic spectra computed for different combinations of 
effective temperature, surface gravity, and projected rotational velocity. 
Theoretical spectra were synthesized with \textsc{Synplot}, an IDL-based wrapper 
of  the \textsc{Synspec} package for spectral synthesis~\citep{2011ascl.soft09022H}.  
We use ``OSTAR2002’’ grid of pre-computed models of 
~\citet{2003ApJS..146..417L}.

 In the observed spectrum, we identified multiple neutral and ionized helium, 
 carbon, and oxygen  lines (Figure  \ref{fig_ob_model}). Among them, two helium 
 lines He I $\lambda4471$ and  He II $\lambda4541$ are good indicators of 
 effective temperature. The intensity ratio of these lines equals to 1 for 
 the  spectral class O7~\citep{1990PASP..102..379W} and it varies between $>$ 1 
 and $<1$ for the spectral range O5$-$O9. This intensity ratio for star \#1 
 is estimated to be $<1$. By fitting synthetic spectrum, we derive the 
 effective temperature $T_{eff} = 33,000\pm1000$ K and surface gravity 
 $\log g = 3.8\pm0.5$~(Figure \ref{fig_ob_model}), which allowed us to 
 classify this object as an O8.5V$\pm$0.5 star. The synthetic spectrum was 
 computed for $T_{eff} = 33,000\pm1000$ K,  $\log g = 3.8\pm0.5$,  
 and $v\sin i=100\pm50$ km/s for the whole wavelength range, which agrees well 
 with the observed lines in other spectral regions and thereby verifies 
 our estimates. 

The spectral types of the two fainter stars are less certain. Their spectra do 
not contain any measurable lines of helium, which are typical of OB stars. 
Instead, from the comparison with the solar spectrum we identified  neutral 
iron, chromium and titanium  (see Figure~\ref{final_spec}) lines. Detailed 
modelling of these spectra was carried out using  the NEMO grids of 
stellar  atmospheres computed with  a modified version of the ATLAS9 
code~\citep{2002A&A...392..619H}. The modelling results show that the star 
\#2 is hotter than the Sun and has an effective temperature $7000\pm1000$ K 
whereas the star \#3 looks comparatively cooler with  a probable effective 
 temperature of 6400\,K. The existing data do not allow us to determine the 
 surface gravity (and thus the evolutionary status) of these stars with 
 higher accuracy. Our spectral analysis suggests that stars \#2  and \#3 are 
 late A to F type stars. Indeed, the {\it Gaia} data confirmed that stars 
 \#2 and \#3 are the foreground objects, as discussed in Section \ref{s28_dis}.

\subsubsection{Physical Extent}\label{s28_phys}
\begin{figure}
\centering{
\includegraphics[width=7.0cm, trim=0.2cm 0.1cm 0.2cm 0.4cm, clip=true]{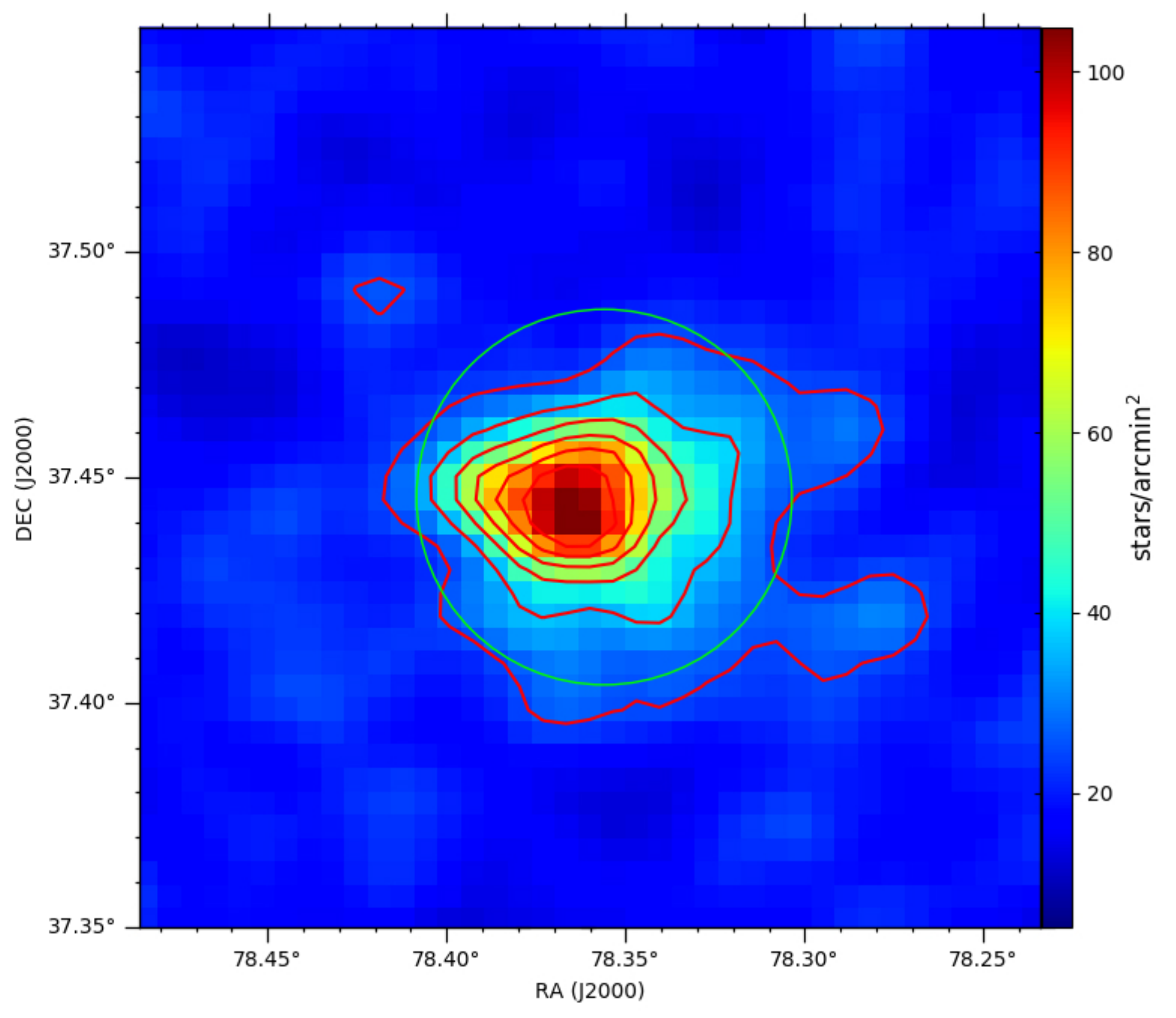}\\
\includegraphics[width=7.5cm, trim=0.2cm 0.4cm 0.2cm 0.4cm, clip=true]{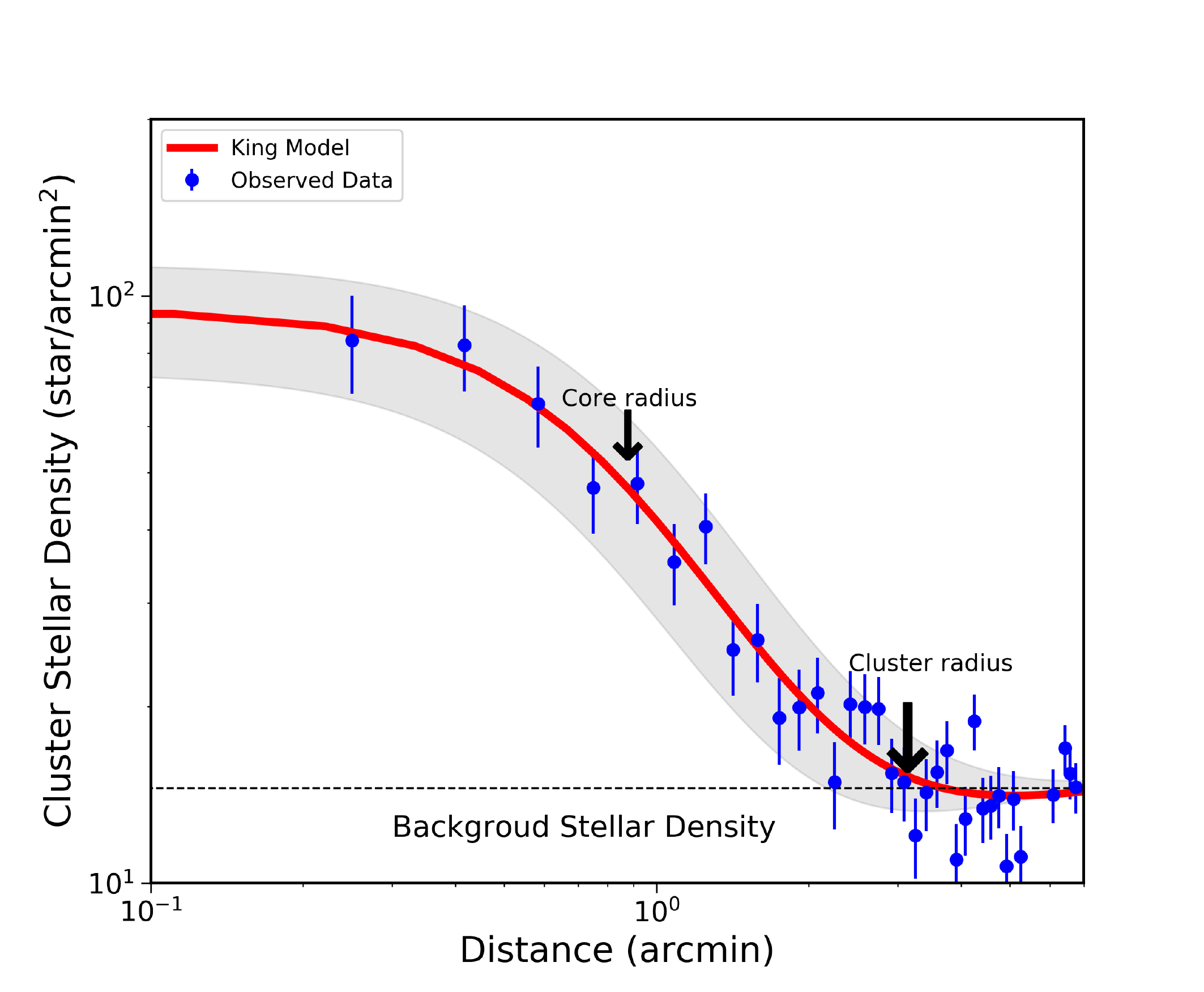}}
\caption{Upper Panel: Stellar surface density map of S228,  made using the 
nearest neighbors method. The color bar represents the stellar surface density in 
the units of stars per \arcm$^2$. Contours correspond to stellar surface 
densities above $3\sigma$  of  the mean background density, where $\sigma$ is 
the standard deviation of the background density. The green circle represents 
the cluster region whose radius was  obtained using the radial density 
profile fitting method shown in the lower panel. Lower Panel: The observed 
stellar surface density (filled circles) of S228 as a function of radius. The 
  horizontal line represents the mean density level of the field stars. 
  The continuous curve shows the fitted King's profile to the observed data, 
  while the shaded area represents the $1\sigma$ error associated to the 
  fitted profile. The error bars represent the Poisson uncertainties. }
\label{fig_ssnd}
\end{figure} 

We derived the physical extent of the cluster by generating the stellar 
surface density map (SSDM) using UKDISS point sources. In order to generate the 
SSDM map, we use the nearest neighbors algorithm as described in 
\citet{gut05}. Succinctly, at each sample position [$i, j$] in a uniform grid, 
we measured $r_N$($i, j$), the projected radial distance to the $N^{th}$ 
nearest star. $N$ is allowed to vary to the desired smallest scale structures 
of interest. We generated the map using $N$=20, which, after a series of tests, 
was found to be a good compromise between the resolution and signal-to-noise ratio 
of the map. The resultant map is shown in Figure \ref{fig_ssnd} 
(upper panel) along with stellar surface density contours. We then considered 
the peak (at \ra = 78.$\!\!$\degree366083 \dec = +37.$\!\!$\degree444889) 
of the SSDM map as the cluster center. 

We also constructed a radial density profile (RDP) of the cluster. 
The RDP is generated by plotting the annular stellar density against 
the corresponding radius (for details, see \citet{pan19}). In order to 
parametrize the RDP, the observed RDP is fitted with the empirical King’s 
profile \citep{kin62}, which is of the form: 
\begin{equation}
\rho (r) \propto b_0 + \frac{\rho_0}{\left(1+\left(\frac{r}{r_c}\right)^{2}\right)}
\end{equation}
where b$_0$, $\rho_0$ and $r_{c}$ are the background stellar density, peak 
density, and core radius, respectively. The fitted King’s profile, shown in the 
lower panel of Figure \ref{fig_ssnd}, yields a central density of 
$\rho_0$ = 116 stars arcmin$^{-2}$, a core radius of $r_c$ = 0.8 \arcm\, and 
 a background density $b_0$ = 13 stars arcmin$^{-2}$. From Figure \ref{fig_ssnd}, 
we note that the model profile merges with the background density at 2.5 
arcmin and is almost constant beyond the radius of $\sim$3.0 \arcm. 
We thus considered the radius of the cluster to be 2.5 arcmin. The estimated 
radius is in agreement with the cluster size of 5.2 \arcm\, (or radius of 
2.6 \arcm) reported in \citet{yu18}. A circle of thus estimated radius is also 
over plotted on the SSDM, which is in  good agreement with the size of the SSDM 
map. 

 \subsubsection{Distance}\label{s28_dis}
 \begin{figure}
     \centering{
\includegraphics[width=6.5cm,trim=0.2cm 0.1cm 0.2cm 0.4cm]{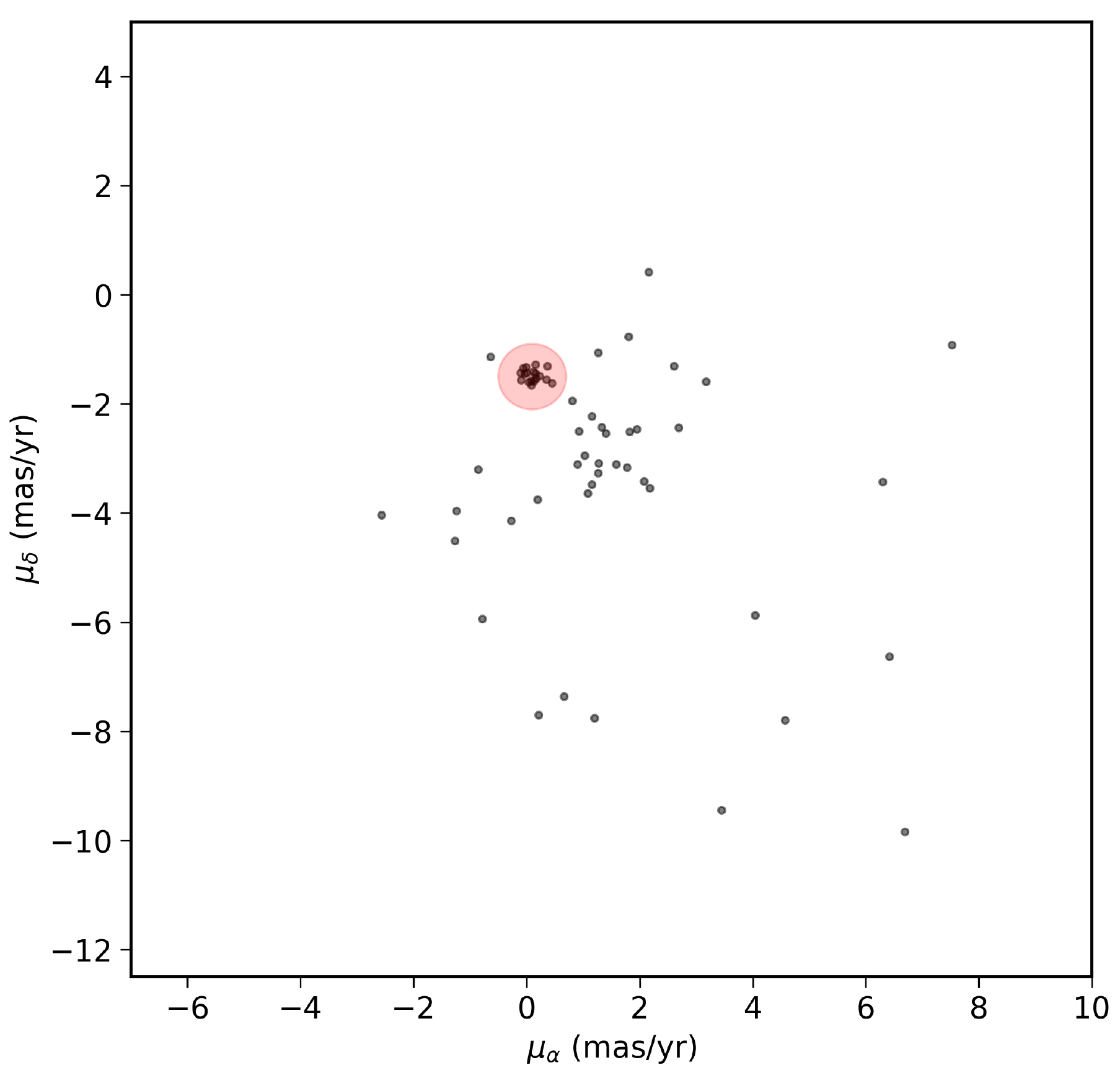}}
     \caption{Proper-motion vector diagram (PVD) for all the sources toward 
     S228 within the cluster region. Sources distributed in the red zone 
     represent the  likely cluster members.}
     \label{fig_pm}
 \end{figure}
 
The distance of S228 is quite uncertain, ranging from 2.2 to 6.8 kpc  \citep{lah85,hun90,bor03,bals11,kha13}. 
With the aim to tighten the distance of the cluster,  we searched for the 
kinematic members of the cluster using the \textit{Gaia} EDR3 catalog. 
Figure \ref{fig_pm} shows the proper-motion vector diagram (PVD)  of 
relatively bright sources ($G$ $<$ 18 mag)  within the cluster radius. 
As can be seen, there are two likely distributions in the PVD, i) a compact 
group consisting of sources of similar motion (marked with red zone), thus 
are considered as cluster members, and ii) a loose group of sources showing 
scattered motion, thus are considered as field sources. A clear separation 
between the cluster and field stars motion enables us to select member stars of 
S228 relatively reliably. Among the compact group sources, we reject a few 
outliers, by  keeping only those sources whose parallax ($\varpi$) values are 
within one standard deviation  of the median parallax of the group and have 
good relative parallax uncertainties ($\varpi$/$\sigma_\varpi$ $>$ 5). The 
latter condition is primarily motivated by the fact that, if fractional 
parallax errors are less  than about 20\%, then the posterior 
probability  distribution of parallax is nearly symmetric \citep{bai15}, 
hence, distance can  simply be computed by inverting the parallax.  With 
these constraints, 7 sources are found to have common proper motions (PM) 
and parallaxes. We find that among these sources, the ionizing source 
(i.e. star \#1) is one of the members, while the other bright spectroscopic 
sources (i.e. star \#2 and \#3) are not members, confirming our 
spectroscopic membership.  
The PM in Right Ascension ($\mu_{\alpha}$), PM in Declination ($\mu_{\delta}$) 
and  parallax ($\varpi$) values of these sources lie in the range $-0.10$ to 0.37 
mas yr$^{-1}$, $-1.45$ to $-1.30$  mas yr$^{-1}$ and 0.28 to 0.39 mas,  
respectively, with median -0.01 $\pm$ 0.15 mas yr$^{-1}$, $-$1.40 $\pm$ 0.06 
mas yr$^{-1}$, and 0.31 $\pm$ 0.04 mas. From the median parallax value,  we 
estimated the distance of the cluster to be 3.2 $\pm$ 0.4 kpc, which is in 
agreement with the kinematic distance of 3.5 
kpc, very recently derived by \citet{meg21} using velocity analysis of the 
molecular gas associated to the region.

\subsubsection{Pre-Main-Sequence Members}\label{s28_pms}
In the absence of spectroscopic or kinematic information of low-mass sources, 
optical and infrared color-color (CC) diagrams are often used as diagnostic tools 
to identify the likely PMS members in a star-forming region 
\citep[e.g.][] {lada92,bar11}. In this work, we used IPHAS, UKIDSS-GPS, and 
{\it Spitzer}-IRAC  point source catalogs above the completeness levels and employed 
CC diagrams to search for the PMS stars with either excess NIR emission due to 
the presence of circumstellar disc or excess \hal emission due to accretion onto 
the stellar photosphere. The details about the CC diagrams and the identified 
sources are given below.

\begin{figure}
\centering{
\includegraphics[width=8.5cm]{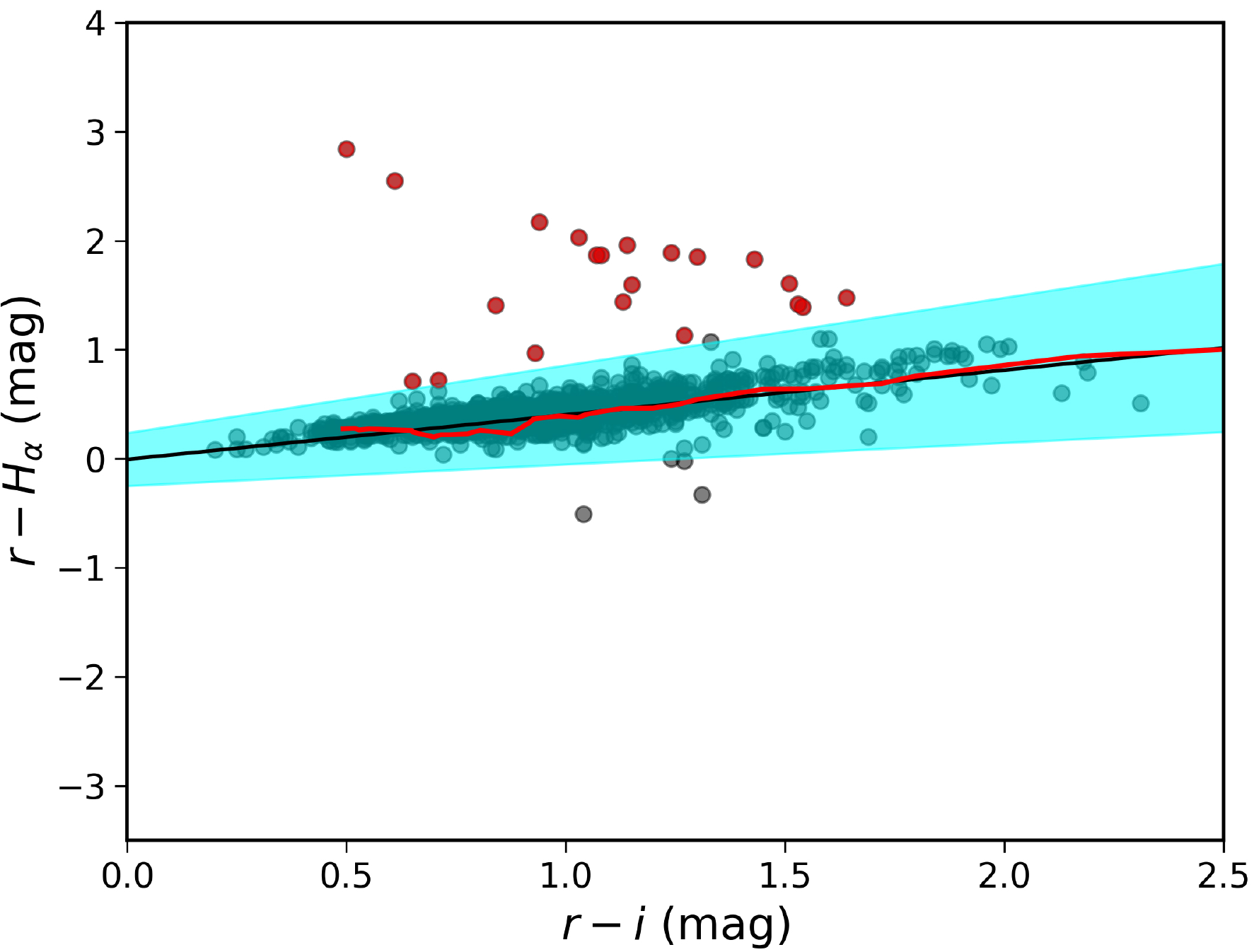}}
\caption{The IPHAS CC ($r -$ \hal, $r - i$) diagram of the point sources in 
the direction of the cluster. The median ($r - $\hal) color of the field sources 
and 10$\sigma$ limits around the median color line (blue) is shown with a shaded  
 area. The red curve represents the synthetic MS track taken from 
 \citet{2005MNRAS.362..753D} corresponding 	to \textit{E(B-V)} = 1.0 mag 
 (or \av $\sim$3.1 mag). The filled red circles are the probable \hal 
 excess candidates within the cluster area. } 
\label{fig_iphas}
\end{figure}

\paragraph{\hal Excess Emission Sources}
The ($r - $\hal$\!$) color measures the strength of the \hal line relative to 
the $r-$band photospheric continuum.  Since most main-sequence stars do not 
show \hal emission, their ($r - $\hal$\!$) color, which is linked to spectral 
type, provides a template, against those whose ($r - $\hal$\!$) color excess 
caused by \hal emission. Moreover, in the ($r -$\hal$\!$, $r - i$) CC space, 
 interstellar reddening has  minimal effect on ($r - $\hal$\!$) color as 
 the reddening  moves only the unreddened MS track almost with a right angle 
to the ($r - $\hal$\!$) color. Thus, in star-forming complexes, the 
($r - $\hal$\!$, $r - i$) diagram is often used to discern \hal emission line 
stars \citep[][]{bar11,dut15}. The ($r -$\hal$\!$, $r - i$) CC diagram for the 
stars in the direction of S228  is shown in   Figure \ref{fig_iphas}. The running 
 average ($r - $\hal$\!$) color of the stars as a function of their ($r-i$) color 
 is shown with a solid line, while its 10$\sigma$ uncertainty  is shown with a 
 shaded area. Figure \ref{fig_iphas} shows the ($r - $\hal$\!$, $r - i$) color 
 of the main-sequence (MS) track taken from \citet{2005MNRAS.362..753D} for 
 \av $\sim$3.1 mag. Using this diagram, those sources with ($r - $\hal$\!$)  
 color excess greater than  10$\sigma$ of the average color of the stars 
 are considered as probable emission line sources, and  are marked with red 
 circles.  With this approach, we identified 21 likely  \hal excess sources.

\paragraph{Near-Infrared Excess Emission Sources}
 \begin{figure*}
\centering{
\includegraphics[width=8.5cm, trim=0.2cm .2cm 0.2cm 0.2cm, clip=true]{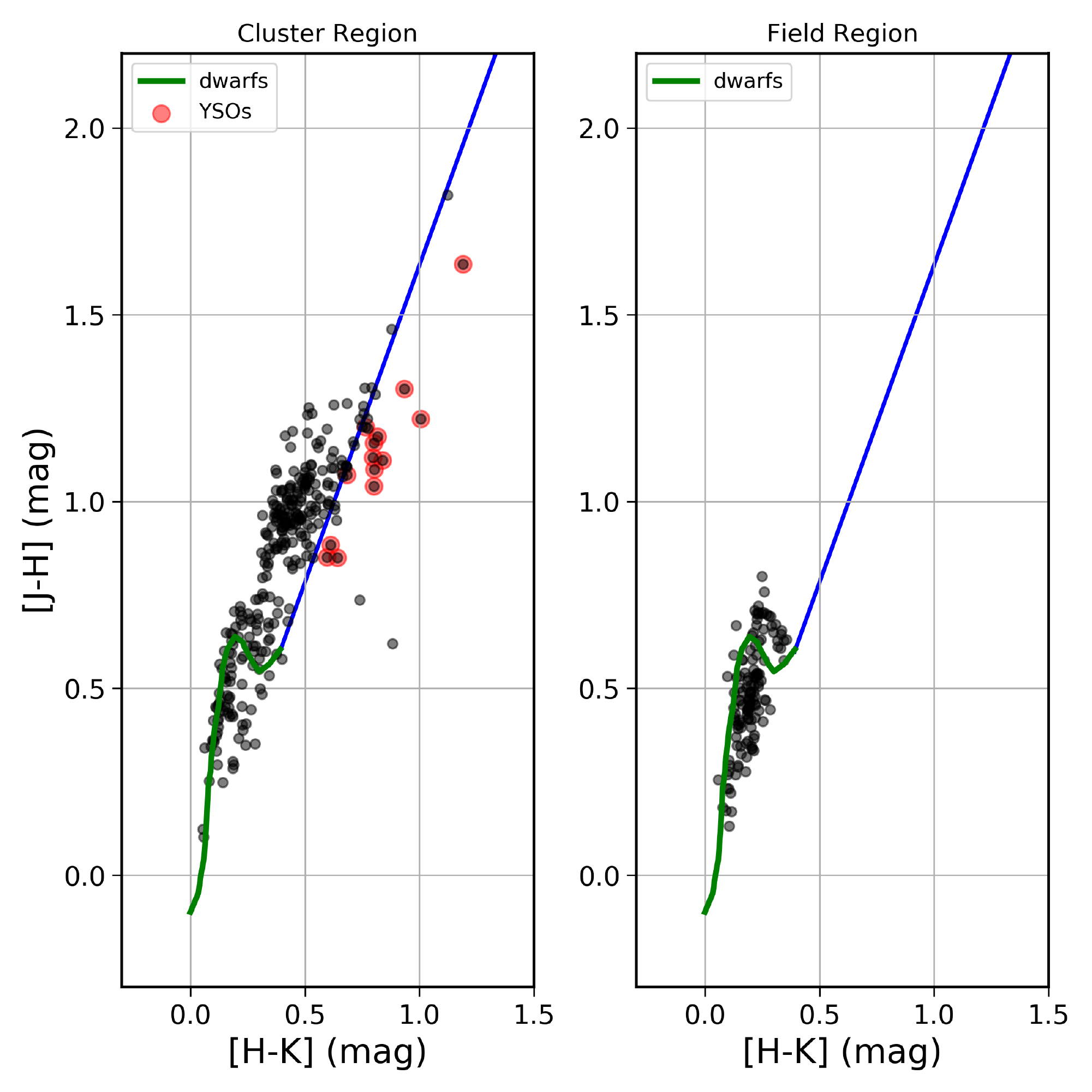}
\includegraphics[width=8.5cm, trim=0.2cm .2cm 0.2cm 0.2cm, clip=true]{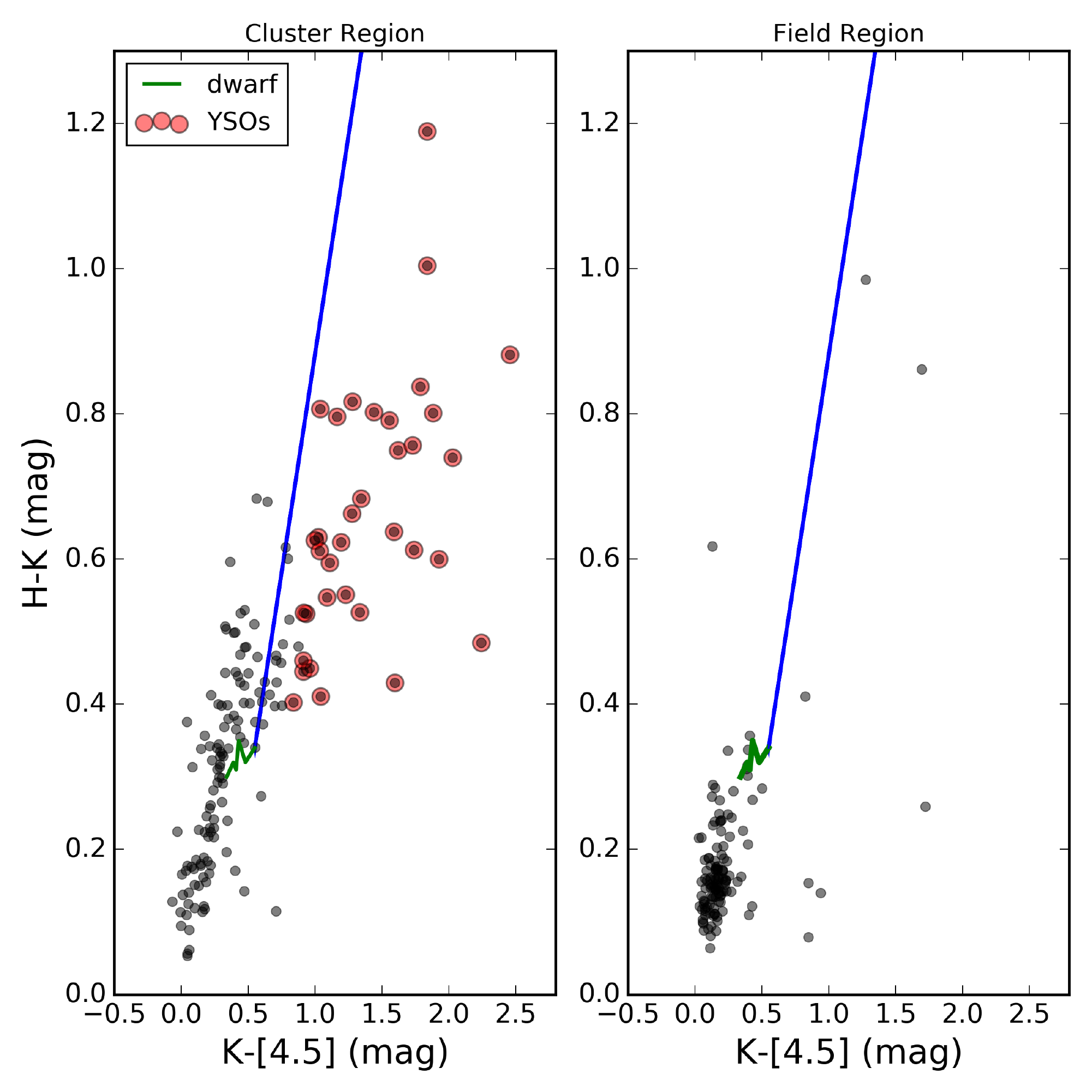}}
\caption{Left: The ($J - H$, $H - K$) CC diagram for the cluster region and the  
control field region. The green curves are the intrinsic dwarf locus from 
 \citet{bes88}. Right: The ($H - K$,  $K -$ [4.5])  CC diagram for the cluster 
 region and the control field region. The green curves are  the intrinsic dwarf 
 locus of late M-type dwarfs \citep{pat06}. In all the plots, the blue lines 
  represent the reddening vectors drawn from the location of M6 dwarf, while the 
  red circles represent likely NIR$-$excess sources.} 
\label{fig_jhk1}
\end{figure*}

The NIR ($J - H$, $H - K$)  diagram is a useful tool to identify PMS sources 
 exhibiting NIR$-$excess emission. However, other dusty objects along the line of 
sight may also appear as NIR$-$excess sources in the CC diagram. One possible way 
to separate out PMS sources is to compare the CC diagram of the cluster with that 
of  a nearby control field of the same area and photometric depth.
The left panel of Figure \ref{fig_jhk1} shows NIR CC diagrams of the cluster as well 
as a control field (centered at \ra = 78.$\!\!$\degree382386, \dec = 
37.$\!\!$\degree17977) located $\sim$5 arcmin away from the cluster field.  
In the NIR CC diagram, sources distributed left to the reddening vector (blue 
line) can be field stars or reddened MS stars or weak-line T$-$Tauri stars, 
while sources located right to the reddening vector are considered as PMS 
sources with NIR$-$excess \citep{lada92}. As can be seen in Figure 
\ref{fig_jhk1}, compared to the cluster region, the NIR$-$excess zone of the 
control field is devoid of sources, implying the presence of true 
NIR$-$excess sources in the cluster region. However, in order to separate 
spurious sources, if any, from genuine excess sources we selected sources based 
on the following criteria: i) sources with ($J - H$) color greater than 0.7 
mag because the maximum ($J - H$) color of the control field population is around 
0.7 mag,  and ii) sources that fall to the right of the reddening vector 
with ($H - K$) color excess larger than the uncertainties in their respective 
($H - K$)  colors. With this approach, we identified 14 NIR-excess candidate 
sources within the cluster region. These sources are identified as red circles 
in  Figure  \ref{fig_jhk1}.  
  
 It is well known that circumstellar emission from young stars  dominates at 
 longer  wavelengths,  where the  spectral energy distribution (SED) 
 significantly deviates  from the pure photospheric emission.  We thus use 
 {\it Spitzer}-IRAC observations in combination  with  $JHK$ data to  identify 
  additional  NIR$-$excess sources. For this purpose, we  use the 
  ($H-K$, $K-$[4.5]) CC diagram, shown in the right panel  of Figure 
  \ref{fig_jhk1}. It should be noted that we preferred  4.5 \mum data over 3.6 
  \mum as  it  is less  affected by Polycyclic aromatic hydrocarbon (PAH) 
  emission, which is often present in the \hii region environments 
  \citep[][]{smi10}.  Similar to ($J - H$,  $H - K$) analysis, we compared 
  the distribution of cluster sources with control field sources, then selected 
  NIR$-$excess sources  whose excess emission is more by the error associated to 
  the $K-[4.5]$ color of the sources  and  ($H - K$) color greater than 0.4 mag. 
 With this  approach, we identified 25 additional candidate stars with NIR 
 excess.  This makes 39 NIR excess sources in total. 

 In summary, with the above approaches, we identified  21 likely \hal emission 
 line sources and 39  NIR excess sources. Among 21 \hal sources, 6 sources are 
 found to have NIR excess counterparts. In total,  within the cluster region, 
 we identified 54 PMS sources. Out of the 54 PMS sources, 52 sources have 
 optical  counterparts in the Pan-STARRS1 bands, and are used to derive the age 
 of the cluster.

\subsubsection{Extinction}\label{s28_exti}\label{extin}
We derived visual extinction of the cluster using its likely OB members.  Briefly, 
we select OB stars from the Gaia member sources  \citep[e.g, see][]{sam10} 
identified in Section \ref{s28_dis} by considering only those sources whose Q 
value is $<$ 0.1 showing no NIR excess. This resulted in five sources out of 
the seven Gaia members. We then derived the ($H - K$) color excess, $E(H - K)$, 
 of the members from the observed  and intrinsic ($H - K$) colors. Since the 
 most massive star of the cluster is an O8.5V star, we thus adopted an intrinsic 
 ($H - K$) color of $\sim$0.03 mag for our analysis, which is the mean intrinsic 
  color of O8 to B9 MS stars as tabulated in \citet[][]{pec13}. We then estimated 
  the visual extinction of the sources using the relation, 
  $A_V$ = 15.9 $\times$ $E(H - K)$, adopting the extinction laws of \citet{rie85}. 
This yields a mean visual extinction \av = 3.3 $\pm$ 0.6 mag, which is in 
agreement with the extinction measurements between 3.0 mag \citep{lah85} and 3.9 
mag \citep{hun90} derived for S228. 

\subsubsection{Age}\label{s28_age}
\begin{figure*}
\centering{
	\includegraphics[width=14cm]{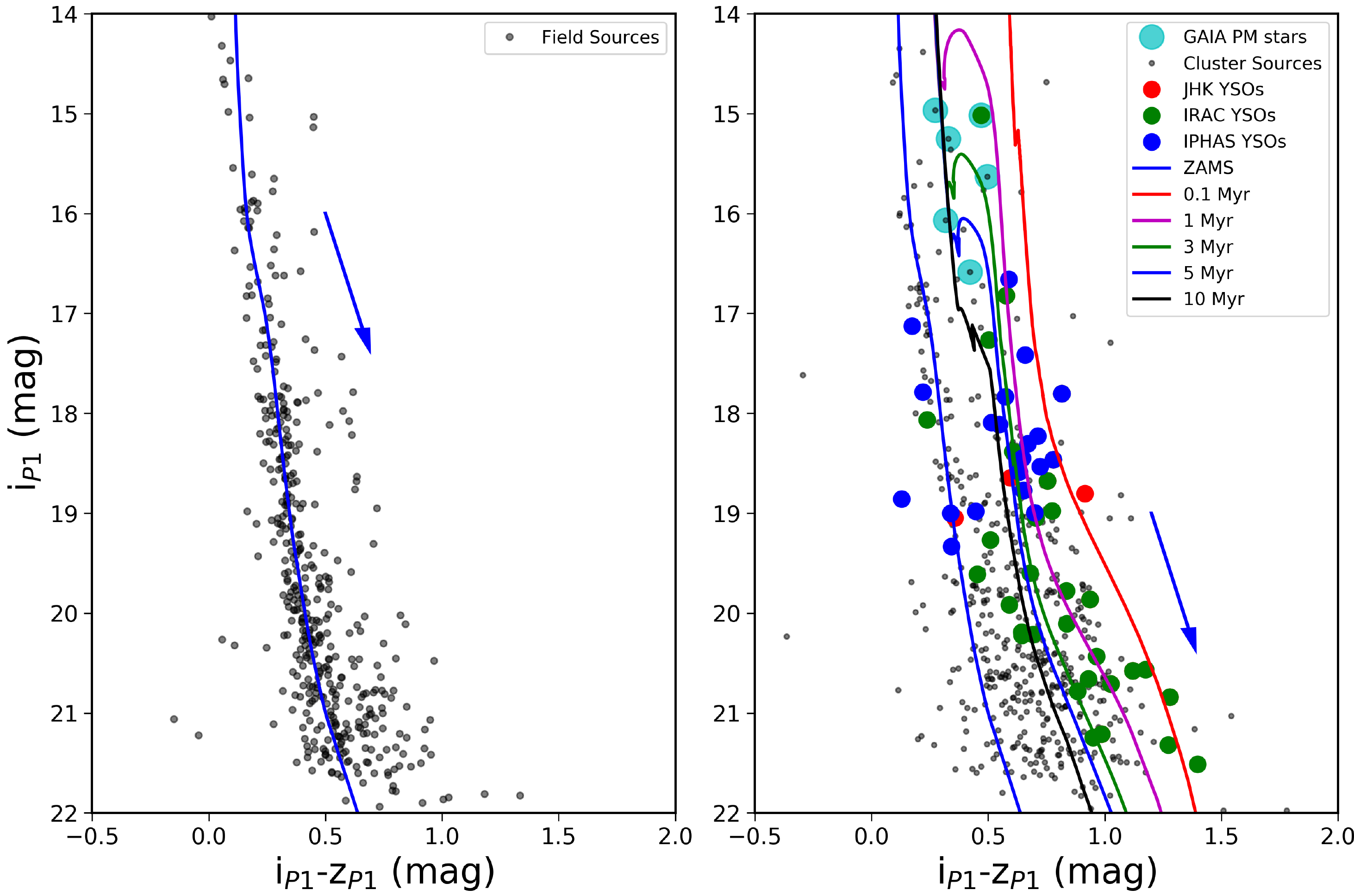}}
\caption{ The optical (\ip, [\ip - \zp]) CMD of the control field  (left panel) 
and cluster region (right panel). In the cluster CMD, the NIR$-$excess sources 
	identified based on $JHK$ and $HK[4.5]$ bands are shown by red and green 
circles, while the H$\alpha$ excess sources are shown by blue circles, and the 
PM   based massive members are shown by cyan circles. In cluster CMD the 
PMS  isochrones of ages 0.1, 1, 3, 5, 7 and 10 Myr from \citet{dot16} are shown 
	from left to right, while the blue solid curves in both the CMDs represent 
	the ZAMS isochrone, to guide the distribution of filed stars in both the 
	plots.  All the PMS isochrones are corrected for a distance of 3.2 kpc and 
	an average reddening of \av = 3.3 mag.  In both the plots, the reddening 
	vectors corresponding \av=0.5 mag are shown by slanted arrows.} 
\label{fig_age}
\end{figure*}

\begin{figure}
\centering{
\includegraphics[width=6.5cm, trim=0.2cm 0.7cm 0.2cm 0.7cm, clip=true]{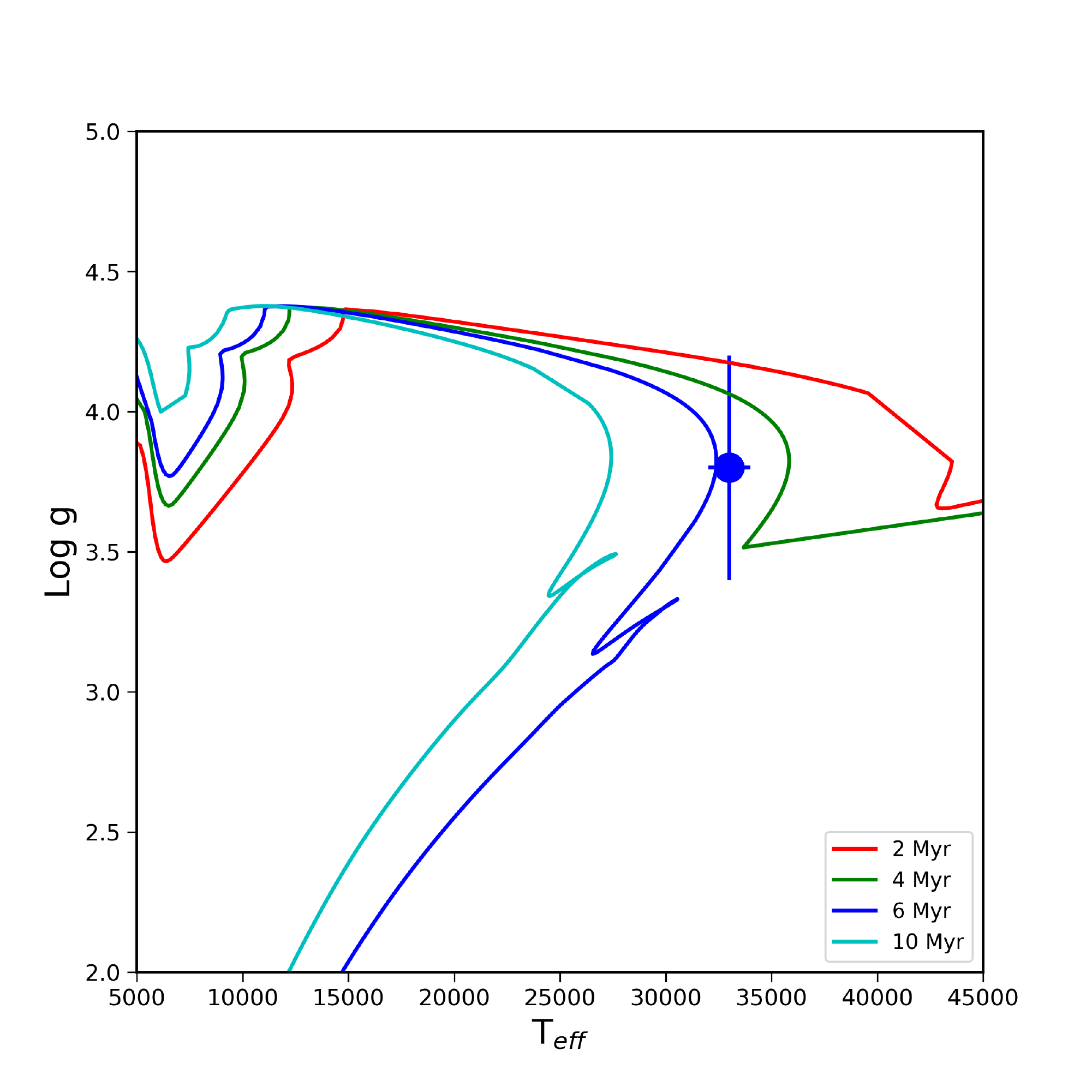}
\includegraphics[width=6.5cm, trim=0.2cm 0.7cm 0.2cm 0.7cm, clip=true]{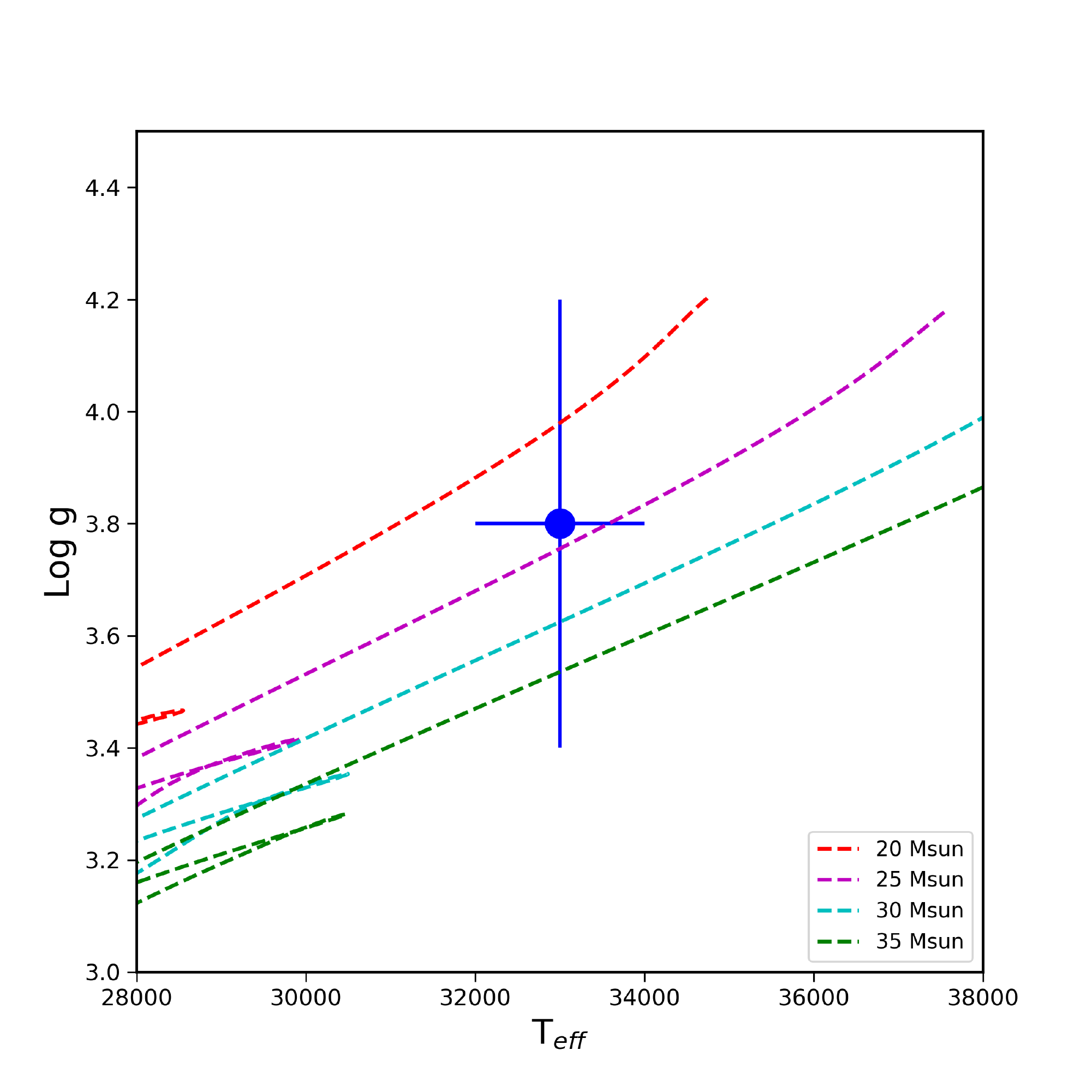}}
\caption{The (gravity ($g$), temperature ($T_{eff}$)) diagrams for the massive star.
In the upper panel, the thin solid curves are the MESA isochrones for ages
 2, 4, 6, and 10 Myr, while in the lower panel  the dashed lines 
are the evolutionary tracks for 20, 25, 30 and 35 \msun.}
\label{fig_age1}
\end{figure}
The ages of young clusters are typically derived  by fitting theoretical 
PMS isochrones to the low-mass contracting population. To derive the age of 
IRAS 05100+3723, we use optical (\ip, [\ip$-$\zp]) CMD of the PMS sources 
identified in Section \ref{s28_pms} as the cluster members. We chose \ip\, 
and \zp\, bands to derive the age, as both of these bands have higher PMS 
 counterparts over other bands of the PS1 survey. Moreover, benefits of using 
 optical bands are that i) it minimizes the effect of the excess emission 
compared to NIR bands, and ii) it minimizes the effect of differential 
reddening as the reddening vectors are nearly parallel to the isochrones. 
Figure \ref{fig_age} shows the (\ip, [\ip$-$\zp]) CMDs of the cluster (right) as 
well as the nearby control field (left). In the cluster CMD, all the 
sources within the cluster area and the likely PMS sources are shown with grey 
dots and filled  circles,  respectively. We find that the average \av of the 
control field population is likely around 2.3 mag, which is estimated by matching 
a reddened theoretical "zero-age main sequence" (ZAMS) with the observed 
population. To guide our analysis, we have also plotted ZAMS at \av = 2.3 mag on 
the cluster CMD. A careful comparison of the distribution of the PMS sources 
and the control field sources on the CMD reveals that the majority of the PMS 
sources are redder compared to the ZAMS isochrone at \av = 2.3 mag. Comparison 
of both the diagrams reveals that the field population is quite significant in 
the direction of the cluster. 

Next, we over-plotted MESA (Modules for Experiments in Stellar Astrophysics)  
isochrones \citep{dot16} on the CMD after correcting them for the adopted 
distance of 3.2 kpc and extinction of \av = 3.3 mag. The adopted extinction 
value is also found to match well with the bright PM based cluster members, 
implying that for the cluster, the average \av = 3.3 mag is a reasonable 
assumption.  We note that compared to the field population, the extra extinction 
of \av = 1 mag observed in the direction of the cluster could be intrinsic to 
the cluster. In fact, we find that the cluster is still associated with gas and 
dust as seen in Figure \ref{fig_sfr} supporting the above hypothesis. As can be 
seen from Figure \ref{fig_age} the location of most of the PMS sources are in the 
age range 0.1$-$10 Myr, implying that the cluster is  young. We note that a 
small fraction of \hal emitting sources are found to be close to the 
field distribution. Such sources could be contaminants such as carbon stars, 
white dwarfs,  and/or interacting binaries \citep[for details, see][]{bar11}. 
In this analysis, we consider all the PMS sources younger than 10 Myr as most 
likely cluster members, because the level of accretion activity for PMS sources 
older than 10 Myr is expected to be very low \citep{will11}.  Then we derive 
the age of the individual PMS sources by comparing their location with 
isochrones having age between 1 and 10 Myr with an interval of 0.1 Myr. We 
assign PMS sources an age equal to the age of the closest 
isochrone. In this process, some sources located to the right of the 0.1 
Myr isochrone are simply considered to have an age of 0.1 Myr. Doing so, we 
estimate the mean age of the PMS sources, thus of the cluster, to 
be $\sim$2.1 $\pm$ 1.3 Myr. Although  the smaller sample of PMS sources 
and variable extinction limit us to derive a very precise age of the cluster, 
yet it agrees well with the age estimates of \citet{bor03}.

\begin{figure}
    \centering
    \includegraphics[width=8.0cm, , trim=0.2cm 0.5cm 0.2cm 0.5cm, clip=true]{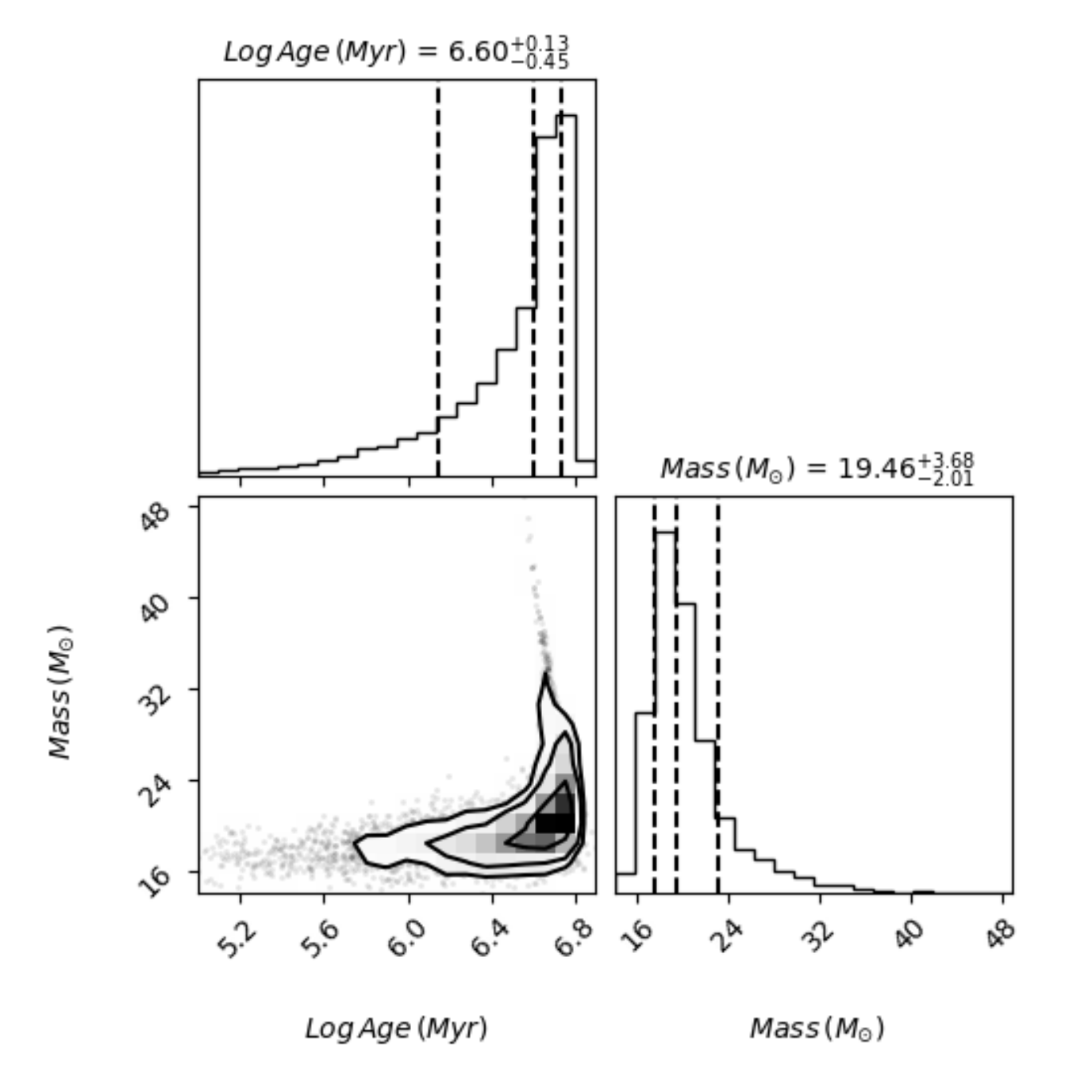}
    \caption{The lower-left plot shows the  posterior
probability distributions of mass and age of the massive star obtained from 
the isochrone fitting. The upper-left and lower-right plots are the probability 
 distribution histograms for the mass and age, 	respectively. The middle line in 
 the histogram plots is the median value of the distribution, while the left 
 and right lines represent the 16\% and 84\% percentile values, respectively. }
    \label{corn_massive}
\end{figure}

To ascertain our result, we further use the properties of the massive star 
derived in Section \ref{s28_ion}. To do so, we compared the derived
$T_{eff}$ and log $g$ of the  star with the MESA isochrones and stellar 
evolutionary tracks as shown in Figure \ref{fig_age1}. As can be seen, because 
of a large error in log $g$ value, the location of the star in Figure 
\ref{fig_age1} spans over a wide range of mass and age. Although accurate 
stellar parameters are needed to constrain the age of the star more 
precisely, nonetheless, we use the Bayesian approach implemented in the 
isochrones package \citep{mor15} to derive the most probable age and mass of 
the star. The isochrones package uses the nested sampling scheme MULTINEST 
 \citep{fer09} to capture the true multi-modal nature of the posteriors. 
 Figure \ref{corn_massive} shows the obtained  posterior probability confidence 
 contours of mass and age with the Bayesian approach. The peak of the likelihood 
 distribution is considered as the most probable value, and the estimated 
 uncertainty is determined by considering 15 and 85 percentile values of 
 the likelihood distribution. With this approach, we find the most probable age 
 of the star to be $\sim$4 Myr, while the most probable mass to be $\sim$20 \msun. 
 If one takes the average uncertainty associated to the most probable value, then 
 the age, 4.0 $\pm$ 2.7 Myr, is in reasonable agreement with the age, 
 2.1 $\pm$ 1.3 Myr, derived using CMD of the low-mass stars. 

 With these two approaches, the average age turns out to be 3.0 $\pm$ 1.5 Myr,  
 which we considered as the age of the cluster for further analyses. Since 
 the cluster is associated with an \hii region that is still bright in optical 
 and radio bands, thus one would expect the cluster to be young. Our analysis  
  confirms this hypothesis. 

\subsubsection{Mass Function and Total Stellar Mass}
\begin{figure}
\centering{
\includegraphics[width=8.7cm]{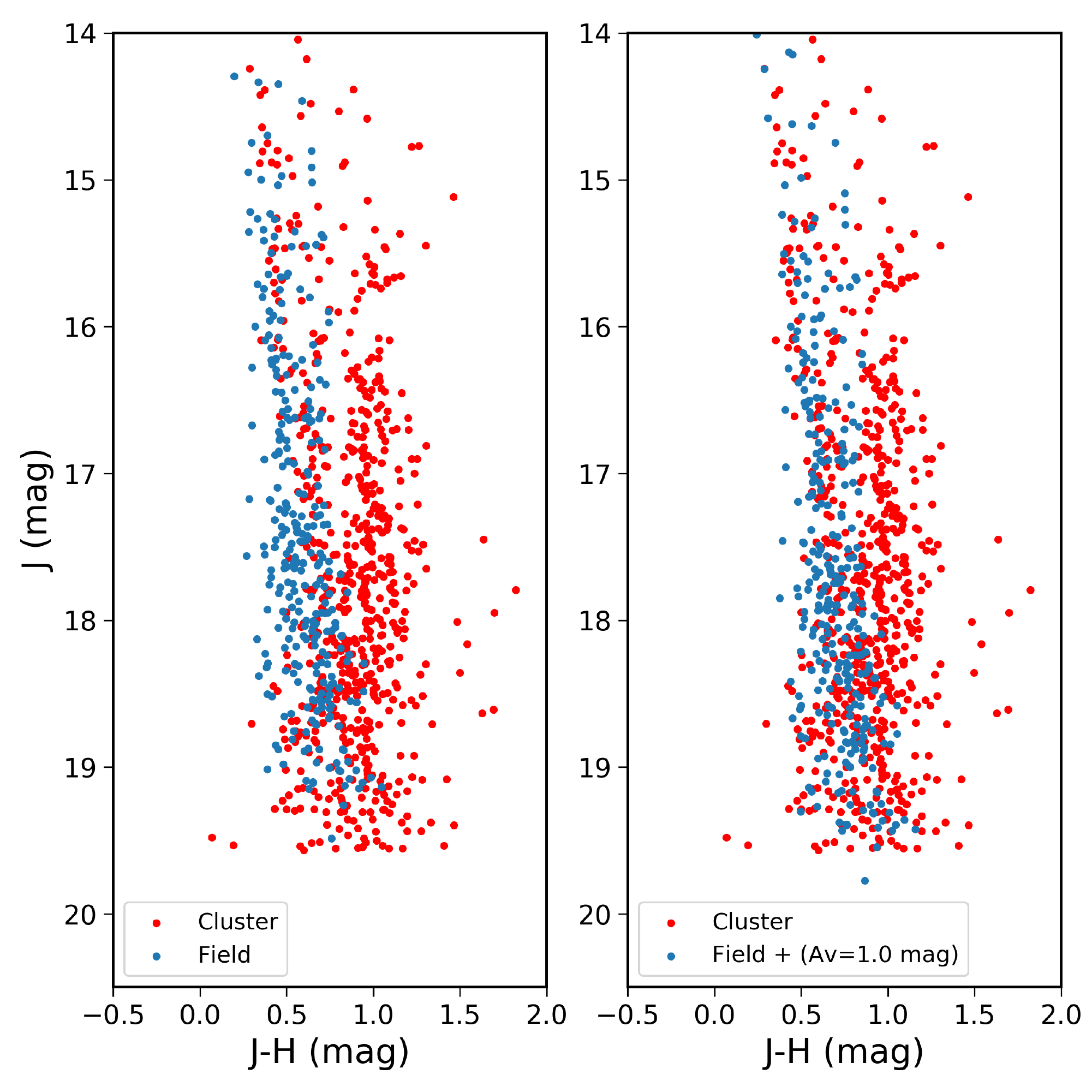}}
\caption{ The ($J$, $J-H$) CMDs of the cluster (red dots) and the control field 
(blue dots). In the left panel, control field sources are plotted without any  
 reddening correction, while in the right panel an extra reddening corresponding 
to \av = 1.0 mag have been applied to control field sources.} 
\label{fig_jjh}
\end{figure}

\begin{figure*}
\centering{
 \includegraphics[width=7.5cm]{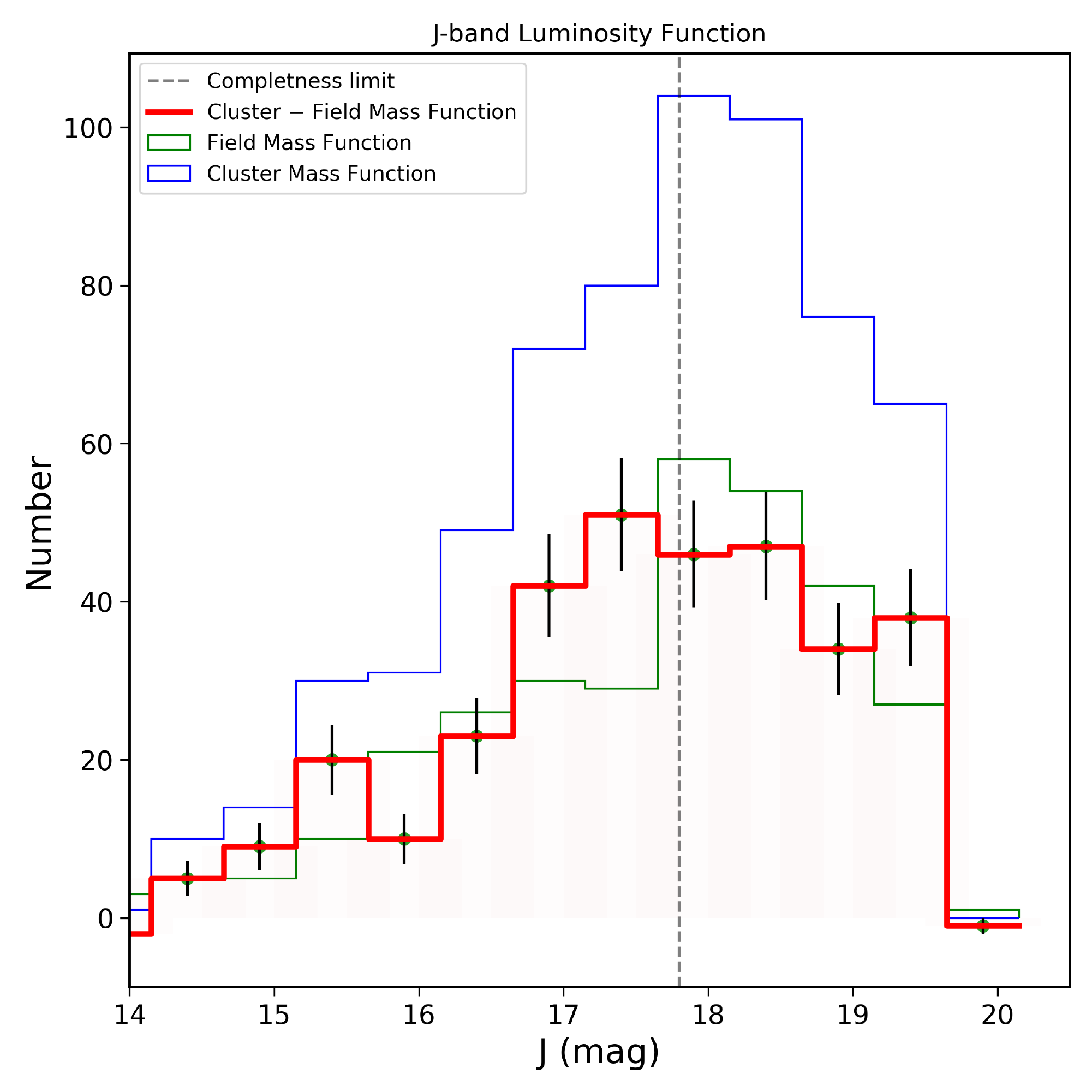}
 \includegraphics[width=7.5cm]{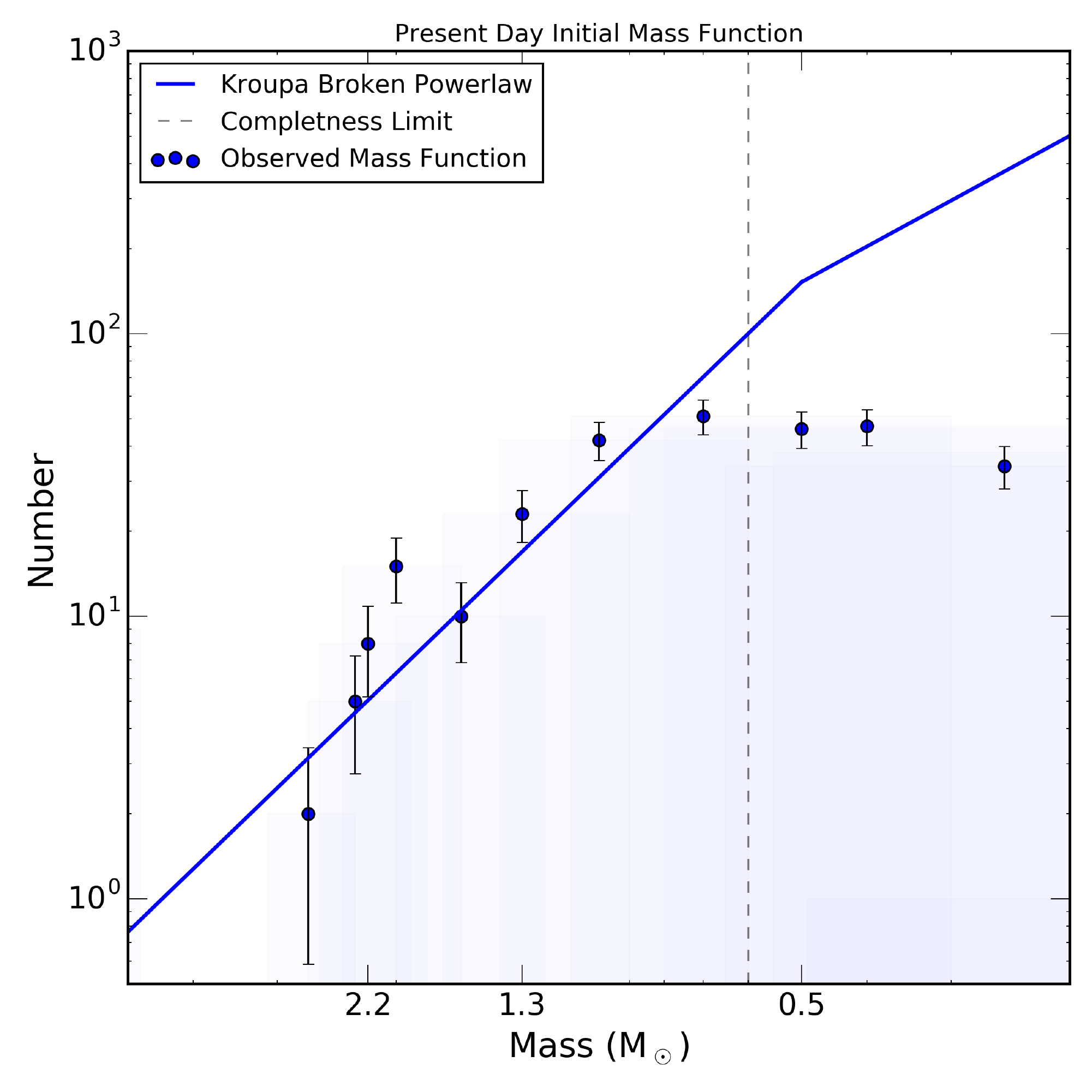}}
\caption{Left Panel: The green and blue histograms represent the $J$-band 
luminosity functions of the cluster and control fields, respectively, while 
the luminosity function after correcting for the field-star contamination is 
shown with the red histogram. The control field is reddened by \av = 1 mag. In 
the plot, the vertical dashed line represents the $\sim$90\% completeness limit 
of the $J$-band, while the error bars represent the Poisson uncertainties. 
Right Panel: Present-day mass function of the cluster computed from  the 
field-star decontaminated $J$-band luminosity function and is shown by dots.  
The 90\% mass completeness limit of the data in terms of mass is shown by a 
vertical dashed line. The fitted power law of the observed mass function is shown 
by a blue line  and is extended towards the low-mass end using  the functional 
form  of the  mass function given by \citet{kro01}.} 
\label{fig_imf}
\end{figure*}

 The stellar initial mass function (IMF) describes the mass distribution of the 
 stars in a stellar system during  birth and is fundamental to several 
 astrophysical concepts. For a young cluster, IMF is in general derived from 
 the luminosity function of member stars  above the completeness limit. In our 
 case, adopting the estimated age and extinction of the cluster, we  find the 
 photometric completeness limits of our $J, H,$ and $K$ bands (see Section
  \ref{s28_anc}) correspond  to mass completeness limits of 0.6 \msun, 0.6 \msun, 
  and 0.7  \msun, respectively. We then used the $J$-band luminosity function (JLF) 
  to derive the IMF. The selection of the $J$-band is motivated mainly by the fact 
   that the  effect of circumstellar excess emission in the $J$-band is minimum 
   compared to the $H$ and $K$ bands (see Section \ref{phot_comp}),  and moreover 
   its mass sensitivity is comparable to these bands. 
Figure \ref{fig_jjh} (left panel) shows the ($J$, $J-H$) CMD for the cluster as 
well as the control field. As can be seen from the figure, the cluster region 
(red dots) appears to have bi-modal color distributions, in  which a  group of 
redder sources with $J$  $>$ 16.0 mag are relatively well separated from a  group 
of bluer sources. This implies that the redder sources are likely the 
cluster members, while the bluer sources are likely the field population in 
the direction of the cluster.  One may also notice that compared to the 
(\ip, [\ip - \zp]) CMD shown in Fig. \ref{fig_age}, the ($J, J-H$) CMD shows 
a richer population of young sources. This could be due to the fact that infrared 
bands are less affected by extinction, and also young sources are intrinsically 
 bright at longer wave bands. In the present case, we have detected 112 extra 
 sources in $J$ and $H$ bands compared to \ip~ and \zp~ bands.
 
One can also  see from the figure that the distribution of  the likely field 
 population of the cluster matches well with the distribution of the control 
 field sources (blue dots), but appears to be slightly redder. To match the 
 control population with the field population of the cluster, we reddened the 
 control population with a  reddening corresponding to \av = 1 mag (3.3$-$2.3 mag), 
  as discussed in section \ref{s28_age}.  Here, we assume that most of the field 
  population in the direction of the cluster is in the background of the cluster. 
  In doing so, we found that the distribution of the field population of the 
  cluster matches well with the control population in both color and photometric 
  depth (see Figure \ref{fig_jjh}, right panel).

Figure \ref{fig_imf} (left panel) shows the JLF of the cluster, the reddened field 
 population, and the field subtracted cluster population. As can be seen that the 
 JLF at the low-luminosity  end, beyond $J$ = 18 mag,  shows a declining trend,   
 which we attribute to the incompleteness of the data beyond $J$ = 17.8 mag. 
 We then constructed the present-day mass function (see Figure \ref{fig_imf} 
 (right panel)) from the field star subtracted luminosity function using the 
 mass-luminosity relation for a 3 Myr MESA isochrone. Although the completeness 
 limit of our $J$-band data keeps us from drawing any conclusions on the peak of 
 the IMF and the shape of the IMF towards the lower mass (M $<$ 0.6 \msun) end,  
 but in general in star clusters, the peak in the stellar distribution lies in 
 the mass range 0.2 $-$ 0.7 \msun\ \citep[e.g.][]{nei15, dam21}. Future deeper 
 observations of the region would shed more light on the shape of the IMF towards 
 the lower-mass end. Nonetheless, a simple power law fit to the data over mass 
 range $\sim$3$-$0.6 \msun\, resulted in a slope of $\alpha$ = $-$2.3 $\pm$ 0.25, 
 which is comparable to the canonical value of $\alpha$ = $-$2.35 given 
 by \citet{sal55} or $\alpha$ = $-$2.3 given by \citet{kro01} for the mass 
 range 0.5 $-$ 10 \msun. This implies that the IMF of the cluster at the 
 high-mass end is similar to other Galactic clusters. We then estimated the 
 total stellar mass of the cluster, assuming that Kroupa's broken power-law 
 (shown in Figure \ref{fig_imf}) holds true down to 0.08 \msun. Since the 
 most massive star of the cluster is of $\sim$20 \msun, thus, we integrated 
 the mass function over the mass range 20$-$0.08 \msun\, which yields a 
 total stellar mass of $\sim$510 \msun. It should be noted that using a 
 mass function slope $\alpha = -2.33$ and integrating over the mass range 20 $-$ 
 0.1 \msun, \citet{bor03} have estimated the mass of the cluster to be 1800 
 \msun. For comparison, if we use a single power-law slope of -2.3 between 
 20$-$0.1 \msun, we find the total cluster mass to be $\sim$900 \msun. Even 
 though our mass estimation is done over a larger area (i.e. over 2.5\arcmin\, 
 radius) yet we obtained stellar mass less by a factor between two and four 
 compared to \citet{bor03}. The exact reason of this discrepancy is not 
 known, however, it is worth noting that \citet{bor03}: i) did not use deep 
control field observations for field star subtraction, instead, they used 2MASS 
data for the field population assessment, and ii) we have used reddened 
control population to match the depth and color of the likely field population in 
the cluster region, and iii) they used $K$-band luminosity function for 
their analysis, while we use $J$-band luminosity function. These factors could be 
the possible reasons for the discrepancy in the total cluster mass estimation.

The empirical relation between the mass of the most massive star 
($ m_\mathrm{max}$) of a cluster and its total mass ($M_\mathrm{ecl}$) is 
given by \cite{bon04} as: 
\begin{equation}
m_\mathrm{max}^\mathrm{Bonnell} = 0.39 \times M_\mathrm{ecl}^{2/3}.
\end{equation}
Using the \cite{bon04} relationship, we estimated that for a cluster like, 
IRAS 05100+3723, whose most massive star is a $\sim$20 \msun\, star. One would 
expect the total cluster mass to be $\sim$400 \msun, consistent with our 
cluster stellar mass estimation. Based on our results, it can be inferred that 
IRAS 05100+3723 is a moderate mass cluster ($100\msun < M_{cl} < 1000\msun$) in 
the classification scheme of \citet[][]{2010MNRAS.401..275W}.

\subsection{Physical Environment and Large-scale Distribution of Gas and Dust}
\subsubsection {Ionized gas Properties and Distribution}\label{sect_ion}
\begin{figure}
\centering{
\includegraphics[width=8.5cm]{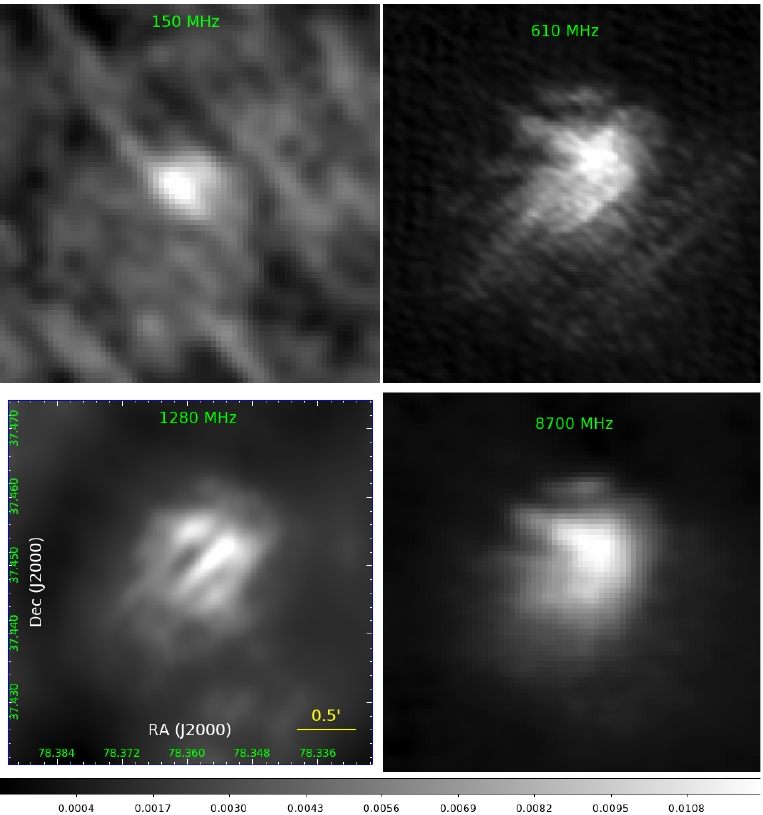}}
\caption{Radio maps of the \hii region S228 at 150, 610, 1280 and 8700 MHz, 
showing the distribution of ionized emission at different bands. The grey scale 
bar shows the intensity of radio emission in the units of mJy/beam.}
\label{radio_maps}
\end{figure} 

\begin{figure}
\centering
	\includegraphics[width=7.0cm]{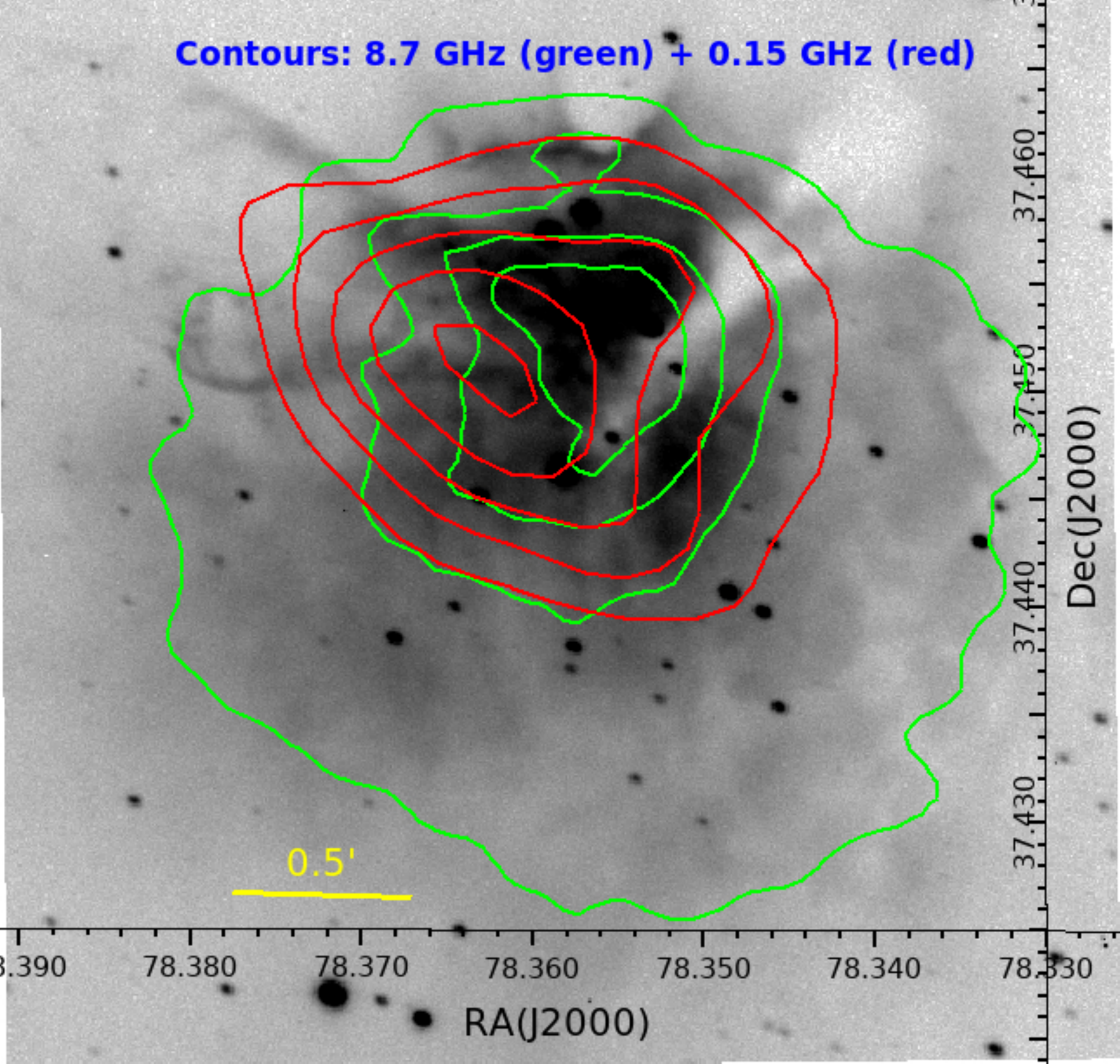}
 \caption{Comparison between emission at 8700 MHz (green contours), 150 MHz 
 (red contours) and optical \hal emission  (grey scale background image) of the 
 \hii region. The contours are drawn above the five times of the rms noise of 
 the respective maps. The 8700 MHz contour levels are at 0.0005, 0.0035, 0.0065, 
 and 0.0095 Jy/beam, whereas the 150 MHz contours are at 0.015, 0.022, 0.029, 
 0.036 and 0.044 Jy/beam.}
 \label{fig_radio_comp}
\end{figure}

\begin{figure}
\centering{
\includegraphics[width=8.0cm]{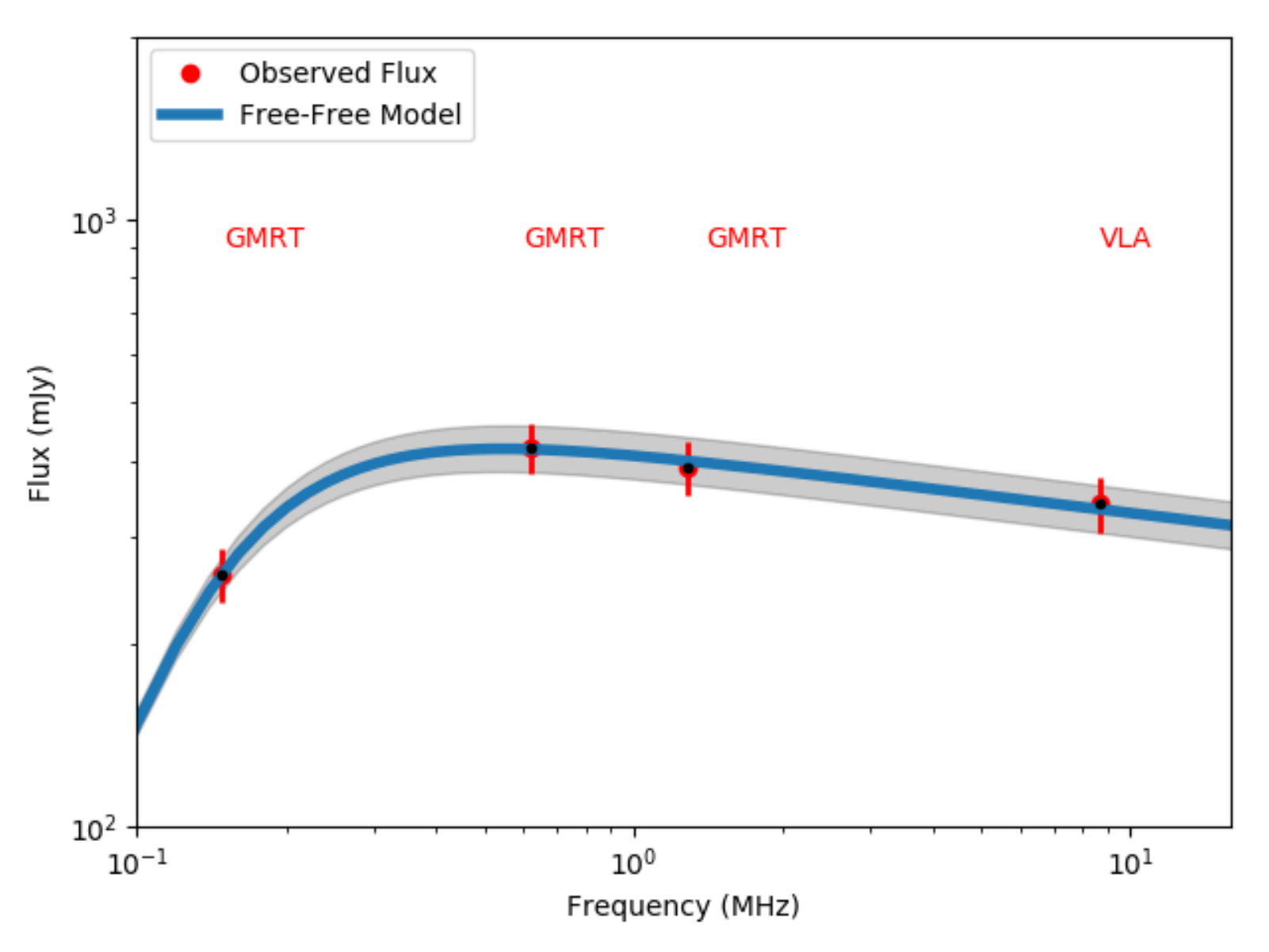}}
\caption{The SED of the \hii region S228 combining flux at 150, 610, 1280 and 
8700 MHz. These measurements represent the fluxes integrated  over a circular 
area of effective radius 0.7 arcmin ($\sim$0.7 pc). The solid line represents 
the fitted free-free emission model, while the 	shaded area represents 
the 1$\sigma$ uncertainty associated to the fit.}
\label{radio_spectrum}
\end{figure} 

As discussed in Section \ref{intro}, the cluster is associated with the \hii 
region S228. Figure \ref{radio_maps} shows the maps of the \hii region at 150, 
610, 1280 and 8700 MHz. These maps include GMRT observations at 610 MHz 
(beam size $\sim$6.6\arcsec $\times$ 3.4\arcsec, rms noise $\sim$0.2 mJy/beam) 
and 1280 MHz (beam size $\sim$10 \arcsec $\times$ 8\arcsec, rms noise $\sim$0.4 
mJy/beam), along with 8700 MHz map (beam  size $\sim$10.5\arcsec $\times$ 
7.5\arcsec, rms $\sim$0.1 mJy/beam) from the VLA archive and 150 MHz map 
(beam size $\sim$25\arcsec $\times$ 25\arcsec, rms $\sim$3.0 mJy/beam) from the 
GMRT TGSS survey. 

In interferometric observations, low-level diffuse emission is often found to 
be missing in high-resolution and/or high-frequency observations. Despite the 
fact that the 150 MHz map is of low-resolution, we find that compared to other 
bands the ionized emission at 150 MHz, however, is seen only in the central 
area (e.g. see Figure \ref{fig_radio_comp}). This could be due to the lower 
 sensitivity of the 150 MHz observations.

Normally, the free-free emission from a homogeneous classical \hii region shows 
a rising SED with flux (S$_\nu) \propto \nu^{2}$ at lower frequencies and almost 
flat SED with S$_\nu \propto \nu^{-0.1}$ at higher frequencies. However, the 
true behavior of S$_\nu$ with respect to $\nu$ strongly depends on the 
evolutionary status of the \hii region and can be well constrained with the 
thermal free-free emission modeling. Since the emission from 150 MHz is 
coming only from the inner region of an effective radius of $\sim$0.7\arcmin, we 
thus integrated fluxes at 610, 1280, and 8700 MHz maps over the same area as we 
did for 150 MHz. We note that before measuring the fluxes, we made 
low-resolution maps similar to the resolution of the 150 MHz map. Then, we 
convolved the maps to the exact resolution as the 150 MHz map. 
Figure \ref{radio_spectrum} shows the radio spectrum of the \hii region along 
with the thermal free-free emission model \citep{ver88} of the form:

\begin{equation}
S_{\nu} = \frac{2 k \nu^2}{c^2} \Omega_s T_e (1-e^{-\tau_{ff}}) 
\end{equation}
where the optical depth, $\tau_{ff}$ is expressed as
\begin{equation}
\tau_{ff}=0.082\times\left[\frac{EM}{{\rm cm^{-6} pc}}\right]\left[\frac{T_e}{{\rm K}}\right]^{-1.35}    \left[\frac{\nu}{{\rm GHz}}\right]^{-2.1}.
          \label{ff_fit}
\end{equation}
In the above equations, $k$ is the Boltzmann constant, $\nu$ is the frequency, $c$ 
is the speed of light in vacuum, $T_e$ is the electron temperature, $EM$ is 
the emission measure, and $\Omega_s$ is the source solid angle. Here, we 
opted $\Omega_s$ to be $\frac{2r\pi}{6}$ for a circular aperture of radius 
$r$ \citep{1967ApJ...147..471M}. The free-free emission model resulted in 
the electron temperature $T_e \simeq 5700$ $\pm$ 400 K and emission measure 
$EM \simeq$3.3$\pm$0.3 $\times 10^4$ cm$^{-6}$ pc. We then estimated the rms 
electron density ($n_e$) using the relation, $EM$ = $n_e^2$ $\times$ $l$, where 
$l$ is the path length and $n_e$ is the electron density. This yields 
$n_e$ $\simeq$ 165 $\pm$ 10 cm$^{-3}$ using source size as the path length. We 
note that these values represent the average properties of the \hii region over 
a radius of $\sim$0.7 \arcm, and if the \hii region is clumpy, as often the 
case, these values can be higher at peak positions of the clumpy 
structures. Nonetheless, we find that our estimates are in agreement with 
the electron temperature in the range 0.78 $\pm$ 0.13 to 1.02 $\pm$ 0.07 K 
\citep{fer17} and electron density in the range 180 $\pm$ 10 to 222 $\pm$ 10 
cm$^{-3}$ obtained by \citet{fer17} using the nebular analysis of the optical 
 emission lines. 

For optically thin free-free emission, the radio flux density is 
directly proportional to the flux of the ionizing photons. And as can be seen 
from Figure \ref{radio_spectrum}, the \hii region is optically thin at 
high frequencies. We thus estimate the total Lyman continuum photons 
$N_{Lyc}$ emitted per second from the ionizing star using the total integrated 
flux $S_{\nu}$ of the \hii region at 8700 MHz as per the  relation given 
in \citet{rub68} 
\begin{equation} 
N_{Lyc} = 4.76 \cdot 10^{48} \left( \frac{S_{\nu}}{\textnormal{Jy}} \right)
\left( \frac{T_e}{\textnormal{K}} \right) ^{-0.45} 
\left( \frac{\nu}{\textnormal{GHz}} \right) ^{0.1} 
\left( \frac{d}{\textnormal{kpc}} \right) ^2
\end{equation}
where  d is the distance of the region, while the meaning of the other terms are 
the same as in Equs. 3 \& 4. Using this relation, we estimate log ($N_{Lyc}$) to 
be $\simeq$ 47.70.

\begin{figure}
\centering{
\includegraphics[width=7.5cm]{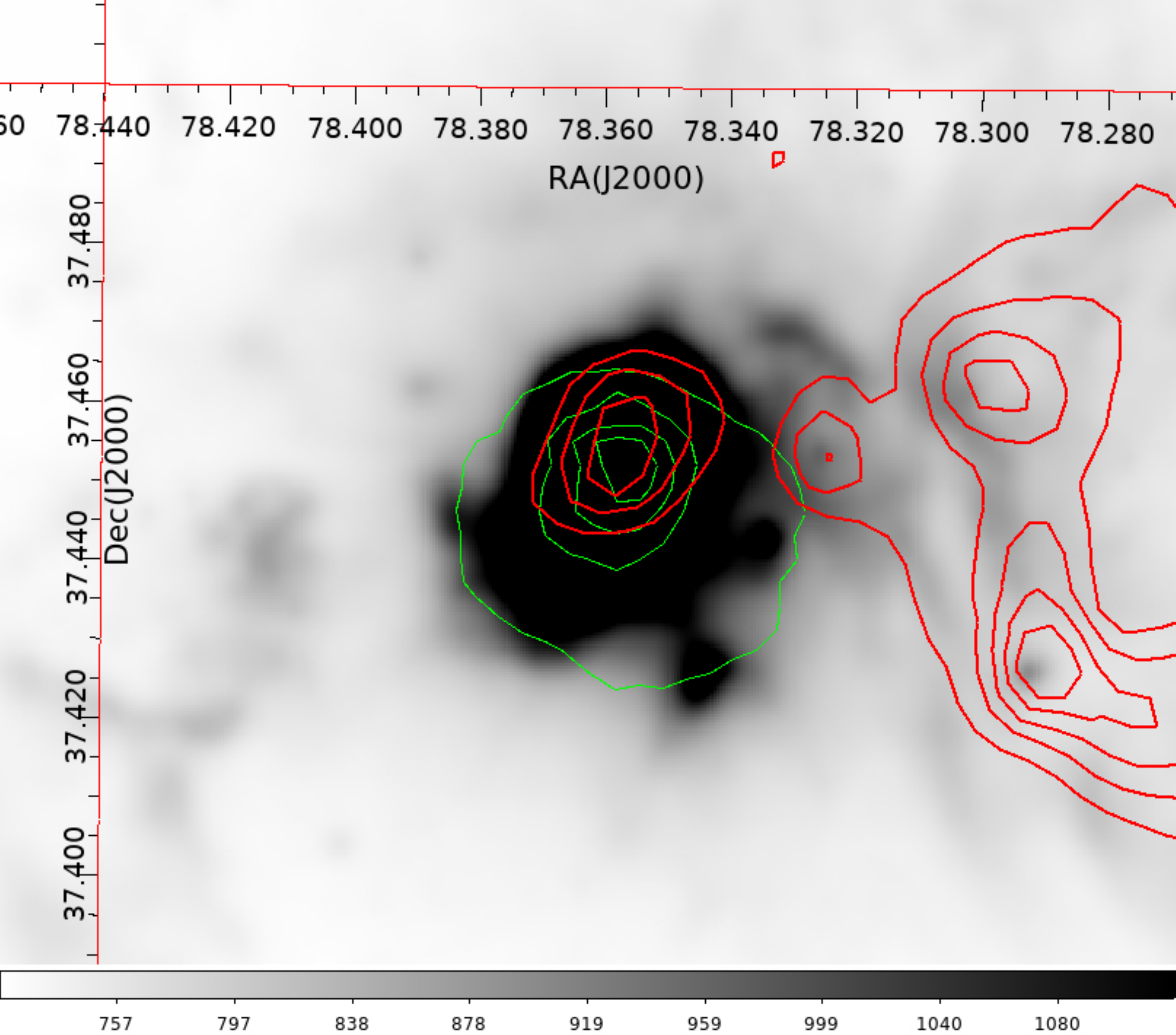}}
\caption{Image showing the distribution of WISE 12 $\mu$m  (background 
image) emission and ionized gas emission at 8700 MHz (green contours)  of  S228.  
The red contours show the distribution of cold dust of column density greater than 
3 $\times$ 10$^{21}$ cm$^{-2}$,  while the green contours reflect the 8700 
MHz emission above 0.0005 Jy/beam.}
\label{fig_pdr}
\end{figure} 

In general, photo-dissociation regions (PDRs) are found at the surface layer 
of molecular clouds surrounding the \hii regions. PAH molecules are strong tracers 
of PDRs \citep[see][and references therein]{sam07}. In this regard, WISE band 3 
with an effective wavelength of $\sim$12 $\mu$m is a good tracer of PDR as 
it contains emission lines of PAH molecules \citep[for details, see][]{and19}. 
We thus consider the bright diffuse 12 $\mu$m emission around the \hii region 
coming from the PDR region.  As can be seen from  Figure \ref{fig_pdr}, the 
8700 MHz emission is mainly distributed within the bright PDR, implying that most 
of the radio emission is coming primarily from the \hii region bordered by PDR. 

From spectroscopic observations, we know that the effective temperature of 
the massive star is $\sim$33000 $\pm$ 1000 K, thus one would expect that the 
minimum log($N_{lyc}$) from such a star to be $\simeq$ 48.10 photons per 
second \citep{mar05} considering the lower limit of the temperature. We find 
that this value is higher than log($N_{Lyc}$) estimated from the 8700 MHz 
emission, implying that the fraction of the photons could have been leaked into 
the ISM along  low-density pathways of the \hii region. We discuss this point 
further in Section \ref{star_form}.

\subsubsection{Dust Distribution and Properties}\label{s28_dust}

To investigate the physical condition of dust around the S228 region, we 
derived column density and dust temperature maps by performing a pixel-to-pixel 
modified black-body fit to the  160, 250, 350, and 500 \mum {\it Herschel} 
images following the procedure outlined in \citet{bat11,mal15}. 
Briefly, prior to performing the modified black-body fit, we converted  all the 
SPIRE images to the PACS flux unit (i.e., Jy pixel\textsuperscript{-1}). Then 
we convolved and regridded all the shorter wavelength images to the resolution 
and pixel size of the 500 \mum map. Next, we minimize the contribution of 
possible excess dust emission from each image along the line-of-sight by 
subtracting the corresponding background flux, which we estimated from a field 
nearly devoid of emission. Finally, we fitted the modified black-body on 
these background-subtracted fluxes. For the spectral fitting, we use a dust 
spectral index of $\beta$=2, and the dust opacity per unit mass column 
density $\kappa_{\nu} = 0.1~(\nu/1000~{\rm GHz})^{\beta}$ cm$^2$gm$^{-1}$  
as given in \citet{beck91}, keeping the dust temperature T$_{dust}$, and the 
dust column density $N$(H$_{2})$ as free parameters.
\begin{figure*}
\centering{
\includegraphics[width=18.5 cm]{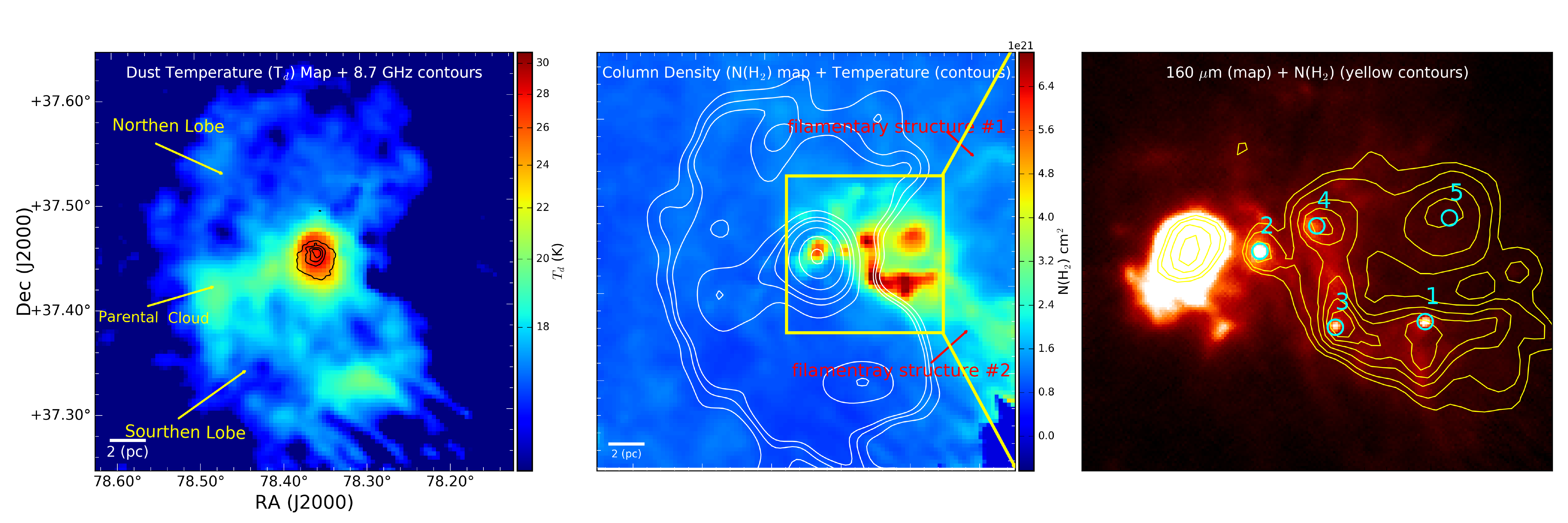}}
\caption {The left panel shows the dust temperature distribution of the 
S228 environment along with the distribution  of ionized gas (black contours). 
The contour levels of the  ionized gas are the same as in Figure 
 \ref{fig_radio_comp}. The warm dusty lobes and the  heated dust of the 
central parental 	cloud are also marked. The middle panel shows the dust 
column density distribution along with the dust temperature contours at 17.2, 
16.4,  17.6, 18.4, and 19.0 K. Two filamentary structures attached to the 
western clump  are also marked. The right panel shows the zoomed view of the 
western clump. The background image is the {\it Hershel} 160 $\mu$m map and 
the yellow contours  correspond to the dense gas above column density greater than 
3 $\times$ 10$^{21}$  cm$^{-2}$. The CUTEX identified compact sources  are labelled as $\#1$ to $\#5$. }
\label{fig_temp_den}
\end{figure*}

Figure \ref{fig_temp_den} shows the beam-averaged low-resolution 
($\sim$36${\arcsec}$)  dust temperature and (left) dust column density maps 
(middle) of the S228 complex, along with the emission at $Herschel$ 160 $\mu$m of 
the high column density region. As can be seen from the figure, the temperature 
of the dust is between 17 and 28 K, and is distributed over a much wider area 
(size $\sim$10 \arcm$\times$ 20 \arcm) than the size of the cluster 
(i.e. radius $\sim$2.5 \arcm). The temperature map shows an almost elongated 
 dust distribution similar to those in bipolar \hii regions \citep{sam18}. 
For example, similar to bipolar \hii regions, S228 displays two dusty lobes, 
each with a size of $\sim$10 \arcm\,  ($\sim$9.6 pc) extending nearly 
perpendicular to a faint warm dust lane seen at the base of the lobes. The peak 
of the temperature map coincides with the radio continuum emission, implying 
the high-temperature zone is primarily created by the intense UV radiation 
coming from the cluster. In contrast to the temperature map, the column density 
map displays a lower dust column density within the bipolar \hii region, but 
exhibits higher column density structures with density in the range 
2$-$7 $\times$ 10$^{21}$ cm$^{-2}$ in the western part of the \hii region. 
In particular, in the immediate vicinity of the \hii region lies a nearly 
a semicircular clumpy structure  with a  radius $\sim$2.8 pc and column 
density $>$ 3 $\times$ 10$^{21}$ cm$^{-2}$. In general, molecular clumps have 
typical sizes of a few parsecs and are sites of active star formation. Thus, it 
could be a site of ongoing star formation as column densities above  
$\sim$5 $\times$ 10$^{21}$ cm$^{-2}$ are, in general,  observed to be sites of 
recent star formation \citep{lad10}. In fact, we found four 70/160 $\mu$m 
point sources within the condensation affirming the above hypothesis (see 
Figure \ref{fig_temp_den} right panel). The average dust temperature of the clump 
is  $\sim$16 K, while the average column density is $\sim$4 $\times$ 10$^{21}$ 
 cm$^{-2}$. The column density map also illustrates that the clump is 
associated with two filamentary structures in the north-western direction, 
and south-western directions, respectively. The north-western filament is 
faint and narrow with a mean temperature around 15.5 K, while the 
south-western filament is slightly structured with temperature in the range 
14.5$-$15.5 K.

We estimated the total mass ($M_{gas}$) of  the clump above column density 
$\sim$3 $\times$ 10$^{21}$ cm$^{-2}$ using the  following equation: 
\begin{equation}
 M = \mu {\mathrm m_{H}} A_{\mathrm pix} \Sigma {\mathrm H_{2}} \, \label{eq:mass}
\end{equation}
 where $\mu$ is the mean molecular weight, ${\mathrm m_{H}}$ is the mass of 
 the hydrogen atom, $\Sigma$H$_2$ is the integrated H$_2$ column density, 
 and $A_{\mathrm pix}$ is the area of a pixel in cm$^{-2}$ at the distance of 
 the region. The above approach yields area, mass, and density of the clump to 
 be  $\sim$28 pc$^2$, $\sim$2700 M$_{\odot}$, and $\sim$350 cm$^{-2}$, 
 respectively. We find that the properties of the clump are similar to the 
 nearby Ophiuchus star forming region which is one of the youngest (age $\sim $1 
 Myr) and closest (distance $\sim$125 pc) star-forming regions having a size of  
 $\sim$29 pc$^2$ and mass $\sim$3100 \msun\, \citep{dun15}. 

 The stability of a clump against gravitational collapse can be evaluated using 
 the virial parameter \citep{kau13} 

\begin{equation}
\mathit{\alpha}=1.2\left(\frac{\sigma_{\upsilon}}{\rm km \ s^{-1}}\right)^{2}\left(\frac{R_{\rm eff}}{\rm pc}\right)\left(\frac{M}{10^{3}M_{\odot}}\right)^{-1}
\end{equation}

where $\sigma_{\upsilon}$ is the one-dimensional velocity dispersion, 
$R_{\rm eff}$ is the effective radius, and $M$ is the mass of the clump. In 
general, $\alpha$ $<$ 1 is suggestive of collapsing clumps while $\alpha$ $>$ 2 is 
suggestive of dissipating clumps and  $\alpha$ $\sim$1$-$2  describes a clump that 
is in approximate equilibrium. However, the external pressure confined clumps 
with $\alpha \leq 2$ are also found to be bound and can live longer with respect to 
the dynamical timescale \citep{ber92}. In the present case, using $\sigma$ = 
1.3 \kms estimated over clump area from the \tco map (discussed in Section 
 \ref{star_form}), we find $\alpha$ $\sim$1.6, implying that clump may still be 
 bound.

\subsubsection{Herschel Compact Sources and Properties}\label{her_ps}
 From the right panel of Figure~ \ref{fig_temp_den}, it appears that the 
 western clump has been fragmented into five compact structures (marked with 
 number \#1 to \#5), four of which (sources \#1 to \#4) are protostellar as each 
 of them is associated with a 70/160 \mum point source, while the source \#5 is 
 not associated with any point like source, thus is likely a prestellar source. 
 The presence of 70/160 \mum point sources within column density peaks suggests 
 a fresh star formation is happening at the western border of the nebula. 

\begin{figure}
    \centering
    \includegraphics[width=8.5cm]{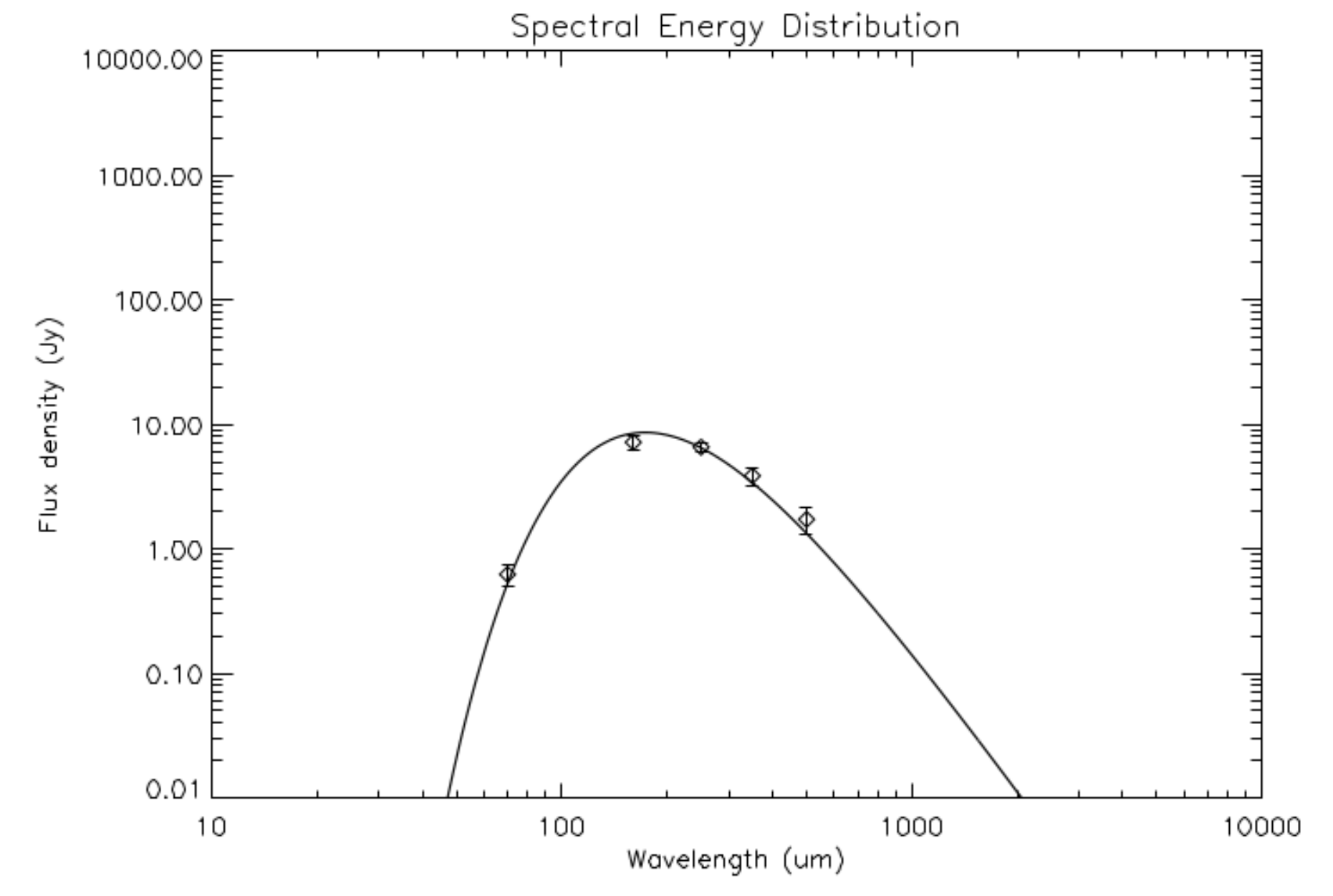}
    \caption{The SED of the embedded source \#4. The circles denote the observed 
    flux values. The black solid curve shows the grey-body fit to the data points, 
    between 70 \mum and 350 $\mu$m. }
    \label{fig_grey}
\end{figure}

In order to understand the properties and evolutionary status of the 160 \mum 
point sources, we extracted  their far-infrared fluxes between 70 and 500 \mum 
using the CUrvature Thresholding EXtractor (CUTEX) software described in Molinari 
et al. (2011). The CUTEX was specifically developed to optimize the source 
detection and extraction in the spatially varying background like the emission 
seen in the {\it Herschel} maps of the star-forming environments 
\citep[e.g.][]{mol11}. In order to estimate the envelope temperatures of 
the identified point sources, we fitted the observed fluxes at 70, 160, 250, 
and 350 \mum with the modified blackbody model. Fluxes at 500 \mum have 
been excluded, owing to the low-resolution of the 500 \mum beam. This is done 
in order to avoid bias in the fitting procedure due to the overestimation of 
fluxes at 500 \mum because of source confusion and the inclusion of excess 
background emission. We also adopted  20 percent error in flux values instead 
of formal photometric errors, in order to avoid any possible bias caused 
by underestimation of the flux uncertainties. We fitted the modified blackbody of 
the form, $S_{\nu} = A\, \nu^{\beta}\, B_{\nu}(T_{\rm dust})$,  where $S_{\nu}$ 
is the observed flux distribution, $A$ is a scaling factor, $B_{\nu}(T_{\rm dust})$ 
 is the Planck function for the dust temperature $T_{\rm dust}$~, and $\beta$ is 
the dust emissivity spectral index. A sample SED is  shown in  Figure \ref{fig_grey}
 for $\beta$=2. 

\begin{figure}
\centering{
\includegraphics[height=8cm, angle=270]{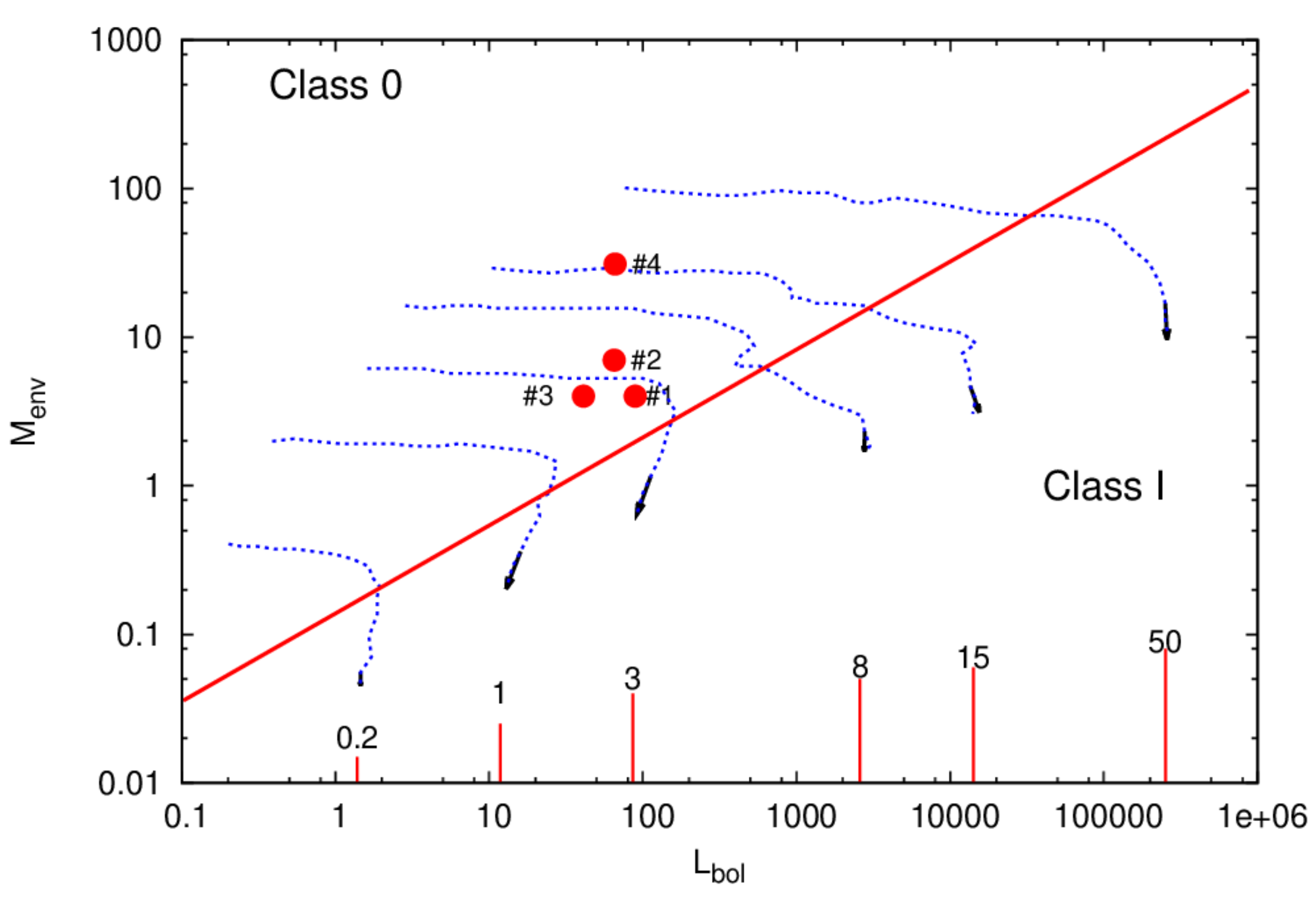}}
\caption{ The (L$_{bol}$, M$_{env}$) diagram of the candidate protostars detected 
in the western clump (ID numbers are the same as in Figure \ref{fig_temp_den}). 
The dashed blue lines represent the evolutionary tracks from \citet{and08}. 
Evolution proceeds from the upper left to the lower right (indicated by arrows at 
the	end of each track). The final stellar masses of these tracks, in solar units, 
are given above the vertical lines. The slanted red line  corresponds to the  
location where 50\% of the initial core mass is converted into stellar mass  \citep{bon96,and00}.}
\label{fig_hr}
\end{figure}

Having derived dust temperature, we then determined the total mass $Mass$ (gas $+$ 
dust) of the  envelope   from the dust continuum flux, $S_{\nu}$, using the 
following equation: 
\begin{equation}
{\rm\it Mass} = \frac{S_{\nu} \,R d^{2}}{\kappa_{\nu} \, B_{\nu}(T_{dust})}
\end{equation}
where $d$ is the distance, $R$ is the gas-to-dust ratio and is considered to be 
100, and $\kappa_{\nu}$ is the dust mass opacity.  We use the same dust opacity 
law as used in Section \ref{s28_dust}.We also estimated  bolometric luminosities 
of the sources by integrating SED between 2 and 1000 \hbox{$\mu$m}. We note 
that, since we have not taken into account the wavelength dependence of $\beta$, 
 therefore, the uncertainty in the mass can be of a factor of two,  as discussed 
in \citet[][]{deh15}. Moreover, if the compact structures contain multiple 
point sources that are unresolved at {\it Herschel} bands, the derived mass 
and luminosity can be even more discrepant. Nonetheless, taking these properties 
of the sources at face value, we infer their likely evolutionary status  by 
plotting them on the M$_{env} -$ L$_{bol}$ diagram. Figure \ref{fig_hr} shows 
the M$_{env}$ $-$ L$_{bol}$ diagram of the sources along with the evolutionary 
tracks of protostellar objects from \citet{and08}. The figure also marks the zones 
of Class 0 and Class I sources. As can be seen, all the sources lie in the 
Class 0 zone of the plot and the evolutionary tracks in the plot indicate that 
these objects would evolve into stars of the mass range $\sim$3 $-$ 15 $\msun$. 
This implies that the condensation is possibly a site of low- to 
intermediate-mass star formation. 

\section{Discussion}\label{s28_disc}

\subsection{Gas Removal and Dynamical Status of the Cluster}
 Stellar feedback plays an important role in the removal of gas from star 
 clusters, and subsequently cluster members dissolve completely into the 
 galactic field \citep{lada03}. It is suggested that by comparing the age of a 
 star in the cluster with its crossing time ($T_{cr}$) one can distinguish 
 expanding clusters from bound clusters. According to \citet{gieles11}, 
 $\frac{T_{cr}}{Age} \sim 1$, separates bound clusters ($T_{cr}$ $<$ Age) from 
 the unbound associations ($T_{cr}$ $>$ Age). The general definition of $T_{cr}$ 
(=  2$R_H/\sigma$) includes half mass radius ($R_H$) and the root-mean-square 
velocity dispersion ($\sigma$) of the member stars \citep{bin08}. Since we do not 
have velocity measurements of stars, we thus use the expression 
$T_{cr} = 2.8({R_V^3}/{GM})^{1/2}$, to estimate the crossing time; where $M$ is 
the total mass of the system,  $G$ is the gravitational constant and is $\simeq$ 
0.0045 pc$^3$ \msun$^{-1}$ Myr$^{-1}$, and $R_V$ is the virial radius of the cluster 
\citep{gie10, wei07}. The latter is related to the $R_H$ as $R_V$ = 1.25$\times 
R_H$. We made an approximate estimate of the $R_H$ from the cluster center as 
the radius where the star count above the photometric completeness level is 
exactly half of the total number of stars within the cluster radius, and it turns 
out to be 1.07 pc. We assume that the mass of each star is roughly $\sim$0.5 
\msun, based on the fact thatf the mass distribution of stars in young clusters 
peaks somewhere between 0.2 to 0.7 \msun\, \citep{dam21}. Using this approach, 
we estimated  $T_{cr}$ to be $\sim$3 Myr, implying that the cluster is 
probably  marginally bound or in its initial stage of expansion, considering its 
age $\sim$3 Myr. It is worth noting that recently \citet[][]{kuhn19} studied  
a sample of 28 clusters and associations with ages $\sim$1-5 Myr using 
PM measurements from {\it Gaia} DR2 and revealed that at least 75\% of these 
systems are expanding. 

\subsection{Comparison to Other Young Clusters}
\label{diss_comp}
\citet{lada03} examined 
a sample of young embedded clusters within 2 kpc from the Sun with deep NIR 
observations and have tabulated their properties. We find that IRAS 05100+3723 is 
more massive than the majority of the nearby embedded  clusters, except the Orion 
Nebula Cluster (ONC). The ONC is one of the nearest ($\sim$450 pc), young 
($\sim$2 Myr), massive ($>$1000 \msun) clusters, and hosts four massive stars 
of 
spectral type between B0 and O7
\citep[for details see][and references therein]{2018AJ....155...44P}. 
Compared to ONC, IRAS 05100+3723 is a slightly more evolved (age $\sim$3 Myr) and 
less massive (mass $\sim$500 \msun) cluster and hosts only a 
single massive star of spectral type O8.5.  We find that in terms of mass, age, and 
size, the studied cluster resembles the cluster Stock 8,  studied by 
\citet{jose17}. Stock 8 is a young cluster of age $\sim$3 Myr with the most massive 
star being a star of spectral type between O8 and O9 and total stellar mass 
 $\sim$580 \msun. We also find that IRAS 05100+3723 lies well below the mass-radius 
 relation ($mass \propto r^{1.67}$) given by \citet{pfa16}, derived for embedded 
 clusters. This could be due to the fact that the cluster is possibly expanding 
 and dispersing into the Galactic ISM. As a result, the cluster is not compact 
 anymore, while embedded clusters are, in general, bound and compact systems. 
 In fact,  we find that IRAS 05100+3723 lies between embedded and loose clusters 
 in the  radius-age plane of \citet{pfa13}.  Although, the total \av in the 
 direction of the cluster is 3.3 mag, but extinction intrinsic to the cluster is 
 only $A_V\sim$1 mag, which also points towards  the fact that the 
 cluster is probably no more an embedded cluster.

\subsection{Star Formation Processes and Activity in the Complex}

\begin{figure*}
    \centering
    \includegraphics[width=14cm]{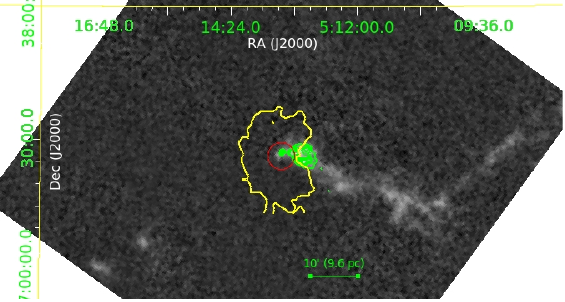}
    \caption{The image showing the distribution of \tco intensity in the gray scale. 
    The outer boundary of the $Herschel$ dust temperature map is shown with a yellow 
    contour with $T_d$ = 17.2 K. The extent of the cluster is shown with a 
    red circle. And the $Herschel$ column density contours above density 
    3 $\times$ 10$^{21}$ cm$^{-2}$ (in the direction of western clump) are shown 
    in green. The contour levels are same as given in the right panel of 
    Figure \ref{fig_temp_den}.  }
    \label{fig_sfr}
\end{figure*}

\label{star_form}
\begin{figure}
    \centering
    \includegraphics[width=8.5cm]{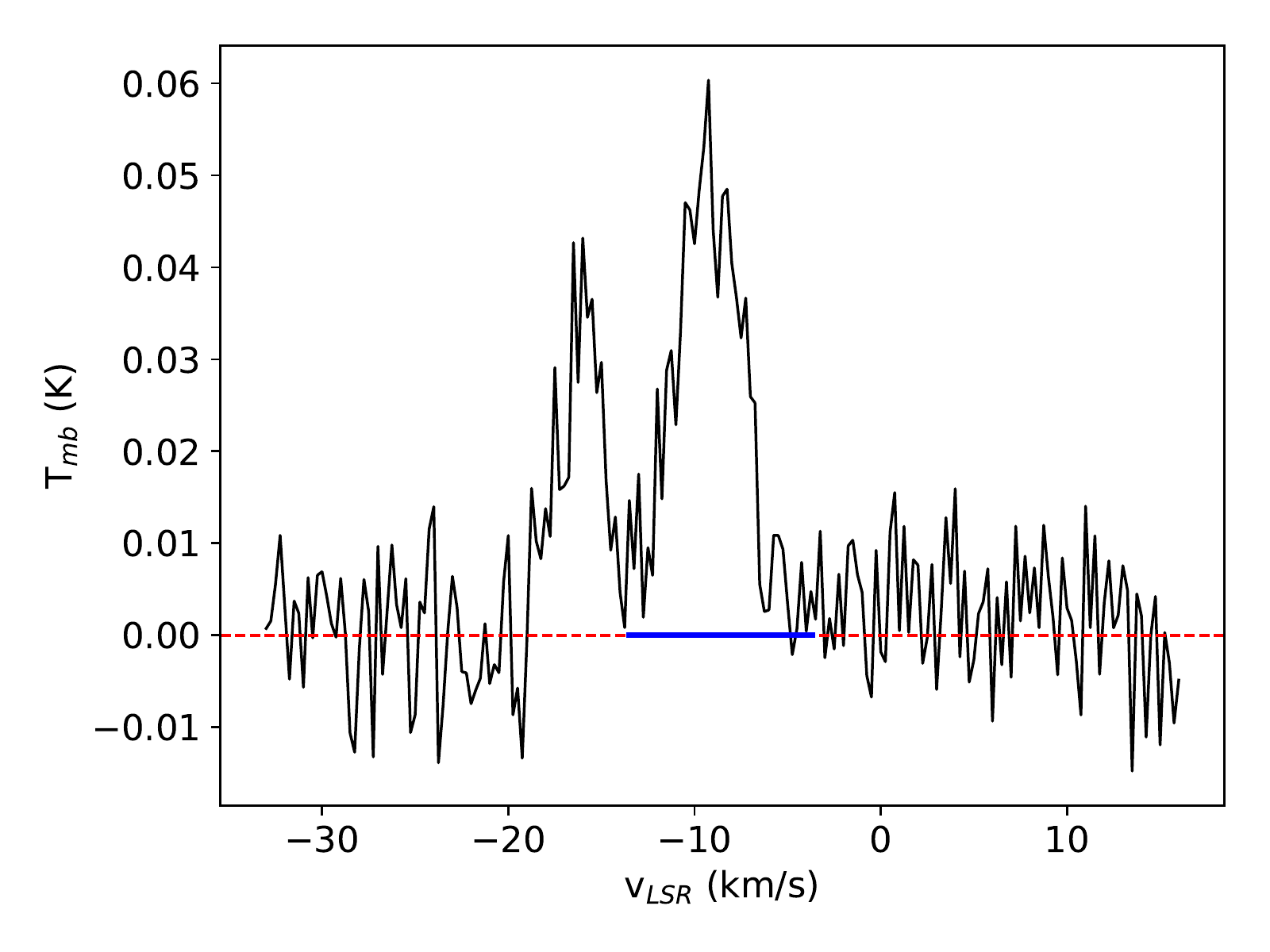}    
    \caption{Average  \tco  spectrum  observed  towards  the  S228 complex. The 
    red and blue lines indicate the entire velocity range and velocity range associated with  the S228 complex. }
    \label{co_comp}
\end{figure}

\subsubsection{Cloud Structure, Cluster Formation, and Leaking fraction of the Ionized Gas}
\label{s28_str}

Figure \ref{fig_sfr} shows the \tco gas distribution in the complex, along with  
warm and cold dust distributions derived from the temperature and column 
density maps. The \tco intensity map was made using the observations taken with 
the FCRAO telescope at a spatial resolution of $\sim$45 \arcs\, and a 
velocity resolution $\sim$0.25 \kms (provided by Mark Heyer, in a 
private communication). In the \tco spectrum, two velocity components are found 
in the direction of S228, as shown in Figure \ref{co_comp}. We obtained the 
moment maps (i.e. intensity, velocity, and velocity dispersion maps) of S228 
region using its velocity component in the range $-$13.5 to $-$3.5 \kms, the 
velocity range corresponding to S228 \citep[e.g.][]{chen20}. As shown in 
Figure \ref{fig_sfr}, the distribution of the \tco intensity suggests that 
the cluster lies at the end of a filamentary cloud. It is generally hypothesized 
that star-forming clouds collapse to lower-dimensional structures, 
producing first sheets and then filaments \citep{lin65, bur04,gom14,nar20}.
Simulations show that such clouds experience highly non-linear 
gravitational acceleration as a function of position, causing the material to 
pile up near the cloud edge, forming a young cluster like Orion  
\citep[e.g.][]{bur04,hart07, hei08}. 

From the \tco moment 1 map, we find that the \vlsr~ of the \tco molecular gas in 
the zone of the ionized gas is $\sim -$5.5 \kms, while the \vlsr\, of the 
ionized gas lies in the range $-$9.4 to $-$13.6 \kms \citep{isr77,chen20}. The 
mean velocity of the ionized gas differs from that of the molecular gas 
by 4$-$8 \kms, implying that the ionizing gas is possibly streaming away from 
the cloud. This behavior is typical for \hii regions where massive stars form \\
near the very edge of a molecular cloud \citep{ten79} or in the center of a flat 
or sheet-like cloud \citep{bod79a}, resulting in an easy flow of the ionized 
gas along the low-density paths of the parental cloud.

Since the morphology of the warm  dust is more like the morphology of a 
bipolar \hii region, we thus hypothesized that the \hii region might have formed 
at the very end of a sheet-like or flattened  cloud containing a central filament 
as advocated in \citet[][]{deh15}. In such clouds density along the equatorial 
axis is expected to be high, whereas it is expected to be low in the polar 
 directions. Considering that the ionized gas is streaming away from the 
 filamentary cloud at a minimum velocity of $\sim$5 \kms, one would expect that 
 the ionized gas to reach  $\sim$15 pc in 3 Myr (i.e the age of the cluster) from 
 its original location. This is comparable with the projected size, $\sim$10 pc, 
of the warm dusty lobes seen in the temperature map. This also suggests that 
a fraction of the ionizing photons could have leaked into the surrounding ISM 
 heating the dust up to several parsecs. Comparing the Lyman continuum photons 
  expected from the ionizing star of the \hii region with the observed 
  Lyman continuum  photons derived from radio observations within the bright 
  PDR zone, we find that approximately 60\% of the  Lyman photons likely have 
  escaped from the \hii region into the diffuse ISM during the lifetime of the 
  star. If we consider the uncertainty in the estimation of temperature and 
  density of the ionized gas, then also the escape fraction is in the range 
  of 40$-$50\%. Our estimated escape fraction is in good agreement with the 
  values found for other \hii regions \citep[e.g.][]{oey97,pel12}. We note, 
  the escape fraction also strongly depends on the age of the Lyman photon 
  emitting source, and structure and geometry of the medium  \citep{bod79a,yor82,how17}; 
  thus, may vary from region to region. 

\subsubsection{Compression and Confinement of Cold Gas, and Formation of Dense Clump} \label{s28_clump}
 Figure \ref{fig_sfr} shows that the western clump (shown by green contours) 
 lies close to the \hii region. The clump displays bow-like morphology with its 
 apex facing the \hii region, as found in numerical simulations of \hii regions
 expanding into collapsing molecular clouds \citep[e.g.][]{wal15}. This suggests 
 that the over-pressured expanding \hii region possibly has compressed and pushed 
 the western clump to its present shape. To evaluate the degree of interaction 
 between the \hii region and the clump, we evaluated various average pressures 
 within both the regions using the equations given below. We estimated pressure 
 due to ionized gas of the \hii region using the following relation: 
\begin{equation}
\begin{split} 
P_{\hii} & =\sqrt{3Q_{\rm H}\over 4\pi\alpha_{rec} r^3}kT \\ 
& \approx 10^{-9}\, 
\left(\frac{Q_{\rm H}}{2\times10^{49}\,\textrm{s}^{-1}}\right)^{1/2}
\left(\frac{0.72\, \textrm{pc}}{r}\right)^{3/2}\, \textrm{dyn cm}^{-2}\,
\end{split}
\end{equation} 
where $r$ is the radius, $Q_{\rm H}$ is the Lyman continuum photon responsible 
for the ionization of the \hii region, and $\alpha_{rec}$ is the 
recombination coefficient \citep[for details see Eq. 6 of][]{mur09}. $Q_{\rm H}$ 
 is taken to be $\sim$5.1 $\times$ 10$^{47}$ photons per second, while for the clump 
it is assumed to be zero as no hyper or ultra-compact \hii regions present in 
our high-resolution 8700 MHz image. We estimated radiation pressure using 
the following relation: 
\begin{equation}
\begin{split}
   P_{rad} & = L / 4 \pi c r^2 \\
   & = 4.4 \times 10^{-15}\, 
    \left(\frac{L}{ \lsun} \right)
    \left(\frac{r}{\textrm{ pc}} \right)^{-2}\, \textrm{dyn cm}^{-2}\,
\end{split}
\end{equation}
where we adopt the ionizing star of the \hii region as the dominant source of 
stellar luminosity ($L$), while in the clump the $L_{bol}$ of the most massive 
 protostar is adopted as a proxy for stellar luminosity. We then estimated the 
  turbulent pressure using the following relation: 
\begin{equation}
\begin{split}
P_{turb} & =  \rho \sigma_{nt}^2 \\ 
      &= 4.7 \times 10^{-14}\, 
    \left(\frac{n}{\textrm{cm}^{-3}} \right)
    \left(\frac{\sigma_{nt}}{\textrm{km\ s}^{-1}} \right)^{2}\, \textrm{dyn cm}^{-2}\,
\end{split}
\end{equation}
within the \hii region and the clump, where $\rho$ is the mass density, $n$ is 
the particle density, and 
\begin{equation}
\begin{split}    
    \sigma_\mathrm{nt} =\sqrt{ \sigma_\mathrm{obs}^{2}-\sigma_\mathrm{th}^{2}}
\end{split}
\end{equation}    
is the non-thermal velocity dispersion of the gas. Assuming a Gaussian 
distribution of the line profiles, $\sigma_\mathrm{obs}$ and $\sigma_\mathrm{th}$ 
can be estimated from line-widths ($\Delta V$) using 
\begin{equation}
    \sigma_\mathrm{obs}  =  \frac{\Delta V_{obs}}{2.35} \mbox{  and  } \sigma_\mathrm{th}= \frac{\Delta V_{th}}{2.35}.
\end{equation}
For the ionized gas within the \hii region, we use the observed line-width of 
the hydrogen radio recombination (RRL) line from \citet{chen20}, while for the 
clump we use the line-width of the \tco line within the clump area.  For 
the molecular gas, we use the expression, 
${\Delta V_{th}} = \sqrt{\frac{8\it{ln2}kT}{ m_\mathrm{co}}}$,  where $k$ is 
the Boltzmann's constant,  $T$ is the kinetic temperature of the molecular gas, 
and $m_\mathrm{co}$ is the mass of \tco molecule in amu. For the ionized gas, we  
use the expression, ${\Delta V_{th}} = 21.4\sqrt{\frac{T_\mathrm{e}}{10^4 K}}$, 
for the hydrogen atom \citep{gar99}, where $T_\mathrm{e}$ is the electron 
 temperature. Lastly, we estimated thermal pressure using the equation: 
\begin{equation}
\begin{split}
P_{them} & =  2nkT \\ & = 2.8 \times 10^{-16}\, 
    \left(\frac{n_e}{\textrm{cm}^{-3}} \right)
    \left(\frac{T}{K} \right)\, \textrm{dyn cm}^{-2}
\end{split}
\end{equation}
where we use the  mean density and mean temperature of the ionized gas and cold dust. 
Doing so, we estimated the total pressure within the ionized region 
\begin{equation}
     P_{ionized}  = P_{\hii} +  P_{\rm {rad}} + P_{\rm {turb}} + P_{\rm {them}}
\end{equation}
 to be $\sim$4.5 \into 10$^{-8}$ dyn cm$^{-2}$, while the total  pressure 
 within the clump
\begin{equation}     
     P_{\rm clump} =  P_{\rm {rad}} +  P_{\rm {turb}} + P_{\rm {therm}}
\end{equation}    
to be $\sim$2.3 \into 10$^{-9}$ dyn cm$^{-2}$, implying that the \hii region must 
still be compressing the clump, as a result the clump may be under external 
pressure confinement. However, since the typical average pressure of the ISM is 
in the range of 10$^{-11}$ $-$ 10$^{-12}$ dyn cm$^{-2}$ \citep{blo87, dra11}; 
thus, we hypothesized that the \hii region must be expanding more rapidly into 
the ISM. In the present case, the expansion may be occurring more preferentially 
in the direction perpendicular to the plane of the cloud, as a result, 
we are observing warm dusty bipolar lobes. 
\begin{figure*}
    \centering
    \includegraphics[width=18.5cm]{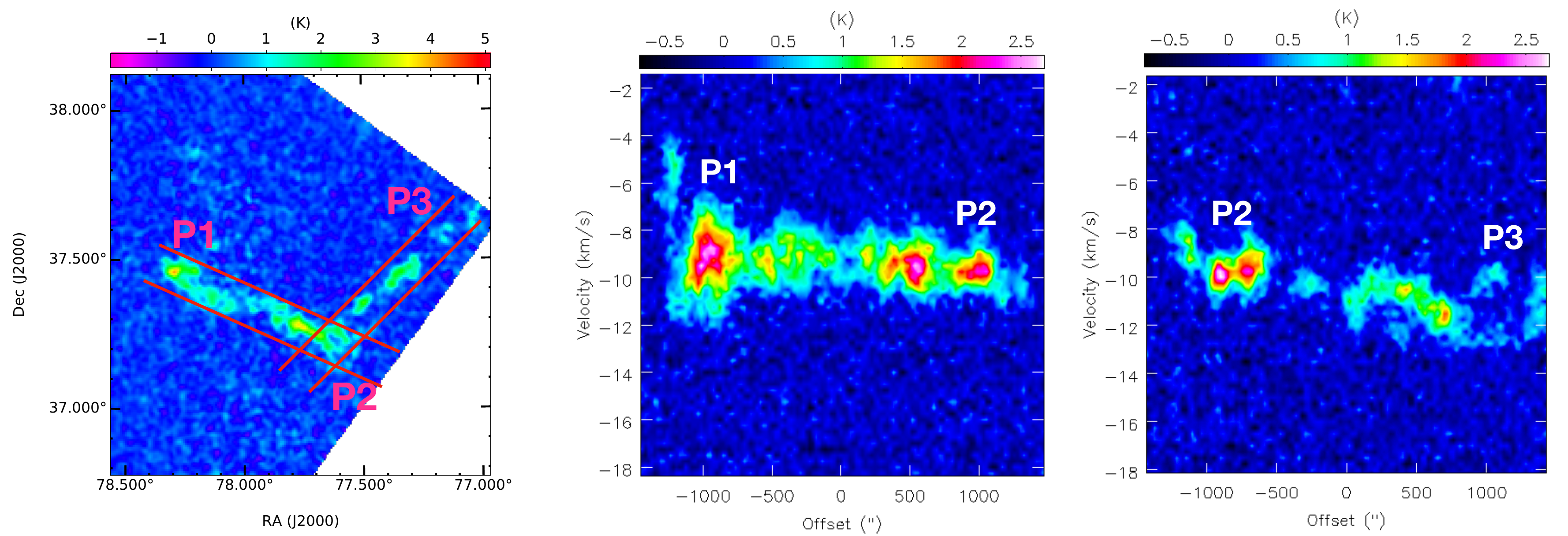}
    \caption{Position-velocity (PV) maps along the long axis of the 
    L-shaped filament. The left panel shows the \tco intensity map of the filament 
    in Galactic longitude and latitude coordinates, the middle panel shows the
    PV map of the P1$-$P2 filament, while the right panel shows the PV map of 
    the P2$-$P3 filament.}
    \label{pv_map}
\end{figure*}

 {\it Herschel} column density map shows (see Section \ref{s28_dust}) that the 
 clump is located at the junction of two filamentary structures, while in the 
 large-scale \tco map, the clump seems to be located at the end of a long 
 L-shaped filamentary structure.  This long filamentary structure corresponds  
 to the  south-western filament seen in the {\it Herschel} image. The non-visibility
 of the  small north-western filament in the \tco map could be due to the 
 lower sensitivity of the \tco map.  We note that the small-scale 
 filamentary structures can be generated due to the self-gravity of the 
 clump. Nonetheless, the situation is very similar to several other 
 star-forming regions where massive clumps have been found at the merger 
 or convergence point of filaments \citep[e.g.][]{mye09,sch12,kum20} or at the end 
of a large-scale filament due to edge collapse \cite[e.g.][]{bur04, pon11, 
yuan20}. The large-scale velocity gradients along the filaments have often 
been interpreted as signatures of mass flow towards star-forming clumps as 
a consequence of the longitudinal collapse of the filaments \citep{kirk13,  
rya18, dut18}.  Although the structure of the filament is not very smooth, 
which could be due to star-formation activity at multiple locations of the 
filament as the filament is presumably older than 3 Myr (i.e. the age of 
the cluster), nonetheless, we search for signatures of underlying large-scale 
flow (i.e. signatures of velocity gradient) along the filament's long axis in 
the position-velocity (PV) map. Figure \ref{pv_map} shows the PV maps of 
the filament, extracted along its spine. To extract the PV maps, we divide 
the L$-$shaped filament into two parts (P1$-$P2 and P2$-$P3) as marked in 
Fig \ref{pv_map} (left panel). Figure \ref{pv_map} (middle panel) and 
Figure \ref{pv_map} (right panel), show the PV maps of P1$-$P2, and P2$-$P3, 
 respectively.  In PV maps,  the velocity gradient in P1$-$P2 is weaker compared 
 to  the one observed for the P2$-$P3 region. Overall, the velocity gradient in 
 the whole  filament is $\sim$0.05 \kms. This weak velocity gradient could be due 
 to the fact that the  filament, being $\sim$3 Myr old (cluster age), star 
 formation along the filament has already distorted its gas kinematics and also 
 has changed the location of star-forming  potential; thus, the kinematics 
 at smaller-scale during its evolution \citep[e.g. see discussions in][]{pere14}.
 Nonetheless, the velocity gradient (i.e $\sim$0.1 \kms pc$^{-1}$) observed 
 within P2$-$P3 is comparable to some of the large-scale filaments such as 
 IRDC G035.39$-$00.33   \citep[$\sim$0.2 \kms pc$^{-1}$;][]{soko17} and W33 
  filamentary system \citep[0.3$-$0.1 \kms pc$^{-1}$;][]{liu21}. This points to 
  the fact that the filament has likely been dynamically active for a few Myr, 
  thus might have supplied cold matter to the cluster location. However, 
  high-resolution spectral observations of the filament close to the clump 
  location will be essential to better understand the filamentary flow.

\subsubsection{Second Generation Star Formation}

In the filamentary environment, numerical simulations by \citet{fuk00} suggest 
that the expansion of \hii region  can generate sequential waves of star-forming 
cores along the long axis of the filament on either side of the \hii 
region. This has particularly been observed in bipolar \hii regions, which 
are thought to be formed due to anisotropic expansion of the \hii region in a flat 
or sheet-like cloud containing filaments \citep{deh15, sam18}. Indeed, 
\citet{sam18}, from the analysis of a sample of bipolar bubbles found that the 
most massive and compact clumps with signatures of massive star formation are 
always located at the waist of the bipolar bubbles and they argue that these 
massive clumps are the possible sites of second-generation massive- to 
moderate-mass star formation. \citet{esw19} using magnetic field geometry, 
strength, and comparing various pressure components, showed that the evolution 
and star formation of the clumps at the waists of bipolar \hii regions 
are indeed strongly influenced by the \hii region feedback.

The morphology of the S228 in WISE 12 $\mu$m and {\it Herschel} temperature map 
appears  to be bipolar with two lobes extending nearly perpendicular to a faint dust 
lane located at the bases of the lobes. At  the western waist of the bubble lies a 
clump whose mean temperature and mean column density are 16 K and 
$\sim$4 $\times$ 10$^{21}$ cm$^{-2}$, respectively, which are conducive for 
the process of star formation \citep{lad10,eden19}. 

Within this dense clump, as discussed in Section \ref{her_ps}, several 
Class 0/I protostars and a starless core, have been identified. The fact that the 
age of Class 0/I sources is of the order of $\sim$10$^{5}$ yrs \citep{evans09}, 
and these sources are located in a clump that lies at the junction point of the 
\hii region and filament, and the clump is under the influence of a  $\sim$3 Myr 
old \hii region, altogether suggest that these protostars are the likley 
second generation stars of the complex.

\section{Summary}\label{s28_conc}
In order to understand the formation of young massive clusters and their 
feedback effects on the parental cloud, we investigate the young cluster 
IRAS 05100+3723 and studied its environment using multiwavelength data sets. 

Our findings conclude that IRAS 05100+3723 is an intermediate-mass 
(mass $\sim$500 \msun)  young cluster formed  around  3 Myr ago at the very end 
of a long  filamentary cloud.  We find that the massive star of the cluster 
has created an \hii region of size $\sim$2.7 pc and temperature $\sim$5,700 K. 
 However, it has heated the dust up to several parsecs and the  distribution of 
 the warm dust on a large-scale resembles a bipolar \hii regions. This 
 implies  that the parental cloud could be sheet-like or flattened in nature 
 containing a central filament as suggested by \citet{deh15} for  molecular 
 clouds that host bipolar \hii regions.

 Although, high-resolution kinematic studies of the filament are needed, nonetheless, 
 from the evidences found in Sects. \ref{s28_str} and \ref{s28_clump},  
 we hypothesized that  the formation of the cluster (and the \hii region) is 
 likely due to the edge or end-dominated global collapse of the filament as 
 advocated in \citet[][]{bur04,pon11} and then the formation of the western 
 clump followed. We suggest that the latter is either induced or facilitated by 
 the compression  of the expanding \hii region (Sect. \ref{s28_clump}) onto 
 the inflowing filamentary material. 

Inside the clump, we observed several far-infrared point sources of class 0/I nature. 
We suggest that these sources are the second-generation stars of the complex as  
such sources are absent in the vicinity of the ionizing star of the \hii region 
(see Sect. \ref{her_ps}), they are significantly younger (age $\sim$ 10$^5$ yr) 
than age of the \hii region (age $\sim$ 3 Myr), and they occupy a distinct 
location (i.e. at the interaction zone of the \hii region and filament) compared 
to optically visible stars. We hypothesize this scenario may be applicable to 
star formation at the border of several \hii region - filament environments and 
may be an efficient process for forming second-generation stars in molecular 
clouds. Future high-resolution studies of a larger sample of young \hii 
region - filament environments would be helpful to support our hypothesis. 

\vspace{1cm}
{\small We thank the anonymous referee for providing valuable comments 
and suggestions that improved the paper.  This paper, in part, is based 
on observations made with MRES mounted on TNT at the Thai National 
Observatory (program ID C06$\_$024). TNT is operated by the National 
Astronomical Research Institute of Thailand (Public Organization). We 
acknowledge Mark Heyer for sharing FCRAO $^{13}$CO observations. This 
research has made use of the SIMBAD database, operated at CDS, Strasbourg, 
France. The {\it Gaia} space mission is operated by the European Space Agency (ESA). 
This publication uses data from the UKIDSS. This work, in part, uses the data 
from the Pan-STARRS1 (PS1) surveys. We acknowledge the data obtained as part of 
the INT Photometric H$\alpha$ Survey of the Northern Galactic Plane (IPHAS). 
This research uses the data obtained with the {\it Spitzer} Space Telescope. 
The GMRT is run by the National Centre for Radio Astrophysics of the 
Tata Institute of Fundamental Research. The National Radio Astronomy Observatory 
is a facility of the National Science Foundation operated under cooperative 
agreement by Associated Universities, Inc. This work has been funded by 
Indo-Thai project, which is supported by Ministry of Higher Education, 
Science, Research and  Innovation (MHESI), Thailand and Department of Science 
and Technology (DST), India (project No. DST/INT/Thai/P-15/2019). AZ thanks 
the support of the Institut Universitaire de France. DKO acknowledges the 
support of the Department of Atomic Energy, Government of India 
(project No. RTI 4002). SP acknowledges the DST-INSPIRE fellowship (No. IF180092) 
of the Department of Science and Technology, India.}

\facilities{FCRAO, $Gaia$, GMRT, $Herschel$, INT,  PANSTARRS1, $Spitzer$, TNT, UKIRT,  VLA.} 
\software  AIPS \citep{van96}, 
CASA \citep{mcm07}, 
APLpy \citep{rob12}, 
Astropy \citep[][]{2013A&A...558A..33A}, 
CUTEX \citep{mol11},
DS9 \citep{joy03,har00}, 
IRAF \citep{tod86,tod93}, 
isochrones \citep{mor15},
STARLINK \citep{cur14}.

\bibliography{myref}{}

\begin{thebibliography}{}
\expandafter\ifx\csname natexlab\endcsname\relax\def\natexlab#1{#1}\fi
\providecommand{\url}[1]{\href{#1}{#1}}
\providecommand{\dodoi}[1]{doi:~\href{http://doi.org/#1}{\nolinkurl{#1}}}
\providecommand{\doeprint}[1]{\href{http://ascl.net/#1}{\nolinkurl{http://ascl.net/#1}}}
\providecommand{\doarXiv}[1]{\href{https://arxiv.org/abs/#1}{\nolinkurl{https://arxiv.org/abs/#1}}}

\bibitem[{{Alam} {et~al.}(2015){Alam}, {Albareti}, {Allende Prieto}, {Anders},
  {Anderson}, {Anderton}, {Andrews}, {Armengaud}, {Aubourg}, {Bailey}, {Basu},
  {Bautista}, {Beaton}, {Beers}, {Bender}, {Berlind}, {Beutler}, {Bhardwaj},
  {Bird}, {Bizyaev}, {Blake}, {Blanton}, {Blomqvist}, {Bochanski}, {Bolton},
  {Bovy}, {Shelden Bradley}, {Brandt}, {Brauer}, {Brinkmann}, {Brown},
  {Brownstein}, {Burden}, {Burtin}, {Busca}, {Cai}, {Capozzi}, {Carnero
  Rosell}, {Carr}, {Carrera}, {Chambers}, {Chaplin}, {Chen}, {Chiappini},
  {Chojnowski}, {Chuang}, {Clerc}, {Comparat}, {Covey}, {Croft}, {Cuesta},
  {Cunha}, {da Costa}, {Da Rio}, {Davenport}, {Dawson}, {De Lee}, {Delubac},
  {Deshpande}, {Dhital}, {Dutra-Ferreira}, {Dwelly}, {Ealet}, {Ebelke},
  {Edmondson}, {Eisenstein}, {Ellsworth}, {Elsworth}, {Epstein}, {Eracleous},
  {Escoffier}, {Esposito}, {Evans}, {Fan}, {Fern{\'a}ndez-Alvar}, {Feuillet},
  {Filiz Ak}, {Finley}, {Finoguenov}, {Flaherty}, {Fleming}, {Font-Ribera},
  {Foster}, {Frinchaboy}, {Galbraith-Frew}, {Garc{\'\i}a},
  {Garc{\'\i}a-Hern{\'a}ndez}, {Garc{\'\i}a P{\'e}rez}, {Gaulme}, {Ge},
  {G{\'e}nova-Santos}, {Georgakakis}, {Ghezzi}, {Gillespie}, {Girardi},
  {Goddard}, {Gontcho}, {Gonz{\'a}lez Hern{\'a}ndez}, {Grebel}, {Green},
  {Grieb}, {Grieves}, {Gunn}, {Guo}, {Harding}, {Hasselquist}, {Hawley},
  {Hayden}, {Hearty}, {Hekker}, {Ho}, {Hogg}, {Holley-Bockelmann}, {Holtzman},
  {Honscheid}, {Huber}, {Huehnerhoff}, {Ivans}, {Jiang}, {Johnson},
  {Kinemuchi}, {Kirkby}, {Kitaura}, {Klaene}, {Knapp}, {Kneib}, {Koenig},
  {Lam}, {Lan}, {Lang}, {Laurent}, {Le Goff}, {Leauthaud}, {Lee}, {Lee},
  {Licquia}, {Liu}, {Long}, {L{\'o}pez-Corre doira}, {Lorenzo-Oliveira},
  {Lucatello}, {Lundgren}, {Lupton}, {Mack}, {Mahadevan}, {Maia}, {Majewski},
  {Malanushenko}, {Malanushenko}, {Manchado}, {Manera}, {Mao}, {Maraston},
  {Marchwinski}, {Margala}, {Martell}, {Martig}, {Masters}, {Mathur},
  {McBride}, {McGehee}, {McGreer}, {McMahon}, {M{\'e}nard}, {Menzel},
  {Merloni}, {M{\'e}sz{\'a}ros}, {Miller}, {Miralda-Escud{\'e}}, {Miyatake},
  {Montero-Dorta}, {More}, {Morganson}, {Morice-Atkinson}, {Morrison},
  {Mosser}, {Muna}, {Myers}, {Nand ra}, {Newman}, {Neyrinck}, {Nguyen},
  {Nichol}, {Nidever}, {Noterdaeme}, {Nuza}, {O'Connell}, {O'Connell},
  {O'Connell}, {Ogando}, {Olmstead}, {Oravetz}, {Oravetz}, {Osumi}, {Owen},
  {Padgett}, {Padmanabhan}, {Paegert}, {Palanque-Delabrouille}, {Pan},
  {Parejko}, {P{\^a}ris}, {Park}, {Pattarakijwanich}, {Pellejero-Ibanez},
  {Pepper}, {Percival}, {P{\'e}rez-Fournon}, {Ṕrez-Ra`fols}, {Petitjean},
  {Pieri}, {Pinsonneault}, {Porto de Mello}, {Prada}, {Prakash},
  {Price-Whelan}, {Protopapas}, {Raddick}, {Rahman}, {Reid}, {Rich}, {Rix},
  {Robin}, {Rockosi}, {Rodrigues}, {Rodr{\'\i}guez-Torres}, {Roe}, {Ross},
  {Ross}, {Rossi}, {Ruan}, {Rubi{\~n}o-Mart{\'\i}n}, {Rykoff},
  {Salazar-Albornoz}, {Salvato}, {Samushia}, {S{\'a}nchez}, {Santiago},
  {Sayres}, {Schiavon}, {Schlegel}, {Schmidt}, {Schneider}, {Schultheis},
  {Schwope}, {Sc{\'o}ccola}, {Scott}, {Sellgren}, {Seo}, {Serenelli}, {Shane},
  {Shen}, {Shetrone}, {Shu}, {Silva Aguirre}, {Sivarani}, {Skrutskie},
  {Slosar}, {Smith}, {Sobreira}, {Souto}, {Stassun}, {Steinmetz}, {Stello},
  {Strauss}, {Streblyanska}, {Suzuki}, {Swanson}, {Tan}, {Tayar}, {Terrien},
  {Thakar}, {Thomas}, {Thomas}, {Thompson}, {Tinker}, {Tojeiro}, {Troup},
  {Vargas-Maga{\~n}a}, {Vazquez}, {Verde}, {Viel}, {Vogt}, {Wake}, {Wang},
  {Weaver}, {Weinberg}, {Weiner}, {White}, {Wilson}, {Wisniewski},
  {Wood-Vasey}, {Ye`che}, {York}, {Zakamska}, {Zamora}, {Zasowski}, {Zehavi},
  {Zhao}, {Zheng}, {Zhou}, {Zhou}, {Zou}, \& {Zhu}}]{sds15}
{Alam}, S., {Albareti}, F.~D., {Allende Prieto}, C., {et~al.} 2015, \apjs, 219,
  12, \dodoi{10.1088/0067-0049/219/1/12}

\bibitem[{{Anderson} {et~al.}(2019){Anderson}, {Makai}, {Luisi}, {Andersen},
  {Russeil}, {Samal}, {Schneider}, {Tremblin}, {Zavagno}, {Kirsanova},
  {Ossenkopf-Okada}, \& {Sobolev}}]{and19}
{Anderson}, L.~D., {Makai}, Z., {Luisi}, M., {et~al.} 2019, \apj, 882, 11,
  \dodoi{10.3847/1538-4357/ab1c59}

\bibitem[{{Andr{\'e}} {et~al.}(2000){Andr{\'e}}, {Ward-Thompson}, \&
  {Barsony}}]{and00}
{Andr{\'e}}, P., {Ward-Thompson}, D., \& {Barsony}, M. 2000, in Protostars and
  Planets IV, ed. V.~{Mannings}, A.~P. {Boss}, \& S.~S. {Russell}, 59.
\newblock \doarXiv{astro-ph/9903284}

\bibitem[{{Andr{\'e}} {et~al.}(2008){Andr{\'e}}, {Minier}, {Gallais},
  {Rev{\'e}ret}, {Le Pennec}, {Rodriguez}, {Boulade}, {Doumayrou}, {Dubreuil},
  \& {Lortholary}}]{and08}
{Andr{\'e}}, P., {Minier}, V., {Gallais}, P., {et~al.} 2008, \aap, 490, L27,
  \dodoi{10.1051/0004-6361:200810957}

\bibitem[{{Ascenso} {et~al.}(2007){Ascenso}, {Alves}, {Beletsky}, \&
  {Lago}}]{asc07}
{Ascenso}, J., {Alves}, J., {Beletsky}, Y., \& {Lago}, M.~T.~V.~T. 2007, \aap,
  466, 137, \dodoi{10.1051/0004-6361:20066433}

\bibitem[{{Astropy Collaboration} {et~al.}(2013){Astropy Collaboration},
  {Robitaille}, {Tollerud}, {Greenfield}, {Droettboom}, {Bray}, {Aldcroft},
  {Davis}, {Ginsburg}, {Price-Whelan}, {Kerzendorf}, {Conley}, {Crighton},
  {Barbary}, {Muna}, {Ferguson}, {Grollier}, {Parikh}, {Nair}, {Unther},
  {Deil}, {Woillez}, {Conseil}, {Kramer}, {Turner}, {Singer}, {Fox}, {Weaver},
  {Zabalza}, {Edwards}, {Azalee Bostroem}, {Burke}, {Casey}, {Crawford},
  {Dencheva}, {Ely}, {Jenness}, {Labrie}, {Lim}, {Pierfederici}, {Pontzen},
  {Ptak}, {Refsdal}, {Servillat}, \& {Streicher}}]{2013A&A...558A..33A}
{Astropy Collaboration}, {Robitaille}, T.~P., {Tollerud}, E.~J., {et~al.} 2013,
  \aap, 558, A33, \dodoi{10.1051/0004-6361/201322068}

\bibitem[{{Bailer-Jones}(2015)}]{bai15}
{Bailer-Jones}, C. A.~L. 2015, \pasp, 127, 994, \dodoi{10.1086/683116}

\bibitem[{{Bailer-Jones} {et~al.}(2018){Bailer-Jones}, {Rybizki}, {Fouesneau},
  {Mantelet}, \& {Andrae}}]{2018AJ....156...58B}
{Bailer-Jones}, C.~A.~L., {Rybizki}, J., {Fouesneau}, M., {Mantelet}, G., \&
  {Andrae}, R. 2018, \aj, 156, 58, \dodoi{10.3847/1538-3881/aacb21}

\bibitem[{{Balser} {et~al.}(2011){Balser}, {Rood}, {Bania}, \&
  {Anderson}}]{bals11}
{Balser}, D.~S., {Rood}, R.~T., {Bania}, T.~M., \& {Anderson}, L.~D. 2011,
  \apj, 738, 27, \dodoi{10.1088/0004-637X/738/1/27}

\bibitem[{{Banerjee} \& {Kroupa}(2015)}]{ban15}
{Banerjee}, S., \& {Kroupa}, P. 2015, \mnras, 447, 728,
  \dodoi{10.1093/mnras/stu2445}

\bibitem[{{Banerjee} \& {Kroupa}(2017)}]{ban17}
---. 2017, \aap, 597, A28, \dodoi{10.1051/0004-6361/201526928}

\bibitem[{{Barentsen} {et~al.}(2011){Barentsen}, {Vink}, {Drew}, {Greimel},
  {Wright}, {Drake}, {Martin}, {Valdivielso}, \& {Corradi}}]{bar11}
{Barentsen}, G., {Vink}, J.~S., {Drew}, J.~E., {et~al.} 2011, \mnras, 415, 103,
  \dodoi{10.1111/j.1365-2966.2011.18674.x}

\bibitem[{{Barentsen} {et~al.}(2014){Barentsen}, {Farnhill}, {Drew},
  {Gonz{\'a}lez-Solares}, {Greimel}, {Irwin}, {Miszalski}, {Ruhland}, {Groot},
  {Mampaso}, {Sale}, {Henden}, {Aungwerojwit}, {Barlow}, {Carter}, {Corradi},
  {Drake}, {Eisl{\"o}ffel}, {Fabregat}, {G{\"a}nsicke}, {Gentile Fusillo},
  {Greiss}, {Hales}, {Hodgkin}, {Huckvale}, {Irwin}, {King}, {Knigge},
  {Kupfer}, {Lagadec}, {Lennon}, {Lewis}, {Mohr-Smith}, {Morris}, {Naylor},
  {Parker}, {Phillipps}, {Pyrzas}, {Raddi}, {Roelofs}, {Rodr{\'\i}guez-Gil},
  {Sabin}, {Scaringi}, {Steeghs}, {Suso}, {Tata}, {Unruh}, {van Roestel},
  {Viironen}, {Vink}, {Walton}, {Wright}, \& {Zijlstra}}]{bar14}
{Barentsen}, G., {Farnhill}, H.~J., {Drew}, J.~E., {et~al.} 2014, \mnras, 444,
  3230, \dodoi{10.1093/mnras/stu1651}

\bibitem[{{Battersby} {et~al.}(2011){Battersby}, {Bally}, {Ginsburg},
  {Bernard}, {Brunt}, {Fuller}, {Martin}, {Molinari}, {Mottram}, {Peretto},
  {Testi}, \& {Thompson}}]{bat11}
{Battersby}, C., {Bally}, J., {Ginsburg}, A., {et~al.} 2011, \aap, 535, A128,
  \dodoi{10.1051/0004-6361/201116559}

\bibitem[{{Beckwith} \& {Sargent}(1991)}]{beck91}
{Beckwith}, S. V.~W., \& {Sargent}, A.~I. 1991, \apj, 381, 250,
  \dodoi{10.1086/170646}

\bibitem[{{Bertoldi} \& {McKee}(1992)}]{ber92}
{Bertoldi}, F., \& {McKee}, C.~F. 1992, \apj, 395, 140, \dodoi{10.1086/171638}

\bibitem[{{Bessell} \& {Brett}(1988)}]{bes88}
{Bessell}, M.~S., \& {Brett}, J.~M. 1988, \pasp, 100, 1134,
  \dodoi{10.1086/132281}

\bibitem[{{Bica} {et~al.}(2003){Bica}, {Dutra}, \& {Barbuy}}]{bica03}
{Bica}, E., {Dutra}, C.~M., \& {Barbuy}, B. 2003, \aap, 397, 177,
  \dodoi{10.1051/0004-6361:20021479}

\bibitem[{{Binney} \& {Tremaine}(2008)}]{bin08}
{Binney}, J., \& {Tremaine}, S. 2008, {Galactic Dynamics: Second Edition}
  (Princeton University Press)

\bibitem[{{Bloemen}(1987)}]{blo87}
{Bloemen}, J.~B.~G.~M. 1987, \apj, 322, 694, \dodoi{10.1086/165765}

\bibitem[{{Bodenheimer} {et~al.}(1979){Bodenheimer}, {Tenorio-Tagle}, \&
  {Yorke}}]{bod79a}
{Bodenheimer}, P., {Tenorio-Tagle}, G., \& {Yorke}, H.~W. 1979, \apj, 233, 85,
  \dodoi{10.1086/157368}

\bibitem[{{Bonnell} {et~al.}(2004){Bonnell}, {Vine}, \& {Bate}}]{bon04}
{Bonnell}, I.~A., {Vine}, S.~G., \& {Bate}, M.~R. 2004, \mnras, 349, 735,
  \dodoi{10.1111/j.1365-2966.2004.07543.x}

\bibitem[{{Bontemps} {et~al.}(1996){Bontemps}, {Andre}, {Terebey}, \&
  {Cabrit}}]{bon96}
{Bontemps}, S., {Andre}, P., {Terebey}, S., \& {Cabrit}, S. 1996, \aap, 311,
  858

\bibitem[{{Borissova} {et~al.}(2003){Borissova}, {Pessev}, {Ivanov}, {Saviane},
  {Kurtev}, \& {Ivanov}}]{bor03}
{Borissova}, J., {Pessev}, P., {Ivanov}, V.~D., {et~al.} 2003, \aap, 411, 83,
  \dodoi{10.1051/0004-6361:20034009}

\bibitem[{{Burkert} \& {Hartmann}(2004)}]{bur04}
{Burkert}, A., \& {Hartmann}, L. 2004, \apj, 616, 288, \dodoi{10.1086/424895}

\bibitem[{{Chambers} {et~al.}(2016){Chambers}, {Magnier}, {Metcalfe},
  {Flewelling}, {Huber}, {Waters}, {Denneau}, {Draper}, {Farrow}, {Finkbeiner},
  {Holmberg}, {Koppenhoefer}, {Price}, {Rest}, {Saglia}, {Schlafly}, {Smartt},
  {Sweeney}, {Wainscoat}, {Burgett}, {Chastel}, {Grav}, {Heasley}, {Hodapp},
  {Jedicke}, {Kaiser}, {Kudritzki}, {Luppino}, {Lupton}, {Monet}, {Morgan},
  {Onaka}, {Shiao}, {Stubbs}, {Tonry}, {White}, {Ba{\~n}ados}, {Bell},
  {Bender}, {Bernard}, {Boegner}, {Boffi}, {Botticella}, {Calamida},
  {Casertano}, {Chen}, {Chen}, {Cole}, {Deacon}, {Frenk}, {Fitzsimmons},
  {Gezari}, {Gibbs}, {Goessl}, {Goggia}, {Gourgue}, {Goldman}, {Grant},
  {Grebel}, {Hambly}, {Hasinger}, {Heavens}, {Heckman}, {Henderson}, {Henning},
  {Holman}, {Hopp}, {Ip}, {Isani}, {Jackson}, {Keyes}, {Koekemoer}, {Kotak},
  {Le}, {Liska}, {Long}, {Lucey}, {Liu}, {Martin}, {Masci}, {McLean}, {Mindel},
  {Misra}, {Morganson}, {Murphy}, {Obaika}, {Narayan}, {Nieto-Santisteban},
  {Norberg}, {Peacock}, {Pier}, {Postman}, {Primak}, {Rae}, {Rai}, {Riess},
  {Riffeser}, {Rix}, {R{\"o}ser}, {Russel}, {Rutz}, {Schilbach}, {Schultz},
  {Scolnic}, {Strolger}, {Szalay}, {Seitz}, {Small}, {Smith}, {Soderblom},
  {Taylor}, {Thomson}, {Taylor}, {Thakar}, {Thiel}, {Thilker}, {Unger},
  {Urata}, {Valenti}, {Wagner}, {Walder}, {Walter}, {Watters}, {Werner},
  {Wood-Vasey}, \& {Wyse}}]{pans16}
{Chambers}, K.~C., {Magnier}, E.~A., {Metcalfe}, N., {et~al.} 2016, arXiv
  e-prints, arXiv:1612.05560.
\newblock \doarXiv{1612.05560}

\bibitem[{{Chen} {et~al.}(2020){Chen}, {Chen}, {Wang}, {Shen}, \&
  {Yang}}]{chen20}
{Chen}, H.-Y., {Chen}, X., {Wang}, J.-Z., {Shen}, Z.-Q., \& {Yang}, K. 2020,
  \apjs, 248, 3, \dodoi{10.3847/1538-4365/ab818e}

\bibitem[{{Chini} \& {Wink}(1984)}]{chini84}
{Chini}, R., \& {Wink}, J.~E. 1984, \aap, 139, L5

\bibitem[{{Comer{\'o}n} \& {Pasquali}(2005)}]{2005A&A...430..541C}
{Comer{\'o}n}, F., \& {Pasquali}, A. 2005, \aap, 430, 541,
  \dodoi{10.1051/0004-6361:20041788}

\bibitem[{{Currie} {et~al.}(2014){Currie}, {Berry}, {Jenness}, {Gibb}, {Bell},
  \& {Draper}}]{cur14}
{Currie}, M.~J., {Berry}, D.~S., {Jenness}, T., {et~al.} 2014, in Astronomical
  Society of the Pacific Conference Series, Vol. 485, Astronomical Data
  Analysis Software and Systems XXIII, ed. N.~{Manset} \& P.~{Forshay}, 391

\bibitem[{{Da Rio} {et~al.}(2010){Da Rio}, {Robberto}, {Soderblom}, {Panagia},
  {Hillenbrand}, {Palla}, \& {Stassun}}]{dar10}
{Da Rio}, N., {Robberto}, M., {Soderblom}, D.~R., {et~al.} 2010, \apj, 722,
  1092, \dodoi{10.1088/0004-637X/722/2/1092}

\bibitem[{{Damian} {et~al.}(2021){Damian}, {Jose}, {Samal}, {Moraux}, {Das}, \&
  {Patra}}]{dam21}
{Damian}, B., {Jose}, J., {Samal}, M.~R., {et~al.} 2021, arXiv e-prints,
  arXiv:2101.08804.
\newblock \doarXiv{2101.08804}

\bibitem[{{Das} {et~al.}(2021){Das}, {Jose}, {Samal}, {Zhang}, \&
  {Panwar}}]{das21}
{Das}, S.~R., {Jose}, J., {Samal}, M.~R., {Zhang}, S., \& {Panwar}, N. 2021,
  \mnras, 500, 3123, \dodoi{10.1093/mnras/staa3222}

\bibitem[{{Deharveng} {et~al.}(2010){Deharveng}, {Schuller}, {Anderson},
  {Zavagno}, {Wyrowski}, {Menten}, {Bronfman}, {Testi}, {Walmsley}, \&
  {Wienen}}]{deh10}
{Deharveng}, L., {Schuller}, F., {Anderson}, L.~D., {et~al.} 2010, \aap, 523,
  A6, \dodoi{10.1051/0004-6361/201014422}

\bibitem[{{Deharveng} {et~al.}(2015){Deharveng}, {Zavagno}, {Samal},
  {Anderson}, {LeLeu}, {Brevot}, {Duarte-Cabral}, {Molinari}, {Pestalozzi},
  {Foster}, {Rathborne}, \& {Jackson}}]{deh15}
{Deharveng}, L., {Zavagno}, A., {Samal}, M.~R., {et~al.} 2015, \aap, 582, A1,
  \dodoi{10.1051/0004-6361/201423835}

\bibitem[{{Dotter}(2016)}]{dot16}
{Dotter}, A. 2016, \apjs, 222, 8, \dodoi{10.3847/0067-0049/222/1/8}

\bibitem[{{Draine}(2011)}]{dra11}
{Draine}, B.~T. 2011, {Physics of the Interstellar and Intergalactic Medium}
  (Princeton University Press)

\bibitem[{{Drew} {et~al.}(2005){Drew}, {Greimel}, {Irwin}, {Aungwerojwit},
  {Barlow}, {Corradi}, {Drake}, {G{\"a}nsicke}, {Groot}, {Hales}, {Hopewell},
  {Irwin}, {Knigge}, {Leisy}, {Lennon}, {Mampaso}, {Masheder}, {Matsuura},
  {Morales-Rueda}, {Morris}, {Parker}, {Phillipps}, {Rodriguez-Gil}, {Roelofs},
  {Skillen}, {Sokoloski}, {Steeghs}, {Unruh}, {Viironen}, {Vink}, {Walton},
  {Witham}, {Wright}, {Zijlstra}, \& {Zurita}}]{2005MNRAS.362..753D}
{Drew}, J.~E., {Greimel}, R., {Irwin}, M.~J., {et~al.} 2005, \mnras, 362, 753,
  \dodoi{10.1111/j.1365-2966.2005.09330.x}

\bibitem[{{Dunham} {et~al.}(2015){Dunham}, {Allen}, {Evans},
  {Broekhoven-Fiene}, {Cieza}, {Di Francesco}, {Gutermuth}, {Harvey},
  {Hatchell}, {Heiderman}, {Huard}, {Johnstone}, {Kirk}, {Matthews}, {Miller},
  {Peterson}, \& {Young}}]{dun15}
{Dunham}, M.~M., {Allen}, L.~E., {Evans}, Neal~J., I., {et~al.} 2015, \apjs,
  220, 11, \dodoi{10.1088/0067-0049/220/1/11}

\bibitem[{{Dutta} {et~al.}(2015){Dutta}, {Mondal}, {Jose}, {Das}, {Samal}, \&
  {Ghosh}}]{dut15}
{Dutta}, S., {Mondal}, S., {Jose}, J., {et~al.} 2015, \mnras, 454, 3597,
  \dodoi{10.1093/mnras/stv2190}

\bibitem[{{Dutta} {et~al.}(2018){Dutta}, {Mondal}, {Samal}, \& {Jose}}]{dut18}
{Dutta}, S., {Mondal}, S., {Samal}, M.~R., \& {Jose}, J. 2018, \apj, 864, 154,
  \dodoi{10.3847/1538-4357/aadb3e}

\bibitem[{{Eden} {et~al.}(2019){Eden}, {Liu}, {Kim}, {Juvela}, {Liu},
  {Tatematsu}, {Francesco}, {Wang}, {Wu}, {Thompson}, {Fuller}, {Li},
  {Ristorcelli}, {Kang}, {Hirano}, {Johnstone}, {Lin}, {He}, {Koch},
  {Sanhueza}, {Qin}, {Zhang}, {Goldsmith}, {Evans}, {Yuan}, {Zhang}, {White},
  {Choi}, {Lee}, {Toth}, {Mairs}, {Yi}, {Tang}, {Soam}, {Peretto}, {Samal},
  {Fich}, {Parsons}, {Malinen}, {Bendo}, {Rivera-Ingraham}, {Liu},
  {Wouterloot}, {Li}, {Qian}, {Rawlings}, {Rawlings}, {Feng}, {Wang}, {Li},
  {Liu}, {Luo}, {Marston}, {Pattle}, {Pelkonen}, {Rigby}, {Zahorecz}, {Zhang},
  {B{\H{o}}gner}, {Aikawa}, {Akhter}, {Alina}, {Bell}, {Bernard}, {Blain},
  {Bronfman}, {Byun}, {Chapman}, {Chen}, {Chen}, {Chen}, {Chen}, {Chen},
  {Chrysostomou}, {Chu}, {Chung}, {Cornu}, {Cosentino}, {Cunningham}, {Demyk},
  {Drabek-Maunder}, { doi}, {Eswaraiah}, {Falgarone}, {Feh{\'e}r}, {Fraser},
  {Friberg}, {Garay}, {Ge}, {Gear}, {Greaves}, {Guan}, {Harvey-Smith},
  {Hasegawa}, {He}, {Henkel}, {Hirota}, {Holland}, {Hughes}, {Jarken}, {Ji},
  {Jimenez-Serra}, {Kang}, {Kawabata}, {Kim}, {Kim}, {Kim}, {Kim}, {Koo},
  {Kwon}, {Kuan}, {Lacaille}, {Lai}, {Lee}, {Lee}, {Lee}, {Li}, {Lo}, {Lopez},
  {Lu}, {Lyo}, {Mardones}, {McGehee}, {Meng}, {Montier}, {Montillaud}, {Moore},
  {Morata}, {Moriarty-Schieven}, {Ohashi}, {Pak}, {Park}, {Paladini}, {Pech},
  {Qiu}, {Ren}, {Richer}, {Sakai}, {Shang}, {Shinnaga}, {Stamatellos}, {Tang},
  {Traficante}, {Vastel}, {Viti}, {Walsh}, {Wang}, {Wang}, {Ward-Thompson},
  {Whitworth}, {Wilson}, {Xu}, {Yang}, {Yuan}, {Yuan}, {Zavagno}, {Zhang},
  {Zhang}, {Zhang}, {Zhou}, {Zhou}, {Zhu}, \& {Zuo}}]{eden19}
{Eden}, D.~J., {Liu}, T., {Kim}, K.-T., {et~al.} 2019, \mnras, 485, 2895,
  \dodoi{10.1093/mnras/stz574}

\bibitem[{{Elmegreen} \& {Lada}(1977)}]{elm77}
{Elmegreen}, B.~G., \& {Lada}, C.~J. 1977, \apj, 214, 725,
  \dodoi{10.1086/155302}

\bibitem[{{Eswaraiah} {et~al.}(2019){Eswaraiah}, {Lai}, {Ma}, {Pand ey},
  {Jose}, {Chen}, {Samal}, {Wang}, {Sharma}, \& {Ojha}}]{esw19}
{Eswaraiah}, C., {Lai}, S.-P., {Ma}, Y., {et~al.} 2019, \apj, 875, 64,
  \dodoi{10.3847/1538-4357/ab0a0c}

\bibitem[{{Evans} {et~al.}(2009){Evans}, {Dunham}, {J{\o}rgensen}, {Enoch},
  {Mer{\'\i}n}, {van Dishoeck}, {Alcal{\'a}}, {Myers}, {Stapelfeldt}, {Huard},
  {Allen}, {Harvey}, {van Kempen}, {Blake}, {Koerner}, {Mundy}, {Padgett}, \&
  {Sargent}}]{evans09}
{Evans}, Neal~J., I., {Dunham}, M.~M., {J{\o}rgensen}, J.~K., {et~al.} 2009,
  \apjs, 181, 321, \dodoi{10.1088/0067-0049/181/2/321}

\bibitem[{{Fern{\'a}ndez-Mart{\'\i}n}
  {et~al.}(2017){Fern{\'a}ndez-Mart{\'\i}n}, {P{\'e}rez-Montero},
  {V{\'\i}lchez}, \& {Mampaso}}]{fer17}
{Fern{\'a}ndez-Mart{\'\i}n}, A., {P{\'e}rez-Montero}, E., {V{\'\i}lchez},
  J.~M., \& {Mampaso}, A. 2017, \aap, 597, A84,
  \dodoi{10.1051/0004-6361/201628423}

\bibitem[{{Feroz} {et~al.}(2009){Feroz}, {Hobson}, \& {Bridges}}]{fer09}
{Feroz}, F., {Hobson}, M.~P., \& {Bridges}, M. 2009, \mnras, 398, 1601,
  \dodoi{10.1111/j.1365-2966.2009.14548.x}

\bibitem[{{Fukuda} \& {Hanawa}(2000)}]{fuk00}
{Fukuda}, N., \& {Hanawa}, T. 2000, \apj, 533, 911, \dodoi{10.1086/308701}

\bibitem[{{Gaia Collaboration} {et~al.}(2020){Gaia Collaboration}, {Brown},
  {Vallenari}, {Prusti}, {de Bruijne}, {Babusiaux}, \& {Biermann}}]{gaiadr3}
{Gaia Collaboration}, {Brown}, A.~G.~A., {Vallenari}, A., {et~al.} 2020, arXiv
  e-prints, arXiv:2012.01533.
\newblock \doarXiv{2012.01533}

\bibitem[{{Gaia Collaboration} {et~al.}(2016){Gaia Collaboration}, {Prusti},
  {de Bruijne}, {Brown}, {Vallenari}, {Babusiaux}, {Bailer-Jones}, {Bastian},
  {Biermann}, {Evans}, \& et~al.}]{gaia16}
{Gaia Collaboration}, {Prusti}, T., {de Bruijne}, J.~H.~J., {et~al.} 2016,
  \aap, 595, A1, \dodoi{10.1051/0004-6361/201629272}

\bibitem[{{Gaia Collaboration} {et~al.}(2018){Gaia Collaboration}, {Brown},
  {Vallenari}, {Prusti}, {de Bruijne}, {Babusiaux}, {Bailer-Jones}, {Biermann},
  {Evans}, {Eyer}, \& et~al.}]{gaia18}
{Gaia Collaboration}, {Brown}, A.~G.~A., {Vallenari}, A., {et~al.} 2018, \aap,
  616, A1, \dodoi{10.1051/0004-6361/201833051}

\bibitem[{{Garay} \& {Lizano}(1999)}]{gar99}
{Garay}, G., \& {Lizano}, S. 1999, \pasp, 111, 1049, \dodoi{10.1086/316416}

\bibitem[{{Gieles}(2010)}]{gie10}
{Gieles}, M. 2010, in IAU Symposium, Vol. 266, Star Clusters: Basic Galactic
  Building Blocks Throughout Time and Space, ed. R.~{de Grijs} \& J.~R.~D.
  {L{\'e}pine}, 69--80, \dodoi{10.1017/S1743921309990895}

\bibitem[{{Gieles} \& {Portegies Zwart}(2011)}]{gieles11}
{Gieles}, M., \& {Portegies Zwart}, S.~F. 2011, \mnras, 410, L6,
  \dodoi{10.1111/j.1745-3933.2010.00967.x}

\bibitem[{{G{\'o}mez} \& {V{\'a}zquez-Semadeni}(2014)}]{gom14}
{G{\'o}mez}, G.~C., \& {V{\'a}zquez-Semadeni}, E. 2014, \apj, 791, 124,
  \dodoi{10.1088/0004-637X/791/2/124}

\bibitem[{{Gutermuth} {et~al.}(2009){Gutermuth}, {Megeath}, {Myers}, {Allen},
  {Pipher}, \& {Fazio}}]{gut09}
{Gutermuth}, R.~A., {Megeath}, S.~T., {Myers}, P.~C., {et~al.} 2009, \apjs,
  184, 18, \dodoi{10.1088/0067-0049/184/1/18}

\bibitem[{{Gutermuth} {et~al.}(2005){Gutermuth}, {Megeath}, {Pipher},
  {Williams}, {Allen}, {Myers}, \& {Raines}}]{gut05}
{Gutermuth}, R.~A., {Megeath}, S.~T., {Pipher}, J.~L., {et~al.} 2005, \apj,
  632, 397, \dodoi{10.1086/432460}

\bibitem[{{Harayama} {et~al.}(2008){Harayama}, {Eisenhauer}, \&
  {Martins}}]{har08}
{Harayama}, Y., {Eisenhauer}, F., \& {Martins}, F. 2008, \apj, 675, 1319,
  \dodoi{10.1086/524650}

\bibitem[{{Hartmann} \& {Burkert}(2007)}]{hart07}
{Hartmann}, L., \& {Burkert}, A. 2007, \apj, 654, 988, \dodoi{10.1086/509321}

\bibitem[{{Heiter} {et~al.}(2002){Heiter}, {Kupka}, {van't Veer-Menneret},
  {Barban}, {Weiss}, {Goupil}, {Schmidt}, {Katz}, \&
  {Garrido}}]{2002A&A...392..619H}
{Heiter}, U., {Kupka}, F., {van't Veer-Menneret}, C., {et~al.} 2002, \aap, 392,
  619, \dodoi{10.1051/0004-6361:20020788}

\bibitem[{{Heitsch} {et~al.}(2008){Heitsch}, {Hartmann}, {Slyz}, {Devriendt},
  \& {Burkert}}]{hei08}
{Heitsch}, F., {Hartmann}, L.~W., {Slyz}, A.~D., {Devriendt}, J.~E.~G., \&
  {Burkert}, A. 2008, \apj, 674, 316, \dodoi{10.1086/523697}

\bibitem[{{Hillenbrand}(1997)}]{hill97}
{Hillenbrand}, L.~A. 1997, \aj, 113, 1733, \dodoi{10.1086/118389}

\bibitem[{{Howard} {et~al.}(2017){Howard}, {Pudritz}, \& {Klessen}}]{how17}
{Howard}, C., {Pudritz}, R., \& {Klessen}, R. 2017, \apj, 834, 40,
  \dodoi{10.3847/1538-4357/834/1/40}

\bibitem[{{Hubeny} \& {Lanz}(2011)}]{2011ascl.soft09022H}
{Hubeny}, I., \& {Lanz}, T. 2011, {Synspec: General Spectrum Synthesis
  Program}.
\newblock \doeprint{1109.022}

\bibitem[{{Hunter} \& {Massey}(1990)}]{hun90}
{Hunter}, D.~A., \& {Massey}, P. 1990, \aj, 99, 846, \dodoi{10.1086/115378}

\bibitem[{{Intema} {et~al.}(2017){Intema}, {Jagannathan}, {Mooley}, \&
  {Frail}}]{inte17}
{Intema}, H.~T., {Jagannathan}, P., {Mooley}, K.~P., \& {Frail}, D.~A. 2017,
  \aap, 598, A78, \dodoi{10.1051/0004-6361/201628536}

\bibitem[{{Israel}(1977)}]{isr77}
{Israel}, F.~P. 1977, \aap, 60, 233

\bibitem[{{Jackson} {et~al.}(2006){Jackson}, {Rathborne}, {Shah}, {Simon},
  {Bania}, {Clemens}, {Chambers}, {Johnson}, {Dormody}, {Lavoie}, \&
  {Heyer}}]{jack06}
{Jackson}, J.~M., {Rathborne}, J.~M., {Shah}, R.~Y., {et~al.} 2006, \apjs, 163,
  145, \dodoi{10.1086/500091}

\bibitem[{{Jose} {et~al.}(2017){Jose}, {Herczeg}, {Samal}, {Fang}, \&
  {Panwar}}]{jose17}
{Jose}, J., {Herczeg}, G.~J., {Samal}, M.~R., {Fang}, Q., \& {Panwar}, N. 2017,
  \apj, 836, 98, \dodoi{10.3847/1538-4357/836/1/98}

\bibitem[{{Joye} \& {Mandel}(2003)}]{joy03}
{Joye}, W.~A., \& {Mandel}, E. 2003, in Astronomical Society of the Pacific
  Conference Series, Vol. 295, Astronomical Data Analysis Software and Systems
  XII, ed. H.~E. {Payne}, R.~I. {Jedrzejewski}, \& R.~N. {Hook}, 489

\bibitem[{{Kauffmann} {et~al.}(2013){Kauffmann}, {Pillai}, \& {Zhang}}]{kau13}
{Kauffmann}, J., {Pillai}, T., \& {Zhang}, Q. 2013, \apjl, 765, L35,
  \dodoi{10.1088/2041-8205/765/2/L35}

\bibitem[{{Kharchenko} {et~al.}(2013){Kharchenko}, {Piskunov}, {Schilbach},
  {R{\"o}ser}, \& {Scholz}}]{kha13}
{Kharchenko}, N.~V., {Piskunov}, A.~E., {Schilbach}, E., {R{\"o}ser}, S., \&
  {Scholz}, R.~D. 2013, \aap, 558, A53, \dodoi{10.1051/0004-6361/201322302}

\bibitem[{{King}(1962)}]{kin62}
{King}, I. 1962, \aj, 67, 471, \dodoi{10.1086/108756}

\bibitem[{{Kirk} {et~al.}(2013){Kirk}, {Myers}, {Bourke}, {Gutermuth},
  {Hedden}, \& {Wilson}}]{kirk13}
{Kirk}, H., {Myers}, P.~C., {Bourke}, T.~L., {et~al.} 2013, \apj, 766, 115,
  \dodoi{10.1088/0004-637X/766/2/115}

\bibitem[{{Kroupa}(2001)}]{kro01}
{Kroupa}, P. 2001, \mnras, 322, 231, \dodoi{10.1046/j.1365-8711.2001.04022.x}

\bibitem[{{Kuhn} {et~al.}(2019){Kuhn}, {Hillenbrand}, {Sills}, {Feigelson}, \&
  {Getman}}]{kuhn19}
{Kuhn}, M.~A., {Hillenbrand}, L.~A., {Sills}, A., {Feigelson}, E.~D., \&
  {Getman}, K.~V. 2019, \apj, 870, 32, \dodoi{10.3847/1538-4357/aaef8c}

\bibitem[{{Kumar} {et~al.}(2020){Kumar}, {Palmeirim}, {Arzoumanian}, \&
  {Inutsuka}}]{kum20}
{Kumar}, M.~S.~N., {Palmeirim}, P., {Arzoumanian}, D., \& {Inutsuka}, S.~I.
  2020, \aap, 642, A87, \dodoi{10.1051/0004-6361/202038232}

\bibitem[{{Lada} \& {Adams}(1992)}]{lada92}
{Lada}, C.~J., \& {Adams}, F.~C. 1992, \apj, 393, 278, \dodoi{10.1086/171505}

\bibitem[{{Lada} \& {Lada}(2003)}]{lada03}
{Lada}, C.~J., \& {Lada}, E.~A. 2003, \araa, 41, 57,
  \dodoi{10.1146/annurev.astro.41.011802.094844}

\bibitem[{{Lada} {et~al.}(2010){Lada}, {Lombardi}, \& {Alves}}]{lad10}
{Lada}, C.~J., {Lombardi}, M., \& {Alves}, J.~F. 2010, \apj, 724, 687,
  \dodoi{10.1088/0004-637X/724/1/687}

\bibitem[{{Lahulla}(1985)}]{lah85}
{Lahulla}, J.~F. 1985, \aaps, 61, 537

\bibitem[{{Lanz} \& {Hubeny}(2003)}]{2003ApJS..146..417L}
{Lanz}, T., \& {Hubeny}, I. 2003, \apjs, 146, 417, \dodoi{10.1086/374373}

\bibitem[{{Lawrence} {et~al.}(2007){Lawrence}, {Warren}, {Almaini}, {Edge},
  {Hambly}, {Jameson}, {Lucas}, {Casali}, {Adamson}, {Dye}, {Emerson},
  {Foucaud}, {Hewett}, {Hirst}, {Hodgkin}, {Irwin}, {Lodieu}, {McMahon},
  {Simpson}, {Smail}, {Mortlock}, \& {Folger}}]{law07}
{Lawrence}, A., {Warren}, S.~J., {Almaini}, O., {et~al.} 2007, \mnras, 379,
  1599, \dodoi{10.1111/j.1365-2966.2007.12040.x}

\bibitem[{{Lin} {et~al.}(1965){Lin}, {Mestel}, \& {Shu}}]{lin65}
{Lin}, C.~C., {Mestel}, L., \& {Shu}, F.~H. 1965, \apj, 142, 1431,
  \dodoi{10.1086/148428}

\bibitem[{{Liu} {et~al.}(2021){Liu}, {Xu}, {Wang}, {Yu}, {Zhang}, {Li}, \&
  {Zhang}}]{liu21}
{Liu}, X.-L., {Xu}, J.-L., {Wang}, J.-J., {et~al.} 2021, \aap, 646, A137,
  \dodoi{10.1051/0004-6361/201935035}

\bibitem[{{Longmore} {et~al.}(2014){Longmore}, {Kruijssen}, {Bastian}, {Bally},
  {Rathborne}, {Testi}, {Stolte}, {Dale}, {Bressert}, \& {Alves}}]{long14}
{Longmore}, S.~N., {Kruijssen}, J.~M.~D., {Bastian}, N., {et~al.} 2014, in
  Protostars and Planets VI, ed. H.~{Beuther}, R.~S. {Klessen}, C.~P.
  {Dullemond}, \& T.~{Henning}, 291,
  \dodoi{10.2458/azu_uapress_9780816531240-ch013}

\bibitem[{{Lucas} {et~al.}(2008){Lucas}, {Hoare}, {Longmore}, {Schr{\"o}der},
  {Davis}, {Adamson}, {Bandyopadhyay}, {de Grijs}, {Smith}, {Gosling},
  {Mitchison}, {G{\'a}sp{\'a}r}, {Coe}, {Tamura}, {Parker}, {Irwin}, {Hambly},
  {Bryant}, {Collins}, {Cross}, {Evans}, {Gonzalez-Solares}, {Hodgkin},
  {Lewis}, {Read}, {Riello}, {Sutorius}, {Lawrence}, {Drew}, {Dye}, \&
  {Thompson}}]{luc08}
{Lucas}, P.~W., {Hoare}, M.~G., {Longmore}, A., {et~al.} 2008, \mnras, 391,
  136, \dodoi{10.1111/j.1365-2966.2008.13924.x}

\bibitem[{{Mallick} {et~al.}(2013){Mallick}, {Kumar}, {Ojha}, {Bachiller},
  {Samal}, \& {Pirogov}}]{mal13}
{Mallick}, K.~K., {Kumar}, M.~S.~N., {Ojha}, D.~K., {et~al.} 2013, \apj, 779,
  113, \dodoi{10.1088/0004-637X/779/2/113}

\bibitem[{{Mallick} {et~al.}(2015){Mallick}, {Ojha}, {Tamura}, {Linz}, {Samal},
  \& {Ghosh}}]{mal15}
{Mallick}, K.~K., {Ojha}, D.~K., {Tamura}, M., {et~al.} 2015, \mnras, 447,
  2307, \dodoi{10.1093/mnras/stu2584}

\bibitem[{{Martins} {et~al.}(2005){Martins}, {Schaerer}, \& {Hillier}}]{mar05}
{Martins}, F., {Schaerer}, D., \& {Hillier}, D.~J. 2005, \aap, 436, 1049,
  \dodoi{10.1051/0004-6361:20042386}

\bibitem[{{McMullin} {et~al.}(2007){McMullin}, {Waters}, {Schiebel}, {Young},
  \& {Golap}}]{mcm07}
{McMullin}, J.~P., {Waters}, B., {Schiebel}, D., {Young}, W., \& {Golap}, K.
  2007, in Astronomical Society of the Pacific Conference Series, Vol. 376,
  Astronomical Data Analysis Software and Systems XVI, ed. R.~A. {Shaw},
  F.~{Hill}, \& D.~J. {Bell}, 127

\bibitem[{{M{\`e}ge} {et~al.}(2021){M{\`e}ge}, {Russeil}, {Zavagno}, {Elia},
  {Molinari}, {Brunt}, {Butora}, {Cambresy}, {Di Giorgio}, {Fenouillet},
  {Fukui}, {Lambert}, {Makai}, {Merello}, {Meunier}, {Molinaro}, {Moreau},
  {Pezzuto}, {Poulin}, {Schisano}, \& {Schuller}}]{meg21}
{M{\`e}ge}, P., {Russeil}, D., {Zavagno}, A., {et~al.} 2021, \aap, 646, A74,
  \dodoi{10.1051/0004-6361/202038956}

\bibitem[{{Mezger} \& {Henderson}(1967)}]{1967ApJ...147..471M}
{Mezger}, P.~G., \& {Henderson}, A.~P. 1967, \apj, 147, 471,
  \dodoi{10.1086/149030}

\bibitem[{{Minniti} {et~al.}(2010){Minniti}, {Lucas}, {Emerson}, {Saito},
  {Hempel}, {Pietrukowicz}, {Ahumada}, {Alonso}, {Alonso-Garcia}, {Arias},
  {Bandyopadhyay}, {Barb{\'a}}, {Barbuy}, {Bedin}, {Bica}, {Borissova},
  {Bronfman}, {Carraro}, {Catelan}, {Clari{\'a}}, {Cross}, {de Grijs},
  {D{\'e}k{\'a}ny}, {Drew}, {Fari{\~n}a}, {Feinstein}, {Fern{\'a}ndez
  Laj{\'u}s}, {Gamen}, {Geisler}, {Gieren}, {Goldman}, {Gonzalez}, {Gunthardt},
  {Gurovich}, {Hambly}, {Irwin}, {Ivanov}, {Jord{\'a}n}, {Kerins}, {Kinemuchi},
  {Kurtev}, {L{\'o}pez-Corre doira}, {Maccarone}, {Masetti}, {Merlo},
  {Messineo}, {Mirabel}, {Monaco}, {Morelli}, {Padilla}, {Palma}, {Parisi},
  {Pignata}, {Rejkuba}, {Roman-Lopes}, {Sale}, {Schreiber}, {Schr{\"o}der},
  {Smith}, {}, {Soto}, {Tamura}, {Tappert}, {Thompson}, {Toledo}, {Zoccali}, \&
  {Pietrzynski}}]{minn10}
{Minniti}, D., {Lucas}, P.~W., {Emerson}, J.~P., {et~al.} 2010, New Astronomy,
  15, 433, \dodoi{10.1016/j.newast.2009.12.002}

\bibitem[{{Molinari} {et~al.}(2011){Molinari}, {Schisano}, {Faustini},
  {Pestalozzi}, {di Giorgio}, \& {Liu}}]{mol11}
{Molinari}, S., {Schisano}, E., {Faustini}, F., {et~al.} 2011, \aap, 530, A133,
  \dodoi{10.1051/0004-6361/201014752}

\bibitem[{{Molinari} {et~al.}(2010){Molinari}, {Swinyard}, {Bally}, {Barlow},
  {Bernard}, {Martin}, {Moore}, {Noriega-Crespo}, {Plume}, {Testi}, {Zavagno},
  {Abergel}, {Ali}, {Anderson}, {Andr{\'e}}, {Baluteau}, {Battersby},
  {Beltr{\'a}n}, {Benedettini}, {Billot}, {Blommaert}, {Bontemps}, {Boulanger},
  {Brand}, {Brunt}, {Burton}, {Calzoletti}, {Carey}, {Caselli}, {Cesaroni},
  {Cernicharo}, {Chakrabarti}, {Chrysostomou}, {Cohen}, {Compiegne}, {de
  Bernardis}, {de Gasperis}, {di Giorgio}, {Elia}, {Faustini}, {Flagey},
  {Fukui}, {Fuller}, {Ganga}, {Garcia-Lario}, {Glenn}, {Goldsmith}, {Griffin},
  {Hoare}, {Huang}, {Ikhenaode}, {Joblin}, {Joncas}, {Juvela}, {Kirk},
  {Lagache}, {Li}, {Lim}, {Lord}, {Marengo}, {Marshall}, {Masi}, {Massi},
  {Matsuura}, {Minier}, {Miville-Desch{\^e}nes}, {Montier}, {Morgan}, {Motte},
  {Mottram}, {M{\"u}ller}, {Natoli}, {Neves}, {Olmi}, {Paladini}, {Paradis},
  {Parsons}, {Peretto}, {Pestalozzi}, {Pezzuto}, {Piacentini}, {Piazzo},
  {Polychroni}, {Pomar{\`e}s}, {Popescu}, {Reach}, {Ristorcelli}, {Robitaille},
  {Robitaille}, {Rod{\'o}n}, {Roy}, {Royer}, {Russeil}, {Saraceno}, {Sauvage},
  {Schilke}, {Schisano}, {Schneider}, {Schuller}, {Schulz}, {Sibthorpe},
  {Smith}, {Smith}, {Spinoglio}, {Stamatellos}, {Strafella}, {Stringfellow},
  {Sturm}, {Taylor}, {Thompson}, {Traficante}, {Tuffs}, {Umana}, {Valenziano},
  {Vavrek}, {Veneziani}, {Viti}, {Waelkens}, {Ward-Thompson}, {White},
  {Wilcock}, {Wyrowski}, {Yorke}, \& {Zhang}}]{mol10}
{Molinari}, S., {Swinyard}, B., {Bally}, J., {et~al.} 2010, \aap, 518, L100,
  \dodoi{10.1051/0004-6361/201014659}

\bibitem[{{Morton}(2015)}]{mor15}
{Morton}, T.~D. 2015, {isochrones: Stellar model grid package}.
\newblock \doeprint{1503.010}

\bibitem[{{Motte} {et~al.}(2018){Motte}, {Bontemps}, \& {Louvet}}]{mot18}
{Motte}, F., {Bontemps}, S., \& {Louvet}, F. 2018, \araa, 56, 41,
  \dodoi{10.1146/annurev-astro-091916-055235}

\bibitem[{{Murray}(2009)}]{mur09}
{Murray}, N. 2009, \apj, 691, 946, \dodoi{10.1088/0004-637X/691/2/946}

\bibitem[{{Myers}(2009)}]{mye09}
{Myers}, P.~C. 2009, \apj, 700, 1609, \dodoi{10.1088/0004-637X/700/2/1609}

\bibitem[{{Naranjo-Romero} {et~al.}(2020){Naranjo-Romero},
  {V{\'a}zquez-Semadeni}, \& {Loughnane}}]{nar20}
{Naranjo-Romero}, R., {V{\'a}zquez-Semadeni}, E., \& {Loughnane}, R.~M. 2020,
  arXiv e-prints, arXiv:2012.12819.
\newblock \doarXiv{2012.12819}

\bibitem[{{Neichel} {et~al.}(2015){Neichel}, {Samal}, {Plana}, {Zavagno},
  {Bernard}, \& {Fusco}}]{nei15}
{Neichel}, B., {Samal}, M.~R., {Plana}, H., {et~al.} 2015, \aap, 576, A110,
  \dodoi{10.1051/0004-6361/201425464}

\bibitem[{{Oey} \& {Kennicutt}(1997)}]{oey97}
{Oey}, M.~S., \& {Kennicutt}, R.~C., J. 1997, \mnras, 291, 827,
  \dodoi{10.1093/mnras/291.4.827}

\bibitem[{{Panwar} {et~al.}(2018){Panwar}, {Pandey}, {Samal}, {Battinelli},
  {Ogura}, {Ojha}, {Chen}, \& {Singh}}]{2018AJ....155...44P}
{Panwar}, N., {Pandey}, A.~K., {Samal}, M.~R., {et~al.} 2018, \aj, 155, 44,
  \dodoi{10.3847/1538-3881/aa9f1b}

\bibitem[{{Panwar} {et~al.}(2019){Panwar}, {Samal}, {Pandey}, {Singh}, \&
  {Sharma}}]{pan19}
{Panwar}, N., {Samal}, M.~R., {Pandey}, A.~K., {Singh}, H.~P., \& {Sharma}, S.
  2019, \aj, 157, 112, \dodoi{10.3847/1538-3881/aafbe6}

\bibitem[{{Patten} {et~al.}(2006){Patten}, {Stauffer}, {Burrows}, {Marengo},
  {Hora}, {Luhman}, {Sonnett}, {Henry}, {Raghavan}, {Megeath}, {Liebert}, \&
  {Fazio}}]{pat06}
{Patten}, B.~M., {Stauffer}, J.~R., {Burrows}, A., {et~al.} 2006, \apj, 651,
  502, \dodoi{10.1086/507264}

\bibitem[{{Pecaut} \& {Mamajek}(2013)}]{pec13}
{Pecaut}, M.~J., \& {Mamajek}, E.~E. 2013, \apjs, 208, 9,
  \dodoi{10.1088/0067-0049/208/1/9}

\bibitem[{{Pellegrini} {et~al.}(2012){Pellegrini}, {Oey}, {Winkler}, {Points},
  {Smith}, {Jaskot}, \& {Zastrow}}]{pel12}
{Pellegrini}, E.~W., {Oey}, M.~S., {Winkler}, P.~F., {et~al.} 2012, \apj, 755,
  40, \dodoi{10.1088/0004-637X/755/1/40}

\bibitem[{{Peretto} {et~al.}(2014){Peretto}, {Fuller}, {Andr{\'e}},
  {Arzoumanian}, {Rivilla}, {Bardeau}, {Duarte Puertas}, {Guzman Fernandez},
  {Lenfestey}, {Li}, {Olguin}, {R{\"o}ck}, {de Villiers}, \&
  {Williams}}]{pere14}
{Peretto}, N., {Fuller}, G.~A., {Andr{\'e}}, P., {et~al.} 2014, \aap, 561, A83,
  \dodoi{10.1051/0004-6361/201322172}

\bibitem[{{Pfalzner} \& {Kaczmarek}(2013)}]{pfa13}
{Pfalzner}, S., \& {Kaczmarek}, T. 2013, \aap, 559, A38,
  \dodoi{10.1051/0004-6361/201322134}

\bibitem[{{Pfalzner} {et~al.}(2016){Pfalzner}, {Kirk}, {Sills}, {Urquhart},
  {Kauffmann}, {Kuhn}, {Bhandare}, \& {Menten}}]{pfa16}
{Pfalzner}, S., {Kirk}, H., {Sills}, A., {et~al.} 2016, \aap, 586, A68,
  \dodoi{10.1051/0004-6361/201527449}

\bibitem[{{Pilbratt} {et~al.}(2010){Pilbratt}, {Riedinger}, {Passvogel},
  {Crone}, {Doyle}, {Gageur}, {Heras}, {Jewell}, {Metcalfe}, {Ott}, \&
  {Schmidt}}]{2010A&A...518L...1P}
{Pilbratt}, G.~L., {Riedinger}, J.~R., {Passvogel}, T., {et~al.} 2010, \aap,
  518, L1, \dodoi{10.1051/0004-6361/201014759}

\bibitem[{{Pon} {et~al.}(2011){Pon}, {Johnstone}, \& {Heitsch}}]{pon11}
{Pon}, A., {Johnstone}, D., \& {Heitsch}, F. 2011, \apj, 740, 88,
  \dodoi{10.1088/0004-637X/740/2/88}

\bibitem[{{Preibisch} \& {Mamajek}(2008)}]{pre08}
{Preibisch}, T., \& {Mamajek}, E. 2008, {The Nearest OB Association:
  Scorpius-Centaurus (Sco OB2)}, ed. B.~{Reipurth}, Vol.~5, 235

\bibitem[{{Rieke} \& {Lebofsky}(1985)}]{rie85}
{Rieke}, G.~H., \& {Lebofsky}, M.~J. 1985, \apj, 288, 618,
  \dodoi{10.1086/162827}

\bibitem[{{Robitaille} \& {Bressert}(2012)}]{rob12}
{Robitaille}, T., \& {Bressert}, E. 2012, {APLpy: Astronomical Plotting Library
  in Python}.
\newblock \doeprint{1208.017}

\bibitem[{{Rubin}(1968)}]{rub68}
{Rubin}, R.~H. 1968, \apj, 154, 391, \dodoi{10.1086/149766}

\bibitem[{{Ryabukhina} {et~al.}(2018){Ryabukhina}, {Zinchenko}, {Samal},
  {Zemlyanukha}, {Ladeyschikov}, {Sobolev}, {Henkel}, \& {Ojha}}]{rya18}
{Ryabukhina}, O.~L., {Zinchenko}, I.~I., {Samal}, M.~R., {et~al.} 2018,
  Research in Astronomy and Astrophysics, 18, 095,
  \dodoi{10.1088/1674-4527/18/8/95}

\bibitem[{{Salpeter}(1955)}]{sal55}
{Salpeter}, E.~E. 1955, \apj, 121, 161, \dodoi{10.1086/145971}

\bibitem[{{Samal} {et~al.}(2018){Samal}, {Deharveng}, {Zavagno}, {Anderson},
  {Molinari}, \& {Russeil}}]{sam18}
{Samal}, M.~R., {Deharveng}, L., {Zavagno}, A., {et~al.} 2018, \aap, 617, A67,
  \dodoi{10.1051/0004-6361/201833015}

\bibitem[{{Samal} {et~al.}(2007){Samal}, {Pandey}, {Ojha}, {Ghosh}, {Kulkarni},
  \& {Bhatt}}]{sam07}
{Samal}, M.~R., {Pandey}, A.~K., {Ojha}, D.~K., {et~al.} 2007, \apj, 671, 555,
  \dodoi{10.1086/522941}

\bibitem[{{Samal} {et~al.}(2010){Samal}, {Pandey}, {Ojha}, {Ghosh}, {Kulkarni},
  {Kusakabe}, {Tamura}, {Bhatt}, {Thompson}, \& {Sagar}}]{sam10}
---. 2010, \apj, 714, 1015, \dodoi{10.1088/0004-637X/714/2/1015}

\bibitem[{{Samal} {et~al.}(2015){Samal}, {Ojha}, {Jose}, {Zavagno},
  {Takahashi}, {Neichel}, {Kim}, {Chauhan}, {Pandey}, {Zinchenko}, {Tamura}, \&
  {Ghosh}}]{sam15}
{Samal}, M.~R., {Ojha}, D.~K., {Jose}, J., {et~al.} 2015, \aap, 581, A5,
  \dodoi{10.1051/0004-6361/201322787}

\bibitem[{{Schneider} {et~al.}(2012){Schneider}, {Csengeri}, {Hennemann},
  {Motte}, {Didelon}, {Federrath}, {Bontemps}, {Di Francesco}, {Arzoumanian},
  {Minier}, {Andr{\'e}}, {Hill}, {Zavagno}, {Nguyen-Luong}, {Attard},
  {Bernard}, {Elia}, {Fallscheer}, {Griffin}, {Kirk}, {Klessen}, {K{\"o}nyves},
  {Martin}, {Men'shchikov}, {Palmeirim}, {Peretto}, {Pestalozzi}, {Russeil},
  {Sadavoy}, {Sousbie}, {Testi}, {Tremblin}, {Ward-Thompson}, \&
  {White}}]{sch12}
{Schneider}, N., {Csengeri}, T., {Hennemann}, M., {et~al.} 2012, \aap, 540,
  L11, \dodoi{10.1051/0004-6361/201118566}

\bibitem[{{Sharpless}(1959)}]{sha59}
{Sharpless}, S. 1959, \apjs, 4, 257, \dodoi{10.1086/190049}

\bibitem[{{Sills} {et~al.}(2018){Sills}, {Rieder}, {Scora}, {McCloskey}, \&
  {Jaffa}}]{sills18}
{Sills}, A., {Rieder}, S., {Scora}, J., {McCloskey}, J., \& {Jaffa}, S. 2018,
  \mnras, 477, 1903, \dodoi{10.1093/mnras/sty681}

\bibitem[{{Smith} {et~al.}(2010){Smith}, {Povich}, {Whitney}, {Churchwell},
  {Babler}, {Meade}, {Bally}, {Gehrz}, {Robitaille}, \& {Stassun}}]{smi10}
{Smith}, N., {Povich}, M.~S., {Whitney}, B.~A., {et~al.} 2010, \mnras, 406,
  952, \dodoi{10.1111/j.1365-2966.2010.16792.x}

\bibitem[{{Smithsonian Astrophysical Observatory}(2000)}]{har00}
{Smithsonian Astrophysical Observatory}. 2000, {SAOImage DS9: A utility for
  displaying astronomical images in the X11 window environment}.
\newblock \doeprint{0003.002}

\bibitem[{{Sokolov} {et~al.}(2017){Sokolov}, {Wang}, {Pineda}, {Caselli},
  {Henshaw}, {Tan}, {Fontani}, {Jim{\'e}nez-Serra}, \& {Lim}}]{soko17}
{Sokolov}, V., {Wang}, K., {Pineda}, J.~E., {et~al.} 2017, \aap, 606, A133,
  \dodoi{10.1051/0004-6361/201630350}

\bibitem[{{Swarup} {et~al.}(1991){Swarup}, {Ananthakrishnan}, {Kapahi}, {Rao},
  {Subrahmanya}, \& {Kulkarni}}]{swa91}
{Swarup}, G., {Ananthakrishnan}, S., {Kapahi}, V.~K., {et~al.} 1991, Current
  Science, 60, 95

\bibitem[{{Tenorio-Tagle}(1979)}]{ten79}
{Tenorio-Tagle}, G. 1979, \aap, 71, 59

\bibitem[{{Tody}(1986)}]{tod86}
{Tody}, D. 1986, in Society of Photo-Optical Instrumentation Engineers (SPIE)
  Conference Series, Vol. 627, Instrumentation in astronomy VI, ed. D.~L.
  {Crawford}, 733, \dodoi{10.1117/12.968154}

\bibitem[{{Tody}(1993)}]{tod93}
{Tody}, D. 1993, in Astronomical Society of the Pacific Conference Series,
  Vol.~52, Astronomical Data Analysis Software and Systems II, ed. R.~J.
  {Hanisch}, R.~J.~V. {Brissenden}, \& J.~{Barnes}, 173

\bibitem[{{van Moorsel} {et~al.}(1996){van Moorsel}, {Kemball}, \&
  {Greisen}}]{van96}
{van Moorsel}, G., {Kemball}, A., \& {Greisen}, E. 1996, in Astronomical
  Society of the Pacific Conference Series, Vol. 101, Astronomical Data
  Analysis Software and Systems V, ed. G.~H. {Jacoby} \& J.~{Barnes}, 37

\bibitem[{{V{\'a}zquez-Semadeni} {et~al.}(2019){V{\'a}zquez-Semadeni}, {Palau},
  {Ballesteros-Paredes}, {G{\'o}mez}, \& {Zamora-Avil{\'e}s}}]{vaz19}
{V{\'a}zquez-Semadeni}, E., {Palau}, A., {Ballesteros-Paredes}, J.,
  {G{\'o}mez}, G.~C., \& {Zamora-Avil{\'e}s}, M. 2019, \mnras, 490, 3061,
  \dodoi{10.1093/mnras/stz2736}

\bibitem[{{Verschuur} \& {Kellermann}(1988)}]{ver88}
{Verschuur}, G.~L., \& {Kellermann}, K.~I. 1988, {Galactic and extra-galactic
  radio astronomy} (Springer)

\bibitem[{{Walborn} \& {Fitzpatrick}(1990)}]{1990PASP..102..379W}
{Walborn}, N.~R., \& {Fitzpatrick}, E.~L. 1990, \pasp, 102, 379,
  \dodoi{10.1086/132646}

\bibitem[{{Walch} {et~al.}(2015){Walch}, {Whitworth}, {Bisbas}, {Hubber}, \&
  {W{\"u}nsch}}]{wal15}
{Walch}, S., {Whitworth}, A.~P., {Bisbas}, T.~G., {Hubber}, D.~A., \&
  {W{\"u}nsch}, R. 2015, \mnras, 452, 2794, \dodoi{10.1093/mnras/stv1427}

\bibitem[{{Weidner} {et~al.}(2010){Weidner}, {Kroupa}, \&
  {Bonnell}}]{2010MNRAS.401..275W}
{Weidner}, C., {Kroupa}, P., \& {Bonnell}, I.~A.~D. 2010, \mnras, 401, 275,
  \dodoi{10.1111/j.1365-2966.2009.15633.x}

\bibitem[{{Weidner} {et~al.}(2007){Weidner}, {Kroupa}, {N{\"u}rnberger}, \&
  {Sterzik}}]{wei07}
{Weidner}, C., {Kroupa}, P., {N{\"u}rnberger}, D.~E.~A., \& {Sterzik}, M.~F.
  2007, \mnras, 376, 1879, \dodoi{10.1111/j.1365-2966.2007.11580.x}

\bibitem[{{Whitney} \& {GLIMPSE360 Team}(2009)}]{2009AAS...21421001W}
{Whitney}, B., \& {GLIMPSE360 Team}. 2009, in American Astronomical Society
  Meeting Abstracts, Vol. 214, American Astronomical Society Meeting Abstracts
  \#214, 210.01

\bibitem[{{Williams} \& {Cieza}(2011)}]{will11}
{Williams}, J.~P., \& {Cieza}, L.~A. 2011, \araa, 49, 67,
  \dodoi{10.1146/annurev-astro-081710-102548}

\bibitem[{{Wright} {et~al.}(2010){Wright}, {Eisenhardt}, {Mainzer}, {Ressler},
  {Cutri}, {Jarrett}, {Kirkpatrick}, {Padgett}, {McMillan}, {Skrutskie},
  {Stanford}, {Cohen}, {Walker}, {Mather}, {Leisawitz}, {Gautier}, {McLean},
  {Benford}, {Lonsdale}, {Blain}, {Mendez}, {Irace}, {Duval}, {Liu}, {Royer},
  {Heinrichsen}, {Howard}, {Shannon}, {Kendall}, {Walsh}, {Larsen}, {Cardon},
  {Schick}, {Schwalm}, {Abid}, {Fabinsky}, {Naes}, \& {Tsai}}]{wis10}
{Wright}, E.~L., {Eisenhardt}, P. R.~M., {Mainzer}, A.~K., {et~al.} 2010, \aj,
  140, 1868, \dodoi{10.1088/0004-6256/140/6/1868}

\bibitem[{{Yadav} {et~al.}(2016){Yadav}, {Pandey}, {Sharma}, {Ojha}, {Samal},
  {Mallick}, {Jose}, {Ogura}, {Richichi}, {Irawati}, {Kobayashi}, \&
  {Eswaraiah}}]{yad16}
{Yadav}, R.~K., {Pandey}, A.~K., {Sharma}, S., {et~al.} 2016, \mnras, 461,
  2502, \dodoi{10.1093/mnras/stw1356}

\bibitem[{{Yorke} {et~al.}(1982){Yorke}, {Bodenheimer}, \&
  {Tenorio-Tagle}}]{yor82}
{Yorke}, H.~W., {Bodenheimer}, P., \& {Tenorio-Tagle}, G. 1982, \aap, 108, 25

\bibitem[{{Yu} {et~al.}(2018){Yu}, {Yu}, {Lee}, {Lin}, {Hsia}, {Chang}, {Chen},
  {Ngeow}, {Ip}, {Chen}, {Laher}, {Surace}, \& {Kulkarni}}]{yu18}
{Yu}, P.-C., {Yu}, C.-H., {Lee}, C.-D., {et~al.} 2018, \aj, 155, 91,
  \dodoi{10.3847/1538-3881/aaa45b}

\bibitem[{{Yuan} {et~al.}(2020){Yuan}, {Li}, {Zhu}, {Liu}, {Wang}, {Liu},
  {Kim}, {Tatematsu}, {Yuan}, \& {Wu}}]{yuan20}
{Yuan}, L., {Li}, G.-X., {Zhu}, M., {et~al.} 2020, \aap, 637, A67,
  \dodoi{10.1051/0004-6361/201936625}

\end{thebibliography}
\bibliographystyle{aasjournal}

\end{document}